\documentclass{aa}  

\usepackage{graphicx}
\usepackage{txfonts}
\usepackage{xcolor}

\begin{document}

   \title{Rocky planet formation in compact disks around M dwarfs}

\titlerunning{Rocky planet formation around M dwarfs}

   \author{M. Sanchez,
          \inst{1}
          N. van der Marel,\inst{1}
          M. Lambrechts,\inst{2}
          G. D. Mulders,\inst{3,4}
          O. M. Guerra-Alvarado \inst{1}
          }

   \institute{Leiden Observatory, Leiden University, P.O. Box 9513, 2300 RA Leiden,
The Netherlands\\
           \email{msanchez@strw.leidenuniv.nl}
         \and
             Center for Star and Planet Formation, GLOBE Institute, University of Copenhagen, Øster Voldgade 5–7, 1350 Copenhagen, Denmark
         \and
             Facultad de Ingeniería y Ciencias, Universidad Adolfo Ibáñez, Av. Diagonal las Torres 2640, Peñalolén, Santiago, Chile
          \and
              Millennium Institute for Astrophysics, Chile
             }

   \date{}

 
  \abstract
   {Due to the improvements in radial velocity and transit techniques, we know that rocky or rocky-icy planets, in particular close-in super-Earths in compact configurations, are the most common ones around M dwarfs. On the other hand, thanks to the high angular resolution of ALMA we know that many disks around very low-mass stars (between 0.1 and 0.5 M$_\odot$) are rather compact and small (without observable substructures and radius less than 20 au), which favors the idea of an efficient radial drift that could enhance planet formation in the terrestrial zone.}
   {Our aim was to investigate the potential formation paths of the observed close-in rocky exoplanet population around M dwarfs, especially close-in super-Earths, assuming that planet formation could take place in compact disks with an efficient dust radial drift.}
   {We developed N-body simulations that include a sample of embryos growing by pebble accretion exposed to planet-disk interactions, star-planet tidal interactions, and general relativistic corrections that include the evolution of the luminosity, radius, and rotational period of the star. For a star of 0.1 M$_\odot$, we considered different gas disk viscosities and initial embryo distributions. We also explored planet formation by pebble accretion around stars of 0.3 M$_\odot$ and 0.5 M$_\odot$. Lastly, for each stellar mass, we ran simulations that include a sample of embryos growing by planetesimal accretion instead of pebble accretion.}
   {Our main result is that the sample of simulated planets that grow by pebble accretion in a gas disk with low viscosity ($\alpha=10^{-4}$) can reproduce the close-in low-mass exoplanet population around M dwarfs in terms of multiplicity, mass, and semi-major axis. Furthermore, we found that a gas disk with high viscosity ($\alpha=10^{-3}$), and thus lower pebble accretion rates, cannot reproduce the observed planet masses as no planet more massive than 0.5 M$_\oplus$ could be formed in our simulations. In addition, we show that planetesimal accretion favors the formation of smaller planets than pebble accretion does. Whether this planet population truly exists remains unknown with the current instrumental sensitivity.}
   {Rocky planet formation around M dwarfs can take place in compact and small dust disks driven by an efficient radial drift in a gas disk with low viscosity ($\alpha=10^{-4}$). This result points toward a new approach in the direction of the disk conditions needed for rocky planet formation around very low-mass stars.}

   \keywords{super-Earth formation --
                 very low mass stars --
                numerical simulations
               }

   \maketitle
%

\section{Introduction}

M dwarfs are the most abundant objects in our solar neighborhood, representing almost 70$\%$ of the stars in our Galaxy \citep{Chabrier2003}. Thanks to the improvements in radial velocity and transit detection techniques, we know that they are hosts to closely packed planetary systems with multiple rocky planets \citep[e.g.,][]{Shields2016,Mulders2018}. To date, rocky planets have been found to be the most abundant around M dwarfs \citep{Sabotta2021}, frequently found close in to the star \citep{Ment2023}. These planets are of great interest as many of them are located in the habitable zone of their systems \citep[e.g.,][]{Kasting1993,Kopparapu2013}, which makes M dwarfs ideal targets for the search for life in the Universe. 
\\

Even though several authors have addressed the study of Earth-like planets around M dwarfs with both population synthesis models and N-body simulations \cite[e.g.,][]{Dugaro2016,Coleman2019,Miguel2019,Liu2020,Sanchez2022}, the formation of super-Earths (planets with sizes between Earth and Neptune) around M dwarfs is still under debate; it is challenging to form such planets in situ with the classic planet formation theory and based on the low disk masses typically derived from observations \citep[e.g.,][]{Raymond2007,Ciesla2015,Schli2018}. Thus, it is necessary to address the problem taking into account different premises. For example, \cite{Liu2019} developed a pebble-driven core accretion model in a population synthesis code to study the formation and evolution of planets around stars from the substellar mass limit at 0.08 M$_{\odot}$ up to 1 M$_\odot$. They found that the characteristic mass of a super-Earth is set by the pebble isolation mass, and that it increases linearly with the mass of the stellar host, and that planets with masses lower than 20 M$_\oplus$ can be formed around stars with a wide range of metallicities.However, they did not consider gravitational interactions among the sample of embryos, which is of great importance as collisions among embryos play a primary role in defining the final masses and orbital configurations of planetary systems \citep[e.g.,][]{Chambers2001, Obrien2006,Raymond2007,deElia2013,Ronco2018,Sanchez2022}. Thus, their simulations may not be realistic when compared to the final exoplanet properties. Other studies, such as \cite{Mulders2021}, proposed a pebble-drift and accretion model to explain the occurrence rate of transiting planets around M dwarfs. They found that the fraction of close-in super-Earths is higher around lower-mass stars and matches the exoplanet occurrence rates from \textit{Kepler}. However, their results are based on the study of a two-planet system, and they do not take into account planet-disk interactions. The inclusion of gas disk interactions with a sample of embryos is crucial, as is defining the early dynamical evolution and orbital configuration of planetary systems \cite[e.g.,][]{Tanaka2004,Paardekooer2011,Ida2020}. Moreover, the inclusion of star-planet tidal interactions is also relevant in the long-term dynamical evolution of close-in planets around very low-mass stars due to the fast evolution of the stellar radius and rotation period \citep[e.g.,][]{Bolmont2011,Sanchez2020}.\\

Generally, the assumptions regarding the disk properties in planet population models may not be representative when compared with the latest insights from disk observations. Recently, the Atacama Large
Millimeter/sub-millimeter Array (ALMA) disk surveys have shown the existence of compact dust disks around very low-mass stars \citep[e.g.,][]{Facchini2019,Kurtovic2021,Vandermarel2022}. A handful of disks are truly as small as 3-5 au in radius, whereas many of these disks remain unresolved in current ALMA images and their radius is only constrained to be $\lesssim$20 au. Such compact disks, lacking large-scale substructures, are more often found around low-mass stars than high-mass stars \citep{VanderMarel2021} and are expected to be dominated by radial drift and thus a high pebble flux \citep{Pinilla2013,Banzatti2020,Banzatti2023}. In such disks, 
super-Earths are expected to form, consistent with a formation through pebble accretion in drift-dominated disks \citep{Mulders2021,VanderMarel2021}. Recently, \cite{Chachan2023} used a simple framework that combines the theory of pebble accretion with the measurements of dust masses in protoplanetary disks around late M dwarfs. They demonstrated that the
fraction of disks that can create inner planetary cores around such stars, is in agreement with the observed enhancement in the occurrence
rate of inner super-Earths around M dwarfs compared to
FGK dwarfs \citep[e.g.,][]{Dressing2015,Mulders2015}, in agreement with \cite{VanderMarel2021}. Additionally, by modeling the evolution of a stellar cluster, \cite{Appelgren2023} investigated the efficiency of radial drift in clearing the disk of millimeter-sized particles. Throughout the disk's lifetime, they found an agreement between the mass locked in pebbles and the evolution of dust masses and ages inferred from nearby star-forming regions. They stated that disks are born with dust masses of up to around 50 M$_\oplus$ for low-mass stars of 0.1 M$_\odot$, and quickly decreases to $\sim$1 M$_{\oplus}$ at 1-2 Myr due to radial drift, which suggests that at the earliest stages there could be enough material to form inner planetary cores. Such high disk dust masses have indeed been inferred for embedded disks in the Class I stage \citep{Tychoniec2020}.\\

Motivated by the fact that the formation of planets more massive than the Earth is hard to explain with classical planet formation models, and that dust disks around very low-mass stars are rather small and compact, dominated by radial drift and with an initial dust mass high enough to form inner planetary cores, we investigated different rocky planet formation paths around M dwarfs. In order to do that, we assumed that planet formation takes place in compact dust disks, dominated by radial drift rather than the well-studied gapped disks \citep[e.g.,][]{Andrews2018} where dust traps are thought to halt the radial drift \citep{Pinilla2012}. Based on this assumption, we developed different sets of N-body simulations that include a sample of small embryos growing by pebble accretion that experience planet-disk interactions as well as star-planet tidal interactions that incorporate the evolution of the luminosity, radius and rotational period of the star. All simulations run for 50 Myr to take into account late collisions among embryos, considering a gas disk lifetime of 10 Myr. We explored different planet formation scenarios around stars of 0.1~M$_\odot$, 0.3~M$_\odot$ and 0.5~M$_\odot$. 


This work is structured as follows. In Section 2 we describe the dust disk model used based on compact dust disk observations. In Section 3 we describe the planet formation model included in the N-body code, and present different formation scenarios. Appendix A and B summarize the details of the gas disk evolution model and external forces on the N-body code. In Section 4 we show the resulting simulated planetary systems and explain their orbital history and final configurations. In Section 5 we compare the simulated planets with the confirmed close-in low-mass exoplanet population around M dwarfs. Finally, in Section 6 we summarize some points of discussion and in Section 7 we give the main conclusions of our work.

  .

\section{Compact dust disks: Models and observations}

\begin{figure}[!b]
    \centering
    \includegraphics[width=7.5cm]{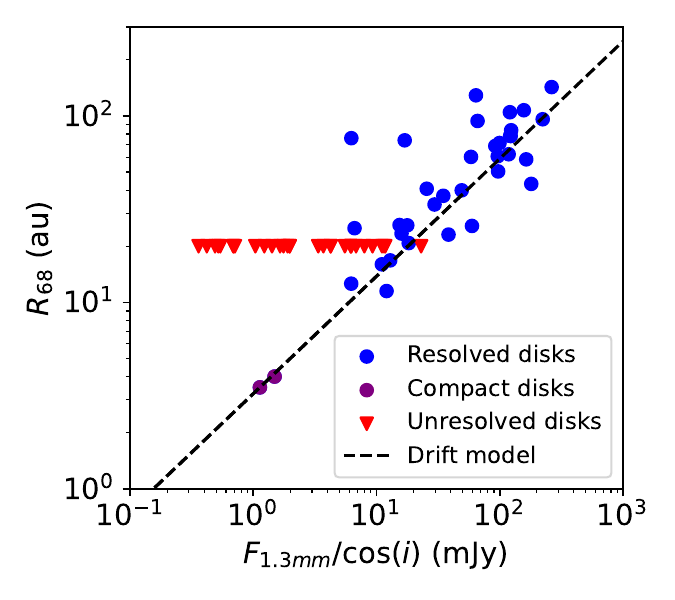}
    \caption{Size-luminosity diagram of all Class II disks in Lupus, based on data from \citet{Ansdell2018}, \citet{Tazzari2021}, and \citet{Vandermarel2022} (disks around M dwarfs, present fluxes less than $\sim 90$~mJy). The $R_{68}$ parameter refers to the size of the dust disk, which is typically measured with a curve-of-growth method, encircling 68\% of the total flux \citep{Hendler2020}. $F_{1.3mm}$ is the observed flux and $i$ represent the inclination of the disks with respect to the observer. The expected trend for drift-dominated disk models from \citet{Rosotti2019} is shown with a dashed line. Other than two resolved compact disks of 3-5 au radius from high-resolution observations in \citet{Vandermarel2022} (purple dots), the majority of compact disks are unresolved with an upper limit of $<$20 au on their radius (red triangles). Considering the trend in the plot as well as the predictions from drift models, in reality most of these disks are expected to be 2-10 au in radius.}
    \label{fig:size-lum}
\end{figure}

Current ALMA observations of protoplanetary disks around M dwarfs at $\sim$1-2 Myr primarily show compact dust continuum disks that remain unresolved at $\sim$0.2" resolution, which means that their dust disk radii are $<$20 au \citep{Ansdell2018}. For a handful of these compact disks, high-resolution images were taken, resolving very small compact disks with a radius of only 3-5 au \citep{Vandermarel2022}. This is also seen in compact disks in Taurus \citep{Kurtovic2021}. It is currently unclear whether all unresolved disks are equally compact, and actual sizes may have a range between 2-20 au. However, the dust disk fluxes at 1.3mm in Lupus of these unresolved disks range between 0.4-12 mJy \citep{Ansdell2018}. An extrapolation of the size-luminosity relation in Lupus predicts dust disk radii between 2-10 au for this flux range \citep[][]{Hendler2020,Tazzari2021}, which is similar to the range predicted by drift-dominated disk models \citep[][Appelgren et al., subm.]{Rosotti2019}, as demonstrated in Figure \ref{fig:size-lum}. Dust evolution models also predict a decrease in observable dust mass of at least one order of magnitude in 1 Myr in disks without pressure bumps \citep{Pinilla2020,Appelgren2023} due to radial drift (i.e., by a decrease in the dust disk size).

Based on these findings, 
we make the assumption that due to efficient radial drift, a compact dust disk of 20 au, or larger, could be reduced to just a few au in size in 
1-2 Myr. This is possibly the moment that the disks are observed in ALMA disk surveys, and considering the Lupus fluxes, they are expected to be just a few au in size at 1-2 Myr of age. Recent high-resolution ALMA observations of a larger disk sample appear to confirm this trend (Guerra-Alvarado et al. in prep.).

\subsection{Pebble mass inventory in observed compact protoplanetary disks}
\label{sec:dustobservations}

In this work, we assume that planet formation takes place in compact dust disks of $<$20 au radius. This means that we have to review carefully what is known about the dust masses in these disks to be able to use a representative pebble mass content for our pebble accretion model. In most observational disk studies, the dust mass is derived directly from the integrated millimeter flux at $\sim$1 mm, using ALMA observations \citep[e.g.,][]{Ansdell2018,Manara2023}. The dust mass is derived under the assumption of optically thin emission and a constant dust temperature of 20 K, using the following relation:

 \begin{eqnarray}
M_{\rm dust}=\frac{F_{\nu}d^{2}}{\kappa_{\nu}B_{\nu}(T_{\rm dust})} \approx 2.03 \times 10^{-6}\left(\frac{d}{150 {\rm pc}}\right)^{2}F_{1.3mm}.
 \label{eq: Masses at submm}
 \end{eqnarray}
Here the dust grain opacity $\kappa_{\nu}$ is 2.3 cm$^{2}$g$^{-1}$, B$_{\nu}$(T$_{\rm dust}$) is the Planck function for a dust temperature of 20 K, and $d$ is the distance in pc. This dust mass relies on several assumptions which may not be representative for compact disks, as emission is likely more optically thick and the dust temperature is higher as the dust content may be located much closer to the star  \citep{vanderMarelPinilla2023}. Therefore, one cannot simply use such dust masses from the literature as actual measurements of the available pebble mass in planet formation models. One of the main problems is that for the majority of these disks, the ALMA images remain spatially unresolved at moderate resolution of 0.2" of disk surveys, so that only an upper limit on the dust disk radius can be derived, which means that optical depth and dust temperature remain uncertain. The handful of compact disks imaged at high angular resolution ($\sim$0.05") do reveal dust disk radii as small as $\sim$3-5 au \citep{Facchini2019,Kurtovic2021,Vandermarel2022}. As explained above, we assume that the majority of unresolved dust disks in Lupus are similarly small.

Rather than attempting to derive more accurate dust masses from unresolved disk images, we use the available integrated millimeter fluxes and the limits on their dust disk size for the compact disks in Lupus around low-mass stars \citep{Ansdell2018,Vandermarel2022} to compare with the expected millimeter flux of a forward model of the simulated disks in this study, to demonstrate that our pebble mass inventory is representative for observed disks at 1-2 Myr. We emphasize that dust disks in Lupus (which are considered to be $\sim$1-2 Myr old) have likely already experienced significant mass loss due to radial drift \citep{2020A&A...635A.105P,Appelgren2023} and the initial disk dust masses and dust disk radii right after disk formation are expected to be much larger \citep{Appelgren2024}. Therefore, we assume that the incoming pebble flux originates from a much larger, initial dust disk of 20 au radius, whereas the dust disk of the simulation is only 2-4 au in radius, which is the disk region where the embryos for the planet formation model will be distributed (see Section \ref{sec:simulations}), as a snapshot at 1-2 Myr where pebble drift has already taken place. This means that we forward model the expected fluxes along an evolutionary track of dust disks that are just formed and then quickly decrease in mass and size as the result of radial drift over 1-2 Myr, in two steps. 

Using the pebble density profile in our pebble model (see Section \ref{sec:pebblemodel}), we conducted radiative transfer models at a wavelength of 1.3 mm using RADMC-3D software \citep{2012ascl.soft02015D}. This analysis was performed at an evolutionary stage of 1 Myr.

We utilized a generic protoplanetary disk model in RADMC-3D, configuring the surface density of the dust disk to match the pebble surface density in our model (see Section \ref{sec:dustmodel}). Within the model, we used stellar masses of 0.1 M$_\odot$, 0.3 M$_\odot$, and 0.5 M$_\odot$, accompanied by luminosities of 0.067 L$_\odot$, 0.33 L$_\odot$, and 0.65 L$_\odot$ \citep{Baraffe2015}, and assumed dust outer radii $r_{\rm out}$ of 2 au, 3 au, and 4 au, respectively for the very compact dust disk of the simulation. These luminosities, in conjunction with effective temperatures derived from \citet{2017A&A...600A..20A} corresponding to the stellar masses (3125 K, 3270 K, and 3342 K, respectively), determined both the stellar radius and the inner dust disk radius, assumed to be equal to the dust sublimation radius (T=1500 K). The dust opacities were computed using the \texttt{optool} package \citep{2021ascl.soft04010D}, employing DSHARP dust particle opacities and setting the minimum particle size to a$_{min}$ as 0.050 $\mu$m and the maximum size as a$_{max}$ as 3 mm \citep{Birnstiel2018}. Anisotropic scattering was assumed within RADMC-3D to enhance the fidelity of reproducing observed emissions from compact disks at 1.3 mm. 

The volumetric density distribution in RADMC-3D is defined as:

 \begin{eqnarray}
 \rho(r,z) =\frac{\Sigma_{\rm{dust}}(r)}{H_{d}\sqrt{2\pi}}\exp\left(-\frac{z^{2}}{2H^{2}_{d}}\right),
 \label{eq: Density distribution}
 \end{eqnarray}

\noindent where $r$ is the radial distance to the star from the disk, $H_{d}$(r) is the dust scale height of the disk, and $\Sigma_{\rm dust}(r)$ is the dust surface density, defined as

 \begin{eqnarray}
\Sigma_{\rm dust}(r)= \Sigma_{0}\left(\frac{r}{r_{\rm out}}\right)^{-p},
 \label{eq: Surface Density distribution}
 \end{eqnarray}
where $p$ is the power law index and $\Sigma_0=M_{\rm dust} /[ 2 \pi (2-p)  r_{\rm out}^{2}]$, with $M_{\rm dust}$ and $r_{\rm out}$ the dust mass and outer radius of the dust disk, respectively. The index $p$ is selected to match the model pebble density profile from Section \ref{sec:dustmodel} in Eq. \ref{eq:pebdensity-sim}. 
The input dust mass is then derived by integrating the profile out to an outer dust radius of 2, 3 or 4 au as explained above for stellar masses of 0.1, 0.3, and 0.5 M$_{\odot}$, resulting in dust masses of 1.5, 10, and 12 M$_{\oplus}$, respectively. For such compact disks, at the end of the radial drift process, the index $p=0.3$. To derive the expected millimeter flux, we set a distance of 160 parsecs, similar to the Lupus star-forming region.

In addition, to represent the initial dust disks, three radiative transfer models with higher dust masses and larger dust radii of 20 au are computed, one for each stellar mass, to illustrate the evolutionary trajectory of the dust influenced by inward radial drift. The initial disk dust masses are considered to be representative of the incoming pebble flux. In this case, the density profile is characterized by $p=0.6$. 
The resulting fluxes of both sets of modeled disks are shown in Table \ref{table:Disk models masses} with both their actual dust masses (input dust mass) and the 'observable' dust mass, which is computed using Eq. \ref{eq: Masses at submm} directly from the simulated integrated flux. The latter is added to demonstrate that this simple dust mass derivation for a given flux from a very compact disk can lead to underestimates of the real dust mass by a factor 3-6. This discrepancy is anticipated as compact disks likely have higher dust temperatures exceeding 20 K and may be more optically thick.

\begin{table}
\caption{Disk model dust fluxes and masses.}\label{Disk masses}
\centering
\begin{tabular}{ccccc}
\hline \hline
Stellar mass & Radius & Flux$_{1.3 mm}$ & Input M$_{\rm dust}$  & Obs M$_{\rm dust}$\\
(M$_{\odot}$) & (au) &  (mJy) & (M$_{\oplus}$) & (M$_{\oplus}$)\\
\hline
\hline
0.1 & 2 & 0.59 & 1.5 & 0.45\\ 
0.1 & 20 & 9.9 & 15 & 7.6\\
0.3 & 3 & 1.8 & 10 &  1.4\\
0.3 & 20 & 23.3 & 50 & 17.9\\
0.5 & 4 & 3.2 & 12 & 2.5\\
0.5 & 20 & 30.6 & 85 & 23.5\\
\hline
 \label{table:Disk models masses}
\end{tabular}
\end{table}

\begin{figure} 
\includegraphics[width=8cm]{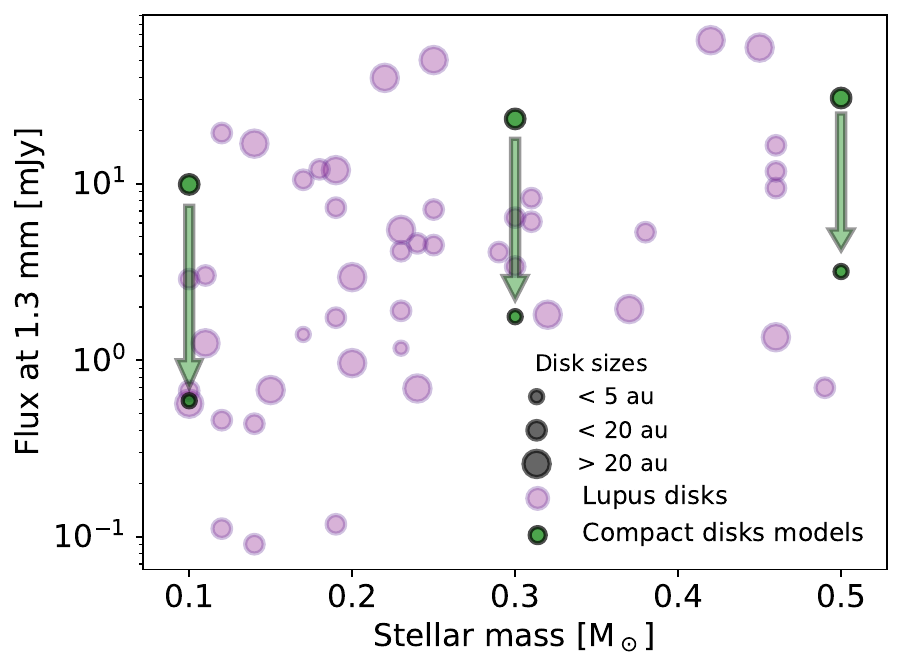}
    \caption{Fluxes at 1.3 mm for both observed Lupus disks with central star masses below 0.5 M$_{\odot}$ (purple), and RADMC-3D forward models (green) of our simulated pebble disks. Observed dust disk sizes in Lupus are categorized into three size groups: $<$5 au; either between 5 and 20 au or unresolved, and thus only constrained to be $<$20 au; and $>$20 au. The simulated disks show both the expected flux at sizes of 20 au, and 2-4 au for the dust masses listed in Table \ref{table:Disk models masses}.
    The green arrow highlights the hypothetical evolutionary trajectory that dust in a disk would follow under the influence of radial drift, 
    leading to more compact disks with lower flux densities.}
    \label{fig:Stellarmassvsmmflux}
\end{figure}

Figure \ref{fig:Stellarmassvsmmflux} shows all the available fluxes from the observed protoplanetary disks in Lupus \citep{Ansdell2018}, with stellar masses less than 0.5 M$_{\odot}$, for direct comparison with our models. 
As explained above, for the majority of these disks, the size can only be constrained to be $<$20 au, and therefore represent a range of dust disk sizes, and dust disk masses along the evolutionary trajectory of radial drift, considering their range of fluxes, and the size-luminosity relation (Figure \ref{fig:size-lum}).

It is evident from Figure \ref{fig:Stellarmassvsmmflux} that both disks with the initial pebble reservoir, and the ones with a more advanced dust evolutionary stage, fall within the typical flux range observed in Lupus disk observations. In the later evolutionary stage, the most outer planet may already be formed, but at orbital radii in the most inner part of the disk well within 1 au.
During the simulation, the embryos in this region of the disk will continue to accrete material until reaching the pebble isolation mass, at which point radial drift is halted outside the most exterior planet. This process may result in a very compact disk $<$3-5 au with one or more pressure bumps depending on the reached planet masses. Such a dust disk would have a significantly lower dust mass than the initial disk of 20 au, but both masses still fall within the range of unresolved disks seen in Lupus (Figure \ref{fig:Stellarmassvsmmflux}). 
These compact disks possibly constitute the majority of disk observations within the submillimeter range for stellar masses $< 0.5$ M$_{\odot}$.

\begin{figure} 
\includegraphics[width=8.8cm]{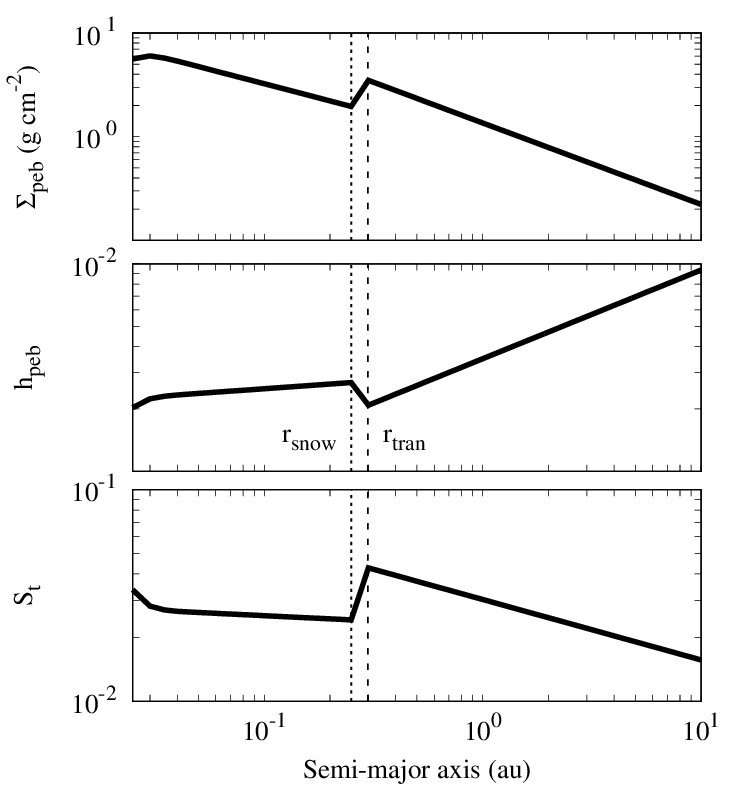}
    \caption{Pebble surface density (top panel), pebble aspect ratio (middle panel), and Stokes number (bottom panel) at 1 Myr, for our standard model in which we assumed a star is 0.1 M$_\odot$, and a turbulent parameter $\alpha_t=10^{-4}$. The transition line associated with the two different heating mechanisms in the gas disk model is displayed together with the snowline (see Appendix A).}
    \label{fig:ev_pebfront}
\end{figure}

\subsection{Dust disk model}
\label{sec:dustmodel}
We parameterize the dust disks following the pebble model proposed by \cite{LJ2014}, with three parameters: the pebble surface density $\Sigma_{\rm peb}$, the Stokes number $S_t$, and the pebble aspect ratio $h_{\rm peb}$. \\
The $\Sigma_{\rm peb}$ is defined as
\begin{equation}
   \Sigma_{\rm peb}=\sqrt{\frac{2F_{\rm {peb}}\Sigma_{g}}{\sqrt{3}\epsilon_p r^{2} \Omega_k}},
\label{eq:pebdensity-sim}
\end{equation}
with $\mathrm{\epsilon}_p=0.5$ the coagulation efficiency between pebbles, $r$ the radial distance to the star, $\Omega_k$ the Keplerian angular velocity, $F_{\rm peb}$ the pebble flux, and $\Sigma_{\textrm{g}}$ the gas surface density. \\
The $\mathrm{S_t}$ is set by equating the pebble growth timescale with the drift timescale, as follows:
\begin{equation}
   S_t = \frac{\sqrt{3}}{8}\frac{\epsilon_p}{\eta}\frac{\Sigma_\textrm{{peb}}}{\Sigma_{\textrm{g}}}.
\end{equation}
Here $\eta=[(1.5-0.5k_T-k_{\Sigma})/2]h_\textrm{g}^{2}$ is the headwind pre-factor that measures the disk pressure gradient \citep{Nakagawa1986} with $\mathrm{k}_{\Sigma}$, and $\mathrm{k}_{T}$ the gradients of the gas disk surface density and temperature and $h_\textrm{g}$ the gas aspect ratio. In order to take into account the ice sublimation of pebbles when they cross the snowline $r_\textrm{snow}$, we multiply $\Sigma_{\rm peb}$ by a factor of 0.5 when $r<r_\textrm{snow}$. Moreover, we considered that pebbles inside the snowline are dry, while outside the snowline have 50$\%$ of water in mass.\\
The $h_{peb}$ is defined as
\begin{equation}
h_{\textrm{peb}}=\sqrt{\alpha_t /(\alpha_t + S_t)}{h_{\textrm{g}}}
\end{equation}
with $\alpha_t$ the turbulent parameter and $h_\textrm{g}$ the gas aspect ratio. \\

We assumed an initial time for our disk model of 1 Myr. We followed the gas disk model proposed by \cite{Ida2016} based on two dominant heating mechanisms: viscous dissipation for the inner disk and stellar irradiation for the outer disk (see Appendix A for a detail description of the model and the parametrization of $\Sigma_\textrm{g}$, $h_\textrm{g}$, and $r_\textrm{snow}$).

Unlike \cite{LJ2014}, who proposed a pebble flux that varies in time depending on the location of the pebble front (radial distance where the pebbles are originated), we assumed a constant pebble flux $F_{\rm peb}$ in time, that varies with the stellar mass. For a star of 0.1 M$_\odot$ we set a pebble flux of $F_{\rm peb}=5\times10^{-5}$~M$_{\oplus}$ yr$^{-1}$, while for a star of 0.3~M$_\odot$ and a star of 0.5~M$_\odot$ we set a pebble flux a factor 6, and a factor 10 higher, respectively, motivated by the fact that the pebble flux scales with the gas density, as suggested in \cite{LJ2014}. This scale is also consistent with the dust mass and stellar mass relations found by \citet{Ansdell2017}. The previous section has demonstrated that these values in compact dust disks correspond to a millimeter dust flux similar to observed values in the unresolved compact disks in Lupus.
We note that such values are also in agreement with the order of magnitude of the pebble fluxes estimated by \cite{Mulders2021} for low-mass stars at 1 Myr, and the value proposed recently by \cite{Pan2024} that studied planet formation around 0.1 M$_\odot$ stars. This pebble flux is also consistent with ALMA observations, as demonstrated in Figure \ref{fig:Stellarmassvsmmflux}.

Figure \ref{fig:ev_pebfront} illustrates the pebble model that we used, showing the evolution of the pebble surface density, the pebble aspect ratio and the Stokes numbers at 1 Myr, in which we assumed a star of 0.1~M$_\odot$, a turbulent parameter $\alpha_t=10^{-4}$ and the pebble flux of F$_{\rm peb}=
5\times10^{-5}$~M$_\oplus~yr^{-1}$. The different slopes of the curves are related to the location of the snowline and to the radial distance that separates the two heating mechanisms proposed in the gas disk model (see Appendix \ref{sec:diskmodel}). We clarify that we assumed the same value for the turbulent parameter as for the gas disk viscosity, in agreement with previous works \citep[e.g.,][]{Izidoro2017,Raymond2018,Sanchez2022}. \\

By considering the pebble fluxes mention above, we corroborated that a planet initially located at a few au reach its isolation mass in less than 200,000 yr for a star of 0.1 M$_\odot$ (see Figure \ref{fig:pebbleaccretion-eg} as an example), and in less than 100,000 yr for a star of 0.3 M$_\odot$ or 0.5 M$_\odot$, as they are also exposed to a fast inward migration in a gas disk with a low viscosity (see Appendix A for a description of the gas model, Appendix B for planet-disk interactions and Appendix C for the test simulations made to validate the pebble accretion model used). We note that if we calculate the pebble mass available during this period of time from the pebble flux that we proposed, we obtain lower values than the total dust mass available in a disk of 20 au (see Section \ref{sec:dustobservations}). Thus, the pebble flux value proposed is a conservative value which is lower than the maximum value needed for all the dust disk mass within 20 au to be available for the rocky planet assembly (see Section \ref{sec:discussion} for further discussion).

\begin{figure*}
    \centering    \includegraphics[width=0.93\textwidth]{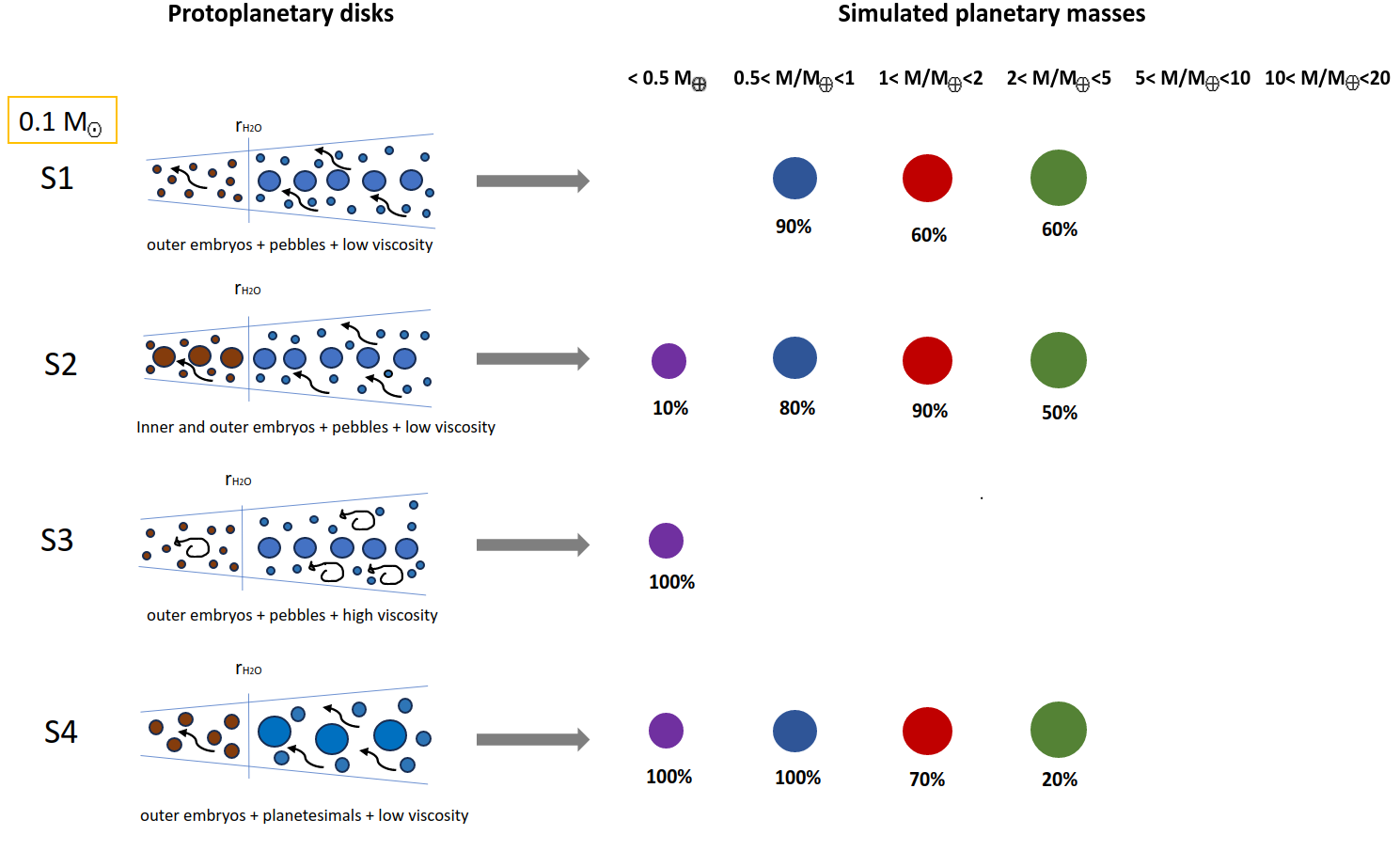}
    \includegraphics[width=0.93\textwidth]{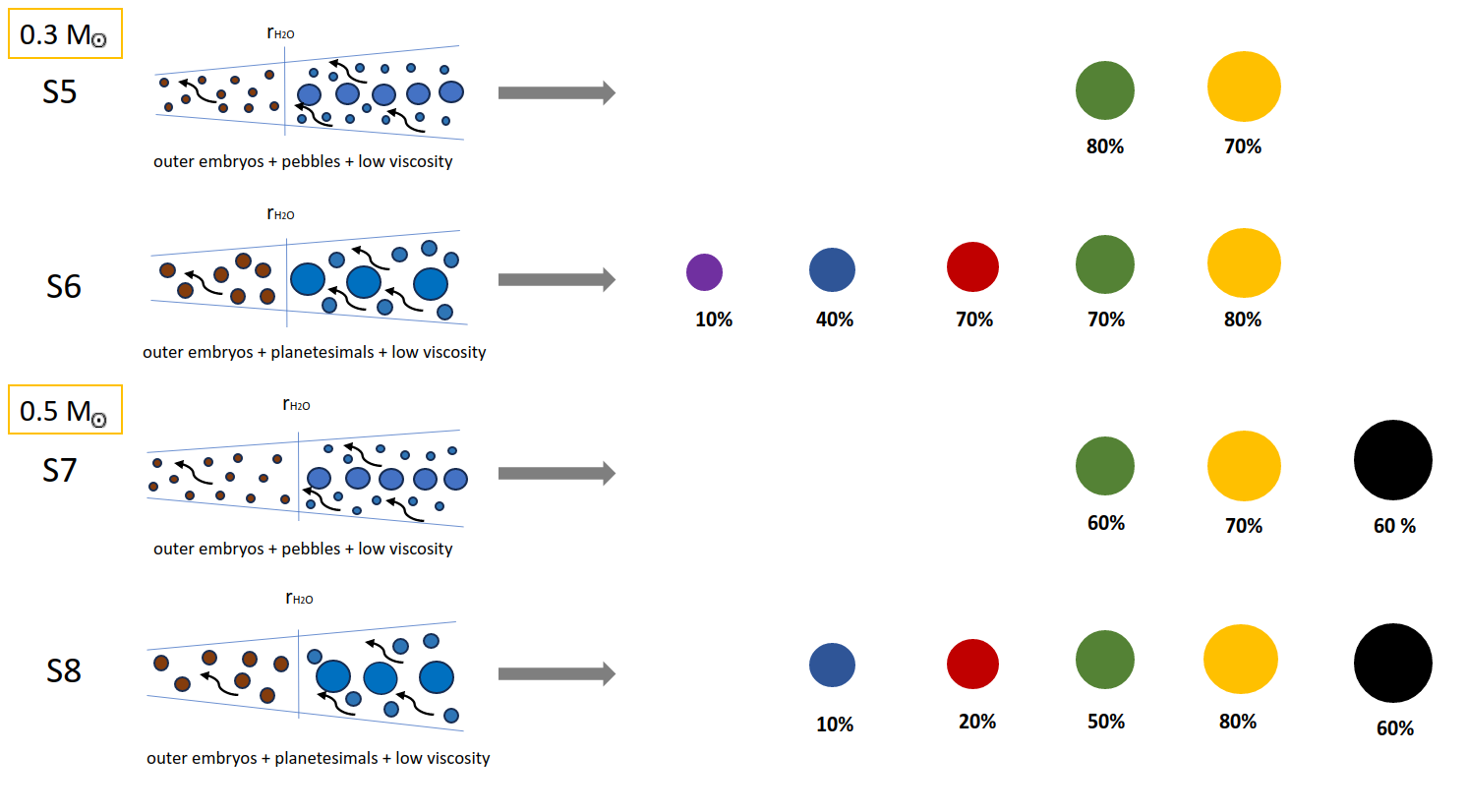}
    \caption{Graphic overview of the disks and final simulated planets in each formation scenario. Left: embryos of different masses (the biggest circles), and pebbles (the smallest circles) or planetesimals (medium circles) located either inside (brown) or outside (blue) the H$_2$O snowline in a gas disk with low viscosity (curved black arrows) or high viscosity (spiral black arrows). Right: planets with masses $<0.5M_\oplus$ (violet circles), $0.5<M/M_\oplus<1$ (blue circles), $1<M/M_\oplus<2$ (red circles), $2<M/M_\oplus<5$ (green circles), $5<M/M_\oplus<10$ (yellow circles) and $10<M/M_\oplus<20$ (black circles). The percentage of simulations in which planets in each mass range were formed is given below the circles.}
    \label{fig:sunmary}
\end{figure*}

\section{Planet formation model}
\label{sec:simulations}

To study rocky planet formation around M dwarfs, we developed different sets of N-body simulations with a modified version of the well known \textsc{Mercury} code \citep{Chambers1999}. We use the same external forces as in the modified version of the code made by \cite{Sanchez2022}, which includes:\\
\begin{itemize}
    \item low-mass planet-disk interactions, following the gas disk evolution model from \cite{Ida2016}, the torques prescriptions from \cite{Paardekooper2010,Paardekooer2011}, and the planetary acceleration corrections prescriptions proposed by \cite{Ida2020}, based on purely dynamical friction \citep[][see Appendix A and B.1 for a detailed description]{Sanchez2022};\\
    \item star-planet tidal interactions based on the tidal equilibrium model proposed by \cite{Hut1981}, and lately adapted by \cite{Bolmont2011} for brown dwarfs and M dwarfs, which includes the evolution of the stellar radius \citep{Baraffe2015}, and rotational period \citep{Bolmont2011} \citep[][see Appendix B.2 for a detailed description]{Sanchez2020,Sanchez2022};\\
    \item general relativistic corrections proposed by \cite{Anderson1975} \citep[][see Appendix B.2 for a detailed description]{Sanchez2020,Sanchez2022}.\\
\end{itemize} 
In this new version of the code, we also include:\\
\begin{itemize}
   \item the pebble accretion model, which was proposed by \cite{LJ2014} with the pebble efficiencies from \cite{LO2018} and \cite{OL2018} for planets in non-circular orbits. In order to do that, we  modified the core of the N-body code, to let a sample of embryos accretes pebbles each few time integration steps, until the outermost embryo reaches its isolation mass (see Section \ref{sec:pebblemodel} below);\\
    \item the evolution of the stellar luminosity, and radius based on the stellar evolution models from \cite{Baraffe2015}, and the evolution of the rotational period of a star of 0.1, 0.3 and 0.5 M$_\odot$ based on the observations of rotational periods of M dwarfs listed in \cite{Bolmont2011}, and in \cite{Scholz2018}, for the gas disk model, and the star-planet interactions (see Appendix A and B.2).\\
\end{itemize}   
We propose different planet formation scenarios, characterize the code, and describe the initial embryo distributions and the two core accretion mechanisms used: pebble and planetesimal accretion.   

\subsection{Formation scenarios}
\label{sec:scenarios}
\begin{table*}
\caption{\label{t7}Formation scenarios}
\centering
\begin{tabular}{cccccc}
\hline\hline
Scenario&M$_\star$&$\alpha$&Embryos&Pebble&Planetesimals\\
 &   &  & Mass and location& Mass & Mass and location\\
\hline
\hline
S1: PA-M01-low$\alpha$-icy           & 0.1 M$_\odot$ & 0.0001 & 0.01 M$_\oplus$ - $0.5<a/\rm au<2$ & 15 M$_\oplus$ & -\\
S2: PA-M01-low$\alpha$-mixed           & 0.1 M$_\odot$ & 0.0001 & 0.01 M$_\oplus$ - $0.02<a/\rm au<1.5$ & 15 M$_\oplus$ & -\\
S3: PA-M01-high$\alpha$-icy           & 0.1 M$_\odot$ & 0.001 & 0.01 M$_\oplus$ - $0.5<a/\rm au<2$ & 5 M$_\oplus$ & -\\
S4: PLA-M01-low$\alpha$-icy           & 0.1 M$_\odot$ & 0.0001 & 0.1 M$_\oplus$ - $0.5<a/\rm au<2$ & - & 0.01 M$_\oplus$ - $0.02<a/\rm au<2$\\ 
S5: PA-M03-low$\alpha$-icy           & 0.3 M$_\odot$ & 0.0001 & 0.01 M$_\oplus$ - $1<a/\rm au<3$ & 25 M$_\oplus$ & -\\
S6: PLA-M03-low$\alpha$-icy           & 0.3 M$_\odot$ & 0.0001 & 0.2 M$_\oplus$ - $1<a/\rm au<3$ & - & 0.02 M$_\oplus$ - $0.065<a/\rm au<5$\\
S7: PA-M05-low$\alpha$-icy           & 0.5 M$_\odot$ & 0.0001 & 0.01 M$_\oplus$ - $1.5<a/\rm au<3.5$ & 35 M$_\oplus$ & -\\
S8: PLA-M05-low$\alpha$-icy           & 0.5 M$_\odot$ & 0.0001 & 0.2 M$_\oplus$ - $1.5<a/\rm au<3.5$& - & 0.05 M$_\oplus$ - $0.075<a/\rm au<6$ \\
\hline
\hline
\label{tab:table2}
\end{tabular}
\end{table*}

We explored different formation scenarios to study rocky planet formation. First, for our standard model we considered a star of 0.1 M$_\odot$ and a sample of embryos with 0.01 M$_\oplus$ initially located beyond the snowline that grow by pebble accretion in a gas disk with low viscosity ($\alpha=10^{-4})$, and a disk lifetime of 10 Myr (see the dust disk model in Section \ref{sec:dustmodel}, and the gas model in Appendix A).\\ 
For the star of 0.1 M$_\odot$ we also proposed a high gas disk viscosity ($\alpha=10^{-3}$) with a lower pebble flux (F$_{\rm peb}=5\times10^{-7}~\mathrm{M_\oplus~yr^{-1}}$, 
as the gas and dust densities are lower in a disk with higher viscosity), to analyze the impact in the resulting planetary masses, and in the migration and orbital decay of the planets. We also assumed a different initial embryo distribution to see the impact in the final fraction of water in planetary masses. Moreover, we let a sample of embryos grow by planetesimal accretion instead of pebble accretion, assuming that the total mass in planetesimals$+$embryos is equal to the pebble mass available in the disk for accretion (see following sections). Furthermore, we considered higher stellar masses to see how rocky planet formation is scaling with the stellar mass. We considered a star of 0.3~M$_\odot$ and a star 0.5~M$_\odot$. In both cases we studied planet formation by pebble accretion and also by planetesimal accretion in a gas disk with low viscosity ($\alpha=10^{-4}$) assuming a sample of embryos initially located outside the snowline. For each scenario proposed, we run 10 N-body simulations. \\

The scenarios are listed in Table \ref{tab:table2}, illustrated in Figure \ref{fig:sunmary} and described as follows:\\
\begin{itemize}
    \item S1: PA-M01-low$\alpha$-icy. It is our standard model, in which we consider a central star of 0.1~M$_\odot$, a gas disk viscosity $\alpha=10^{-4}$, and an initial embryo distribution beyond the snowline that grow by pebble accretion.\\
    \item S2: PA-M01-low$\alpha$-mixed. We consider a central star of 0.1~M$_\odot$, a gas disk viscosity $\alpha=10^{-4}$, and an initial embryo distribution located both inside and outside the snowline.\\
    \item S3: PA-M01-high$\alpha$-icy. We consider a central star of 0.1~M$_\odot$, a gas disk viscosity $\alpha=10^{-3}$, a pebble flux of F$_{peb}=5\times10^{-7}~\mathrm{M_\oplus~ yr^{-1}}$, and an initial embryo distribution beyond the snowline that grow by pebble accretion.\\
    \item S4: PLA-M01-low$\alpha$-icy. We consider a central star of 0.1~M$_\odot$, a gas disk viscosity $\alpha=10^{-4}$, and an initial embryo distribution beyond the snowline that grow by planetesimal accretion.\\
    \item S5: PA-M03-low$\alpha$-icy. We consider a central star of 0.3~M$_\odot$, a gas disk viscosity $\alpha=10^{-4}$, and an initial embryo distribution beyond the snowline that grow by pebble accretion.\\
    \item S6: PLA-M03-low$\alpha$-icy. We consider a central star of 0.3~M$_\odot$, a gas disk viscosity $\alpha=10^{-4}$, and an initial embryo distribution beyond the snowline that grow by planetesimal accretion\\
      \item S7: PA-M05-low$\alpha$-icy. We consider a central star of 0.5~M$_\odot$, a gas disk viscosity $\alpha=10^{-4}$, and an initial embryo distribution beyond the snowline that grow by pebble accretion.\\
    \item S8: PLA-M05-low$\alpha$-icy. We consider a central star of 0.5~M$_\odot$, a gas disk viscosity $\alpha=10^{-4}$, and an initial embryo distribution beyond the snowline that grow by planetesimal accretion.\\
\end{itemize}

For each set of simulations in each scenario, we keep the same inputs for the characterization of the N-body code (see below).   

\subsection{N-body code characterization}

To run the \textsc{Mercury} code, we chosed the hybrid integrator,
which uses a second-order symplectic algorithm to treat interactions between objects with separations greater than 3 R$_{\rm Hill}$, and
the Bulirsch-Stöer method to solve close encounters. We run our simulations for 50 Myr each, in order to study the dynamical stability of the system after the gas disk dissipated at 10 Myr. We adopted an integration time step of 0.1 days, in order to be able to map at least ten times the orbit of a planet with an orbital period of 1 day.
Furthermore, we consider that an embryo is ejected from the system when it
reached a distance of 100 au, and we considered that an embryo had collided with the central star when it is closer than 0.005 au, which corresponds, approximately, to the maximum radius of the host star.
We fixed this value for the entire simulation in order to avoid any
numerical error for small-periastron orbits.

\subsection{Initial embryo distribution}
\label{sec:embryos}
In the scenarios in which we considered pebble accretion, and an initial embryo distribution outside the snowline, we distributed a total amount of 0.25 M$_\oplus$ in 25 seeds of protoplanetary embryos, each of them with a mass of 0.01 M$_\oplus$. The selection of initial embryos with lunar mass is in agreement with several works that consider planetary growth by pebble accretion \citep[e.g.,][]{Liu2019,Coleman2019} and proposed initially by \cite{Kokubo1998}. Embryos are located beyond the snowline: between 0.5 and 2 au for a central star 0.1 M$_\odot$ ($r_\textrm{snow}=0.28$~au), between 1 and 3 au for a central star of 0.3 M$_\odot$ ($r_\textrm{snow}=0.76$~au), and between 1.5 and 3.5 au for a central star of 0.5 M$_\odot$ ($r_\textrm{snow}=1.2$~au). In each scenario, each seed had a random separation between 5 and 15 Hill Radius, according to their classical isolation mass \citep[e.g.,][]{Kokubo2000,Weiss2018}. On the other hand, for the scenario in which we considered embryos located both inside and outside the snowline for a central star of 0.1 M$_\odot$, we distributed 0.5 M$_\oplus$ in 50 seeds of protoplanetary embryos, each of them with a mass of 0.01 M$_\oplus$. In this case they are located between 0.02 and 1.5 au, randomly distributed between 10 and 15 Hill Radius. \\
In the scenarios in which we considered planetesimal accretion, we assumed bigger embryos close to the mass of Mars, as in previous works \citep[e.g.,][]{Ogihara2009,Coleman2019}. For a central star of 0.1 M$_\odot$, we distributed a total amount of 2.5 M$_\oplus$ in 25 seeds of protoplanetary embryos, each of them with a mass of 0.1 M$_\oplus$, located beyond the snowline: between 0.5 and 2 au. For a central star of 0.3 M$_\odot$ and a central star of 0.5 M$_\odot$, we distributed a total amount of 5 M$_\oplus$ in 25 seeds of protoplanetary embryos, each of them with a mass of 0.2 M$_\oplus$, located between 1 and 3 au and between 1.5 and 3.5 au, respectively. In each simulation, we considered that the embryos initially located inside the snowline were dry, and the ones located outside the snowline present 50$\%$ of water in mass\\
In every case, we considered that all embryos initially have $e<0.02$, and $i<0.5$. Those values were taken from uniform distributions. The initial values for the rest of the orbital elements, argument of periastron $\omega$, longitude of the ascending node $\Omega$, and mean anomaly $M$, were randomly determined also from uniform distributions but between 0 and 360 degrees. The physical density of each embryo was fixed to 5 $g~cm^{-3}$ inside the snowline and 1.3 $g~cm^{-3}$ outside the snowline.

\subsection{Core accretion}
\subsubsection{Pebble accretion}
\label{sec:pebblemodel}
 For the sample of planetary embryos immerse in a gas disk that grow by pebble accretion, we followed the pebble model proposed by \cite{LJ2014}. Planetary embryos with different masses accrete pebbles in different regimes. By comparing the pebble accretion radius $\mathrm{r}_{PA}$, and the pebble disk scale height $\mathrm{H_{peb}}=\mathrm{rh_{peb}}$, pebble accretion could be classified into two regimes: the 2D and the 3D regime. If a planetary embryo with mass M$_p$ is large enough ($\mathrm{r_{PA}}>\mathrm{H_{peb}}$), then it would accrete from the complete layer of pebbles, in the 2D regime; while if the embryo is not large enough ($\mathrm{r_{PA}}<\mathrm{H_{peb}}$), it would accrete only a fraction of the pebble layer, in the 3D regime. The $\mathrm{r_{PA}}=\sqrt{GM_p S_t/\Omega_k \Delta v}$ with $\Delta v$ the relative velocity between the pebble and the embryo.

 The pebble accretion rate is defined as $\dot{M}_{\rm acc}=\epsilon F_{\rm peb}$, where, $\epsilon$ represents the efficiency of pebble accretion \citep{Guillot2014,LJ2014}. Considering an embryo with an eccentric orbit and an inclination $\mathrm{i_p}<h_g$, we calculate such efficiencies following \cite{LO2018} and \cite{OL2018}. Thus, the pebble accretion rates in the 2D regime $\dot{M}_{\rm {acc,2D}}$, and in the 3D regime $\dot{M}_{\rm {acc,3D}}$ are defined by:
\begin{equation}
\dot{M}_{\rm {acc,2D}}  =  \epsilon_{2\mathrm{D}} F_{\rm {peb,ac}}\\
\end{equation}
\begin{equation}
\dot{M}_{\rm {acc,3D}}  =  \epsilon_{3\mathrm{D}} F_{\rm {peb,ac}}
\label{eqnarray:accretion}
\end{equation}
where
\begin{equation}
\epsilon_{2\mathrm{D}}=A_{2\mathrm{D}}\sqrt{\frac{q_p}{S_t \eta^{2}}\frac{\Delta v}{r\Omega_k}}f_{set}\\
\end{equation}
and
\begin{equation}
\epsilon_{3\mathrm{D}}=A_{3\mathrm{D}}\frac{q_p}{\eta H_{peb}}r f_{set}^{2}
\label{eq:eff}
\end{equation}
are the pebble efficiencies in the 2D and 3D regime, respectively, with $A_{\rm {2D}}=0.32$ and $A_{\rm {3D}}=0.39$ fitting constant of the model, $q_p=M_p/M_\star$, $f_{set}=\exp[-0.5 (\Delta v/v_\star)^{2}]$, and the relative velocity between the embryo and the pebble $\Delta v = max(v_{cir},v_{ecc})$, with $\mathrm{v}_{cir}$ the relative velocity considering an embryo in a circular orbit and $\mathrm{v}_{ecc}$ the relative velocity considering an embryo in an eccentric orbit, defined as
\begin{eqnarray}
    v_{cir}&=&\left[1+a_\mathrm{{hw/sh}}\left(\frac{q_p}{q_{\mathrm{hw/sh}}}\right)\right]^{-1}v_{hw} + v_{sh}\\ 
    v_{ecc}&=&0.76 e r ~ \Omega_k,
\end{eqnarray}
with $e$ the eccentricity of the embryo's orbit, $a_\mathrm{{hw/sh}}=5.7$ a fitting constant of the model, $q_\mathrm{{hw/sh}}=\frac{\eta^{3}}{S_t}$ the transition mass between the Bondi regime (low impact parameter between the embryo and the pebble) and the Hill regime (high impact parameter) \citep[e.g.,][]{Ormel2017}, $\mathrm{v_{hw}}=\eta r \Omega_k$ relative velocity for a pebble inside Bondi regime, $\mathrm{v}_{sw}=0.52(q_p S_t)^{1/3} r \Omega_k$ the relative velocity for a pebble inside the Hill regime, and $\mathrm{v}_\star=\left(\frac{q_p}{S_t}\right)^{1/3}r \Omega_k$, the transition velocity between the Bondi, and the Hill regimes. 

As each embryo accretes pebbles, the total pebble flux available for a planet at a radial location $r_p$ would be:
\begin{equation}
    F_{\rm {peb,ac}}=F_{\rm peb}-\dot{M}_{\rm acc}(r>r_p),
\end{equation}
where $\dot{M}_{\rm acc}(r>r_p)$ represents the pebble flux accreted by all the embryos exterior to the radial location of the planet. Each embryo will accrete pebbles until they reach the pebble isolation mass. This mass is considered as the necessary mass that an embryo needs to be able to perturb the local disk structure, forming a pressure bump in the disk exterior to its orbit, stopping the inward migration of pebbles, and limiting the amount of pebbles for embryos orbiting in inner orbits \citep{LJM2014}. The isolation pebble mass can be defined as
\begin{equation}
\label{eq:miso}
    M_{\rm iso}=2 M_\oplus \left(\frac{h_g}{0.05}\right)^{3}\left(\frac{M_\star}{0.1 M_\odot}\right).
\end{equation}
following the results from 3D hydrodynamical simulations proposed by \cite{Bitsch2018}.\\
As we are working with a sample of embryos, we clarify that each embryo of the simulation will accrete pebbles until it reaches the isolation mass or the total pebble flux available at the location of the planet is zero. If a planet reaches the isolation mass, then we assume that no more pebbles can migrate inward, and thus no other inner planets can continue accreting pebbles \citep[e.g.,][]{Lambrechts2019,Mulders2021}.\\

We clarify that in this work we evaluate the expressions listed above in the semi-major axis of the embryo's orbit $a$ instead of the radial distance to the star, $r$, as we are considering the eccentricity of the embryo $e$ inside the pebble efficiencies in Eq.~\eqref{eq:eff}.

\subsubsection{Planetesimal accretion}

In the scenarios in which we considered planetesimal accretion, we let the sample of embryos interacts with a sample of 1000 planetesimals. For a star of 0.1 M$_\odot$, each of them has a mass of 0.01 M$_\oplus$, while for a star of 0.3 M$_\odot$ the mass of each of them is 0.02 M$_\oplus$, and for a star of 0.5 M$_\odot$ is 0.05 M$_\oplus$. Their semi-major axis are initially distributed following a power law $\propto r^{-1}$ from the inner edge of the disk until the location where the total amount between embryos, and planetesimals equals the pebble mass available in the disk at 1 Myr, which is between 0.02 au and 2 au for a star of 0.1 M$_\odot$, between 0.065 au and 5 au for a star of 0.3 M$_\odot$ and between 0.075 au and 6 au for a star of 0.5~M$_\odot$. \\
In every case, the rest of the orbital elements: $e$, $i$, $\omega$, $\Omega$, and $M$, were estimated from equivalent distributions as for the sample of embryos. The physical density of the planetesimals was fixed to 3 $g~cm^{-3}$ inside the snowline of the system, and 1.5 $g~cm^{-3}$ beyond the snowline. We consider that planetesimales inside the snowline are dry, while outside the snowline present 50$\%$ of water in mass.\\

\section{Simulated systems: Accretion history and dynamical evolution}

\label{sec:accretionhistory}
We present the differences in the resulting simulated planetary systems  of each formation scenario (see Table \ref{tab:table2}) in terms of their dynamical evolution and accretion history. Moreover, we compare their final configurations (semi-major axis and eccentricity), masses, amount of water in mass, and multiplicity (see Table 3). 

\subsection{Standard model: PA-M01-low$\alpha$-icy}

\begin{figure*}
\centering
\includegraphics[width=8cm]{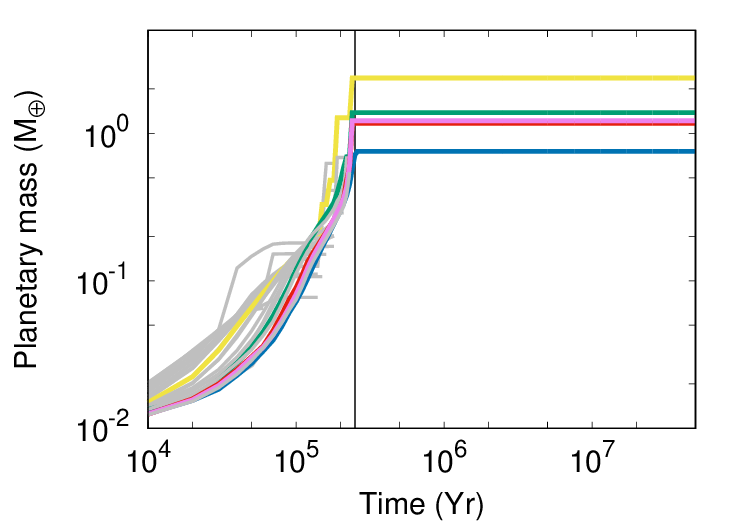}
\includegraphics[width=8cm]{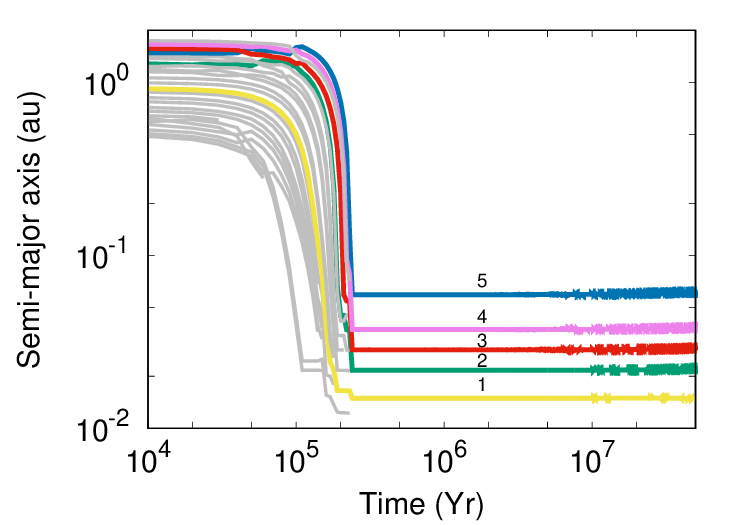}\\
\includegraphics[width=8cm]{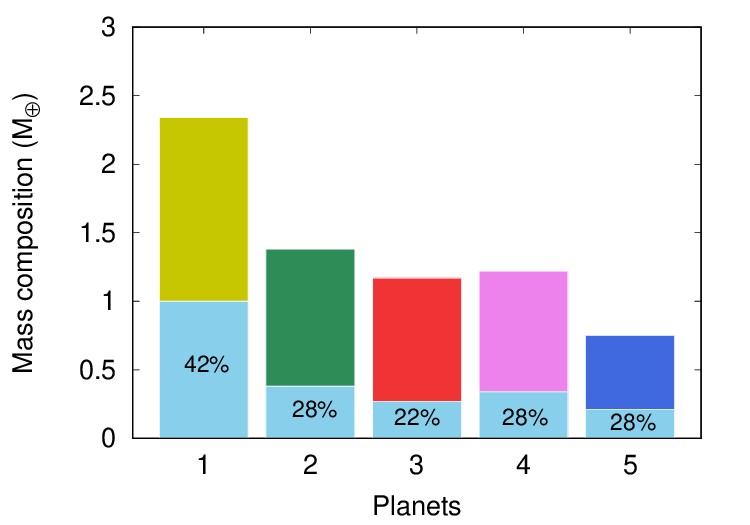}
\includegraphics[width=8cm]{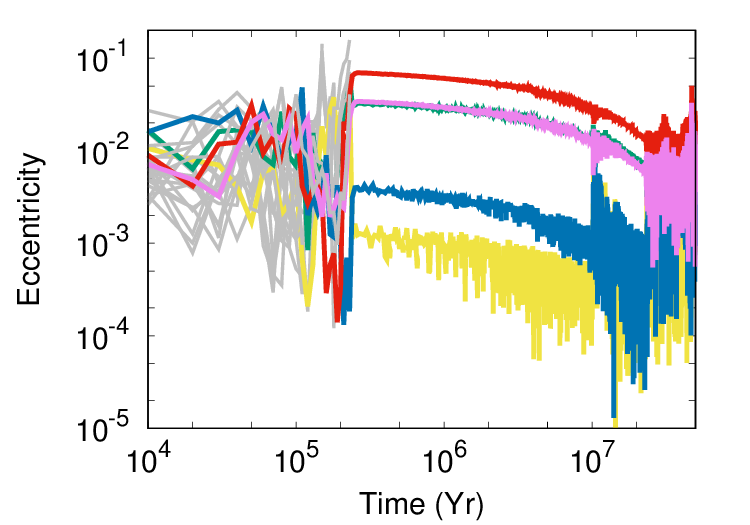}
    \caption{Evolution of the planetary embryo sample of one representative simulation of our standard model. Shown are the embryos that survived until the end of the simulation (colored lines), and the ones that did not survive due to collisions (gray lines). Top panels: Planetary mass evolution (left) and semi-major axis evolution (right) along 50 Myr (integration time). The left vertical line represents the time when the outermost embryo reached the isolation mass. Bottom panels: Final bulk composition of the resulting planets of the simulation (left) and eccentricity evolution along the integration time (right). The light blue area at the bottom of each bar represents the amount of water in mass, while the other colors are associated with the percentage of rock in each resulting planet.}
    \label{fig:ev_sim1}
\end{figure*}

In our standard model we considered a central star of M$_\star=0.1$~M$_\odot$, an initial embryo distribution outside the snowline that grow by pebble accretion and a gas disk with low viscosity ($\alpha=10^{-4}$). In this scenario, the sample of embryos suffered an efficient inward migration in all the simulations along the disk lifetime. The resulting planets ended up located close-in to the star ($a<0.1$~au) in less than 300,000 yr. During this period of time they suffered collisions among embryos and accreted pebbles from outside the snowline for the first $\sim$100,000 yr and then from inside the snowline (r$_{\rm snow}=0.28$~au). Once the outermost planet reached the isolation mass, the pebbles stopped migrating inward and, in most cases, the resulting planets achieved their final masses. The resulting planets stopped migrating inward and reminded in compact configurations close to the inner edge of the disk, in quasi-circular and co-planar orbits. In 60$\%$ of the simulations, either 1 or 2 collisions with the central star occurred within 500,000 yr. In the rest of the simulations, no embryo collided with the central star. On the other hand, in all the simulations between the 60$\%$ and the 88$\%$ of the embryos collided among them. In 70$\%$ of the simulations, all the collisions among embryos occurred within the first 300,000 yr (while the embryos were also accreting pebbles). In the rest of the simulations just 10$\%$ of the embryos collided among them at around 3 Myr, 10.5 Myr and 40 Myr. Thus, in the majority of the cases, the resulting planets achieved their final masses at early stages (within 300,000 yr). \\
The final mass of the simulated planets are between 0.5 and 4~M$_\oplus$. For all the resulting planets, their final masses were a combination of pebble accretion and collisions among embryos. The efficiency in pebble accretion was due to the fact that the available pebble mass was $\sim 15$~M$_\oplus$ and the eccentricity of the embryos was always less than 0.1 (see Section \ref{sec:testsim}). The innermost planets accreted the majority of the pebbles outside the snowline. Between 60$\%$ and 96$\%$ of their mass correspond to pebbles outside the snowline. The rest of the planets received a more equitable contribution of pebbles, both inside and outside the snowline. Between 40$\%$ and 75$\%$ of their mass are from pebbles outside the snowline. Thus, the simulated planets have between 20$\%$ and 50$\%$ of water in mass.\\
As an example of a representative case, in Figure \ref{fig:ev_sim1} we show the evolution of the planetary mass, semi-major axis, and eccentricity of the sample of embryos along the integration time of one of the simulations, together with the final planetary bulk compositions in terms of amount of water in mass. We can see that the planets achieved their final masses and semi-major axis early in their evolution. Once the outermost planets reached its isolation mass, the rest of the planets stopped accreting pebbles and achieve their final configurations. The final masses are between 0.7 and 3 M$_\oplus$, and the semi-major axis between 0.01 and 0.1 au. Their eccentricities decreased along the integration time, being less than 0.02 at the end of the simulation. Once the gas disk dissipated, we can see tiny oscillations in semi-major axis and changes up to one order of magnitude in eccentricity. We expect that if we extend the simulations to a Gyr timescale, the orbits of the planets will be completely circularize due to star-planet tidal interactions. The final planetary composition is always rock+H$_2$O. The innermost planet accreted the major amount of water due to the fact that it received a greater amount of embryo impacts which accreted the majority of the pebbles from outside the snowline, while the rest of the planets received fewer embryo impacts and accreted almost half of the pebbles from inside the snowline and close to the other half from outside the snowline.

\subsection{Changing the initial embryo distribution: PA-M01-low$\alpha$-mixed}
 
 In this scenario, we considered a central star of M$_\star=0.1$~M$_\odot$ with an initial embryo distribution both inside and outside the snowline. We let the embryos grow by pebble accretion also in a gas disk with low viscosity ($\alpha=10^{-4}$). The sample of embryos also suffered a fast inward migration in all simulations early in the disk lifetime. The resulting planets ended up located close-in to the star ($a<0.1$~au) within 200,000 yr. In this case, also the resulting planets remained in compact configurations close to the inner edge of the disk in quasi-circular and co-planar orbits. During this time the embryos suffered most of the collisions between them and accreted all the available pebbles until the outermost planet reached the isolation mass. In just 20$\%$ of the simulations, 1 or 2 embryos collided with the star in an early in the gas disk phase. In all simulations, around 90$\%$ of the embryos collided among them within 200,000 yr. In 60$\%$ of the simulations, few embryos collided between them at a late stage: either after the gas dissipated from the disk or at around 25 Myr. Thus, in more than half of the simulations, the final configurations were achieved after 25 Myr. \\
 The final mass range of the resulting planets is a bit wider than in the standard model: between 0.35 and 4.5 M$_\oplus$ (just one planet at the lower mass limit and one planet at the upper mass limit). For almost all the resulting planets, the major contribution were pebbles from outside the snowline. Just for the innermost planets, located at $a<0.012$~au, the major contribution were pebbles from inside the snowline. Thus, the innermost planets have between $\sim 0\%$ up to 10$\%$ of water in mass, while the outer planets present between 15$\%$ and 35$\%$ of water in mass. \\
 In this scenario, we reproduced analogous planetary systems as in the standard model regarding the compact configuration, number of planets per system and planetary masses, but with different percentage of water in mass, as we could form dry and less water rich inner planets.\\
 
\subsection{Changing the gas disk viscosity and pebble flux: PA-M01-high$\alpha$-icy}

In this scenario we considered a central star of M$_\star=0.1$~M$_\odot$, embryos located outside the snowline growing by pebble accretion in a gas disk with a higher viscosity: $\alpha=10^{-3}$ and also a lower pebble flux F$_{\rm peb}=5\times 10^{-7}~M_\oplus~yr^{-1}$. In this case the sample of embryos migrated inward much slower than in the scenarios with a lower viscosity. The resulting planets located close to the star ($a<0.1$~au) accreted pebbles while migrating inward during the whole disk lifetime. In this scenario, in all simulations there is a sample of small planets located at further distances from the star up to almost 2 au (initial location of the outermost planet). While the close-in planets remind in quasi-circular orbits, the outermost planets present eccentricities $0.08<e<0.25$. The resulting planets present fewer collisions, as just less than 30$\%$ of the embryos collided among them and no one collided with the central star, in any of the simulations. In all the simulations, the collision among embryos occurred both before and after the dissipation of the gas disk at 10 Myr. The latest collisions occurred at around 45 Myr. Thus, the resulting planets achieved the final masses and configurations at around 45 Myr.\\
The final masses of the resulting planets are much smaller than in the previous scenarios. The available pebble mass was just, $\sim 5 M_\oplus$ and the migration rate was so slow that the outer embryo never reached the isolation mass. The masses of the planets located closer to the star ($a<0.1$~au) are between
0.1 and 0.4 M$_\oplus$, while the tail of bodies located further from the star ($0.1<a/\rm au<2$) remind with masses less than 0.1 M$_\oplus$. All the embryos present between 20$\%$ and 45$\%$ of water in mass, regardless of their location in the disk.

\subsection{Changing the accretion mechanism: PLA-M01-low$\alpha$-icy}

In this scenario we considered a star of M$_\star=0.1$~M$_\odot$, a low gas disk viscosity ($\alpha=10^{-4}$), and a sample of embryos located beyond the snowline that grow by planetesimal accretion, which were distributed both inside and outside the snowline. The resulting planets were also located close-in to the star ($a<0.1$~au) but in this case there was also a tail of planetesimals at further distances ($0.1<a/\rm au<2.5$). The planets present compact orbital configurations in quasi-circular, a co-planar orbits. In this case, the embryos also suffered a fast inward migration, where most of the collision among them and with planetesimals occurred within 300,000 yr.  
In all the simulations between 1 and 5 embryos collided with the star early in the gas disk phase, and in half of the simulations late collisions among embryos occurred before and after the gas disk dissipated and up to 40 Myr. Thus, half of the final planetary systems achieve their final configurations between 20 and 40 Myr.\\
The final masses of the resulting planets are between 0.15 and 3 M$_\oplus$. In this case, the final masses are smaller than when considering embryos growing by pebble accretion in a gas disk with low viscosity. In this scenario, more planets can be formed per systems, as in most of the simulations each system has either 5 or 6 planets. Roughly, half of the planets in each system present masses of less than 1 M$_\oplus$. Such masses are always between 45$\%$ and 50$\%$ water; even though the planets accreted a greater number of dry planetesimals, the major contribution was from water-rich embryos together with some water-rich planetesimals.

\subsection{Changing the stellar mass: M$_\star=0.3 M_\odot$ (M03-low$\alpha$-icy) and M$_\star=0.5 M_\odot$ (M05-low$\alpha$-icy)}

We showed the differences in the resulting planetary systems that can be formed around more massive M dwarfs when we let the embryos grow by pebble accretion or planetesimal accretion. 

\subsubsection{Pebble accretion}

As in our standard model, we consider a sample of embryos located outside the snowline growing by pebble accretion in a gas disk with low viscosity. When we consider a star of M$_\star=0.3$~M$_\odot$, the sample of embryos migrated inward within just 70,000 yr, due to mostly the fact that they are immersed in a more massive gas disk (gas density a factor 2 higher than in our standard model). During this time, the outermost planet reached its isolation mass. The planetary systems have between 2 and 5 planets, each of them in compact configurations ($a<0.2$~au), in quasi-circular and co-planar orbits (except just one simulation in which no planet survived). In all the simulations, 90$\%$ of the embryos collided among them, mostly within 70,000 yr. However, in 60$\%$ of the simulations, few outer embryos collided among them just after the gas dissipated from the disk at around 10 Myr. Thus, after that time, they reached their final configurations and masses. Regarding the collisions with the star, just in 1 simulation, 1 embryo collided with the star.\\
The final masses of the resulting planets are bigger than in our standard model, between 2 and 9 M$_\oplus$, as expected due to a bigger pebble mass available of $\sim 25$~M$_\oplus$. Most of the innermost planets ($a<0.04$~au) presented a bigger amount of water in mass, of around 40$\%$, while the outer planets present between 15$\%$ and 25$\%$ of water in mass. \\
If we considered a star of M$_\star=0.5$~M$_\odot$, the sample of embryos experienced even a faster inward migration due to being immersed in a more massive disk (gas surface density a factor 3 higher than in our standard model). They accreted pebbles until the outermost planets reached their isolation masses at $\sim$50,000 yr. In this interval of time they ended up located in compact configurations close to the inner edge of the disk ($a<0.2$~au), in quasi-circular and co-planar orbits. In 30$\%$ of the simulations, either 1 or 2 embryos collided with the central star within 300,000 yr, corresponding to the simulations in which just 1 or 2 planets survived until the end. On the other hand, in all the simulations embryos collided among them, mostly during the first 50,000 yr. More collisions between embryos occurred in simulations in which three planets survived at the end (which represent 70$\%$ of the simulations). In those cases, even though most of the collisions occurred at an early stage, some of them happened just after the gas dissipated from the disk. Thus, just after the gas dissipated from the disk, the resulting planets reached their final masses.
In this case, the final planetary masses were even higher than in the previous scenarios, between 3 and 17 M$_\oplus$, as the available pebble mass was higher, $\sim 3$~M$_\oplus$. In this scenario, just the innermost planet ($a \sim 0.035$~au) accreted $\sim$ 90$\%$ of pebbles beyond the snowline ($\sim 45\%$ water in mass). Planets located at the end of the simulations between $0.04<a/\rm au<0.08$, received between 20$\%$ and 60$\%$ of water-rich pebbles (between 10$\%$ and 30$\%$ of water in mass, respectively), while planets with $a>0.08$~au accreted less than 25$\%$ of pebbles beyond the snowline ($\sim 12 \%$ of water in mass).\\

\subsubsection{Planetesimal accretion}

In the scenarios in which we let the sample of embryos grow by planetesimal accretion instead of by pebble accretion, the simulated planets present almost the same range in semi-major axis and number of planets per system. They are also in compact configurations and in quasi-circular and co-planar orbits. The main difference is that there are smaller planets in each system, which are mostly the outermost planets. Thus, the planetesimal accretion allows the formation of smaller planets, which cannot be formed if we consider pebble accretion. 
For a star of 0.3 M$_\odot$ as for the star of 0.5 M$_\odot$, the embryos suffered a slower inward migration as the majority of the collision between embryos and planetesimal occurred within the first 200,000 yr. In most of the simulations some embryos collided among them during the gas disk, but just in 30$\%$ of the simulations there were late collisions between embryos after the gas disk dissipated and up to 40 Myr. The final masses of the resulting planets are analogous as when considering pebble accretion, but in this case planets smaller than 2 M$_\oplus$ could be formed as well. As for a star of 0.1 M$_\odot$, the planets present almost 50$\%$ of their mass in water. Whereas the smallest planets truly exist cannot be confirmed yet with the current instrumental sensitivity. \\

\begin{table*}
\caption{\label{t7}Characteristics of the simulated planetary systems in each scenario of study}
\centering
\begin{tabular}{cccccc}
\hline\hline
Scenario&M$_\star$& &Simulated planets\\
 & [M$_\odot$]  &  Number per system &Mass [M$_\oplus$]& Semi-major axis [au]&H$_2$O in mass [$\%$]\\
\hline
\hline
S1: PA-M01-low$\alpha$-icy           & 0.1 & 1 - 7 & 0.5 - 4 & 0.01 - 0.1 & 20-50\\
S2: PA-M01-low$\alpha$-mixed           & 0.1 & 1 - 8 & 0.35 - 4.5 & 0.01 - 0.1 & 0-35\\
S3: PA-M01-high$\alpha$-icy           & 0.1 & 8 - 12 &0.1 - 0.4 & 0.01 - 1.5 & 20-45\\
S4: PLA-M01-low$\alpha$-icy          & 0.1 & 2 - 6 & 0.15 - 3 & 0.01 - 0.1 & 45-50\\ 
S5: PA-M03-low$\alpha$-icy           & 0.3 & 2 - 5 & 2 - 9  & 0.02 - 0.2 & 15-40 \\
S6: PLA-M03-low$\alpha$-icy & 0.3  & 2 - 6 &  0.3 - 9.5  & 0.03 - 0.15 & 45-50          \\
S7: PA-M05-low$\alpha$-icy           & 0.5 & 1 - 3 & 3 - 17 & 0.03 - 0.15 & 10-45\\
S8: PLA-M05-low$\alpha$-icy & 0.5 &  2 - 3  &  0.5 - 17.3 & 0.035 - 0.15 & 45-50  \\
\hline
\hline
\label{tab:table3}
\end{tabular}
\end{table*}

\begin{figure*}
\includegraphics[width=18.5cm]{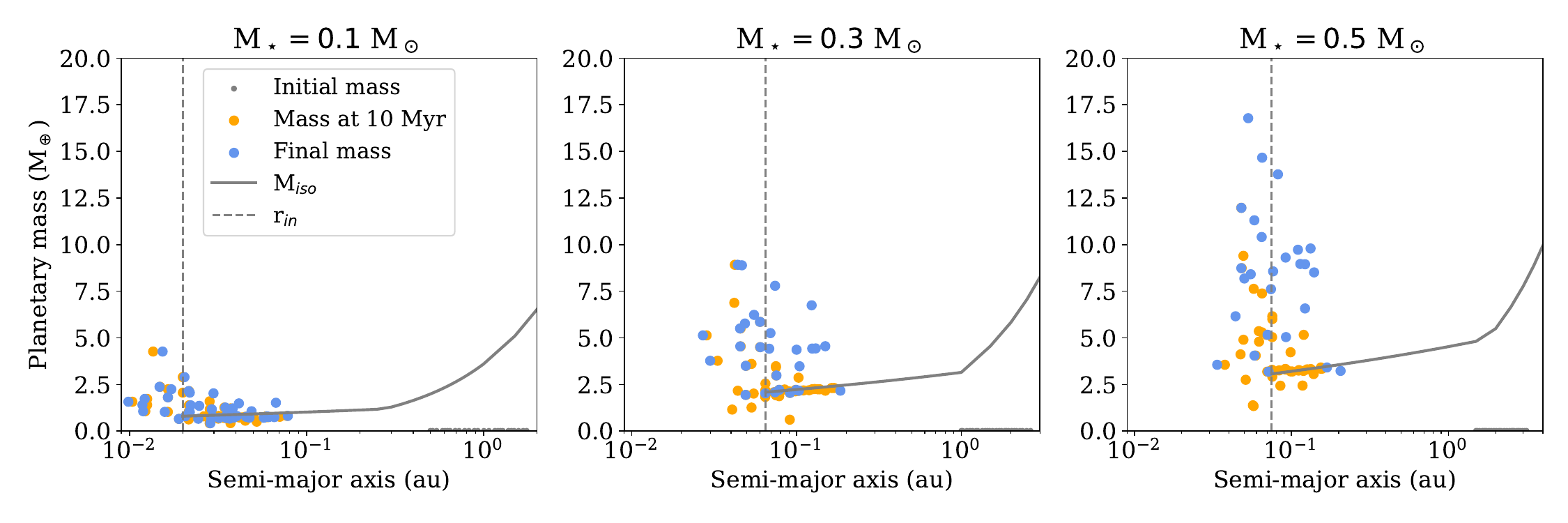}
    \caption{Initial masses of embryos (gray dots), planetary masses at 10 Myr, gas disk lifetime (orange dots), and final planetary masses of the simulated planets (light blue dots) together with their semi-major axis for a star of 0.1~M$_\odot$ (left panel), 0.3~M$_\odot$ (middle panel), and 0.5~M$_\odot$ (right panel), for $\alpha=10^{-4}$. The isolation mass (solid gray line) and the inner edge of the disk for each star (vertical dotted gray line) are overplotted.}
    \label{fig:iso-mass}
\end{figure*}

\subsection{Comparison among the formation scenarios}

We conclude that for a star of 0.1 M$_\odot$, we can form multiple close-in Earth-like and some super-Earth-like planets both by pebble or planetesimal accretion in a gas disk with low viscosity ($\alpha=10^{-4}$). The main difference is that when considering planetesimal accretion, all the planets are water worlds, and we can form also smaller planets. Additionally, changing the initial embryo distribution either inside or outside the snowline has an impact just in the final amount of water in mass. We can form dry worlds in close-in orbits, just if we consider embryos initially located inside the snowline. Moreover, if we consider a gas disk with a high viscosity ($\alpha=10^{-3}$) and lower pebble flux just planets with masses less than 0.5 M$_\oplus$ can be formed, as the migration rate is slower than in a disk with lower viscosity, and the available pebble mass seems not enough to form Earth-like planets. 
For the star of 0.3~M$_\odot$ and 0.5~M$_\odot$ we were able to form more massive planets than for a star of 0.1~M$_\odot$, if we consider either pebble or planetesimal accretion in a gas disk with low viscosity, also located close-in to the star. The main difference is that when considering pebble accretion just super-Earths and potentially mini-Neptunes (see Section \ref{sec:discussion}) can be formed, while when considering planetesimal accretion also Earth-like planets or even smaller planets than the Earth can be formed. In Table \ref{tab:table3} we listed the range of final planetary masses and semi-major axis, together with the number of planets per simulated system in each formation scenario. Furthermore, in Figure \ref{fig:sunmary} we show the final simulated planetary masses per formation scenario.\\

We highlight that in the scenarios in which we let a sample of embryos grow by pebble accretion in a disk with low viscosity ($\alpha=10^{-4}$), the pebble accretion was efficient. For a star of 0.1~M$_\odot$ the average of pebble efficiency was around 30$\%$, while for a star of 0.3 M$_\odot$ and 0.5~M$_\odot$, in each case, the average in efficiency was around 60$\%$. In Figure \ref{fig:iso-mass} we show the semi-major axis and the initial planetary masses of the embryos, the planetary masses at 10 Myr (gas disk lifetime) and the final planetary masses at 50 Myr (integration time) of the simulated planets of the formation scenarios that included pebble accretion in a gas disk with low viscosity. We compare the masses and location of the planets with the isolation masses and the location of the inner edge of the disk for each stellar mass. Despite the fact that the isolation mass can reach 6 and 10 M$_\oplus$ in the outer disk, depending on the stellar mass, the maximum isolation mass that each individual embryo could reach was between 1 and 3 M$_\oplus$, as they rapidly migrated inward. We can see how the final planetary mass of the outer planets is the same as the isolation mass, but the inner planets have final masses between two and five times bigger than the isolation mass in the inner regions. This is mostly due to collisions among embryos in the inner disk during the gas disk lifetime, and in some cases collisions after the gas dissipated from the disk. We conclude that the final planetary masses are mostly determined by the amount of collisions between embryos due to the gravitational interactions between the bodies. Moreover, we found that the mass of super-Earth planets increase almost linearly with the stellar mass, in agreement with previous works \citep[e.g.,][]{Liu2019}.

\section{Simulated planetary systems vs observed low-mass exoplanets around M dwarfs}

In this section we compare the resulting simulated planetary systems from each formation scenario with the sample of confirmed low-mass close-in  exoplanets (M$_p<20$~M$_\oplus$ and $a<0.3$~au) around M dwarfs,\footnote{https://exoplanetarchive.ipac.caltech.edu/} in terms of masses, semi-major axis and orbital periods. We just considered the sample of planets with values of planetary masses, which are 112 exoplanets to date. From the total amount of exoplanets, 62$\%$ were discovered with radial velocity and 38$\%$ with transit technique, with planetary masses estimated by transit timing variations in this last case. They were principally discovered with CARMENES \citep{CArmenes2014}, HARPS \citep{Harps2011} and \textit{TESS} \citep{TESS}, among other instruments. 

\subsection{Planetary systems with M$_\star=0.1$~M$_\odot$}

\begin{figure*}
\sidecaption
\includegraphics[width=12cm]{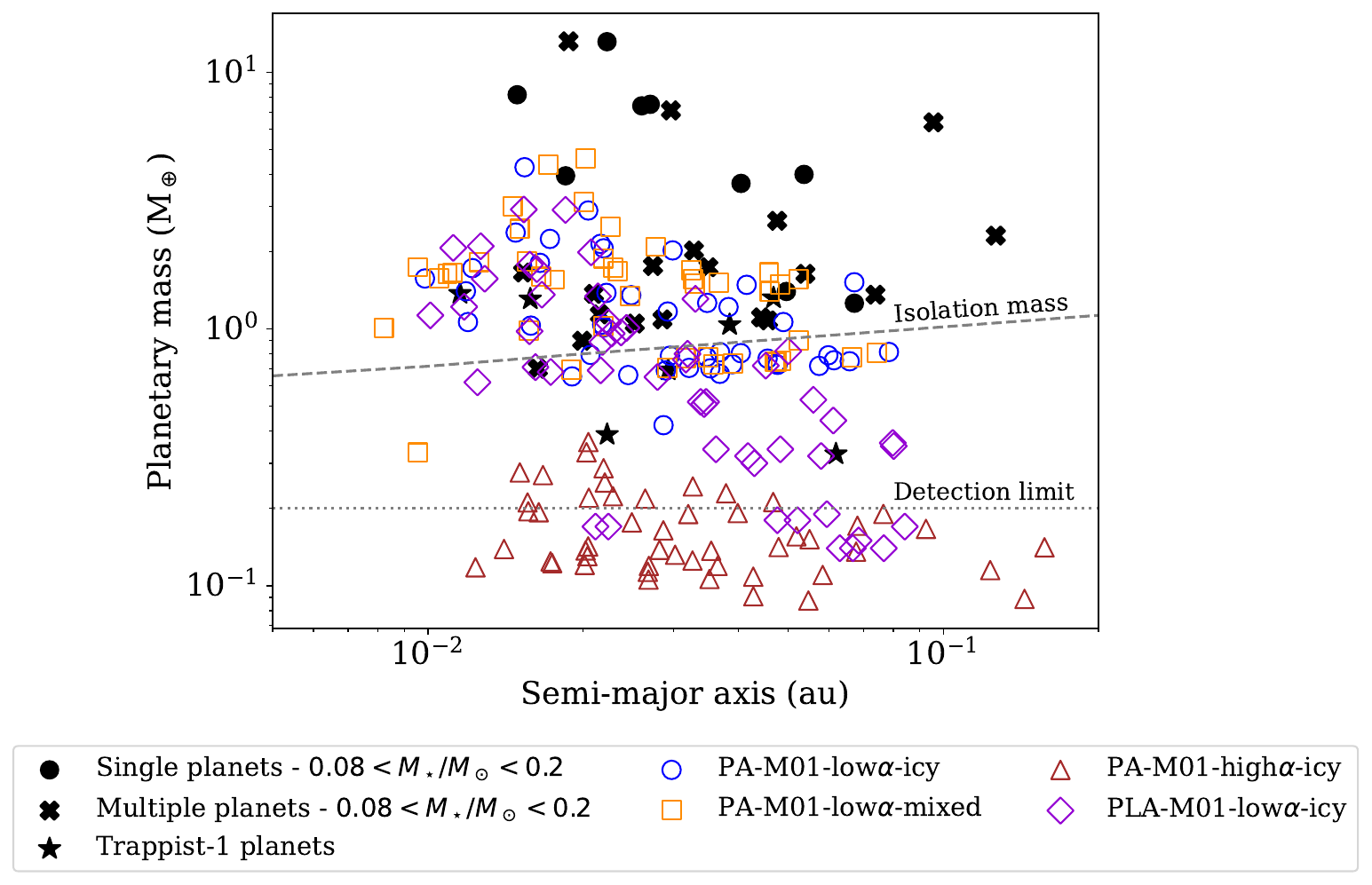}
      \caption{Confirmed exoplanets around stars with masses 0.08$~< $~$M_\star/\rm M_\odot<$ 0.2 and $a<0.2$~au discriminated by observed multiplicity: single planets (black circles), planets in multi-planetary systems (black crosses) and \textit{Trappist-1} planets (black stars), together with the simulated planets from the scenarios in which the star is 0.1 M$_\odot$: PA-M01-low$\alpha$-icy (blue circles), PA-M01-low$\alpha$-mixed (orange squares), PA-M01-high$\alpha$-icy (brown triangles) and PLA-M01-low$\alpha$-icy (violet diamonds). The isolation mass for a star of 0.1 M$_\odot$ at 1 Myr calculated from Eq. \ref{eq:miso} and extrapolated up to the location of the star for visual purposes (dashed gray line) and the detection limit in mass for planets around M dwarfs (dotted gray line) are overplotted. } 
    \label{fig:exo-sample-S1S2S3S4}
\end{figure*}

\begin{figure*}        \centering\includegraphics[width=1.\textwidth]{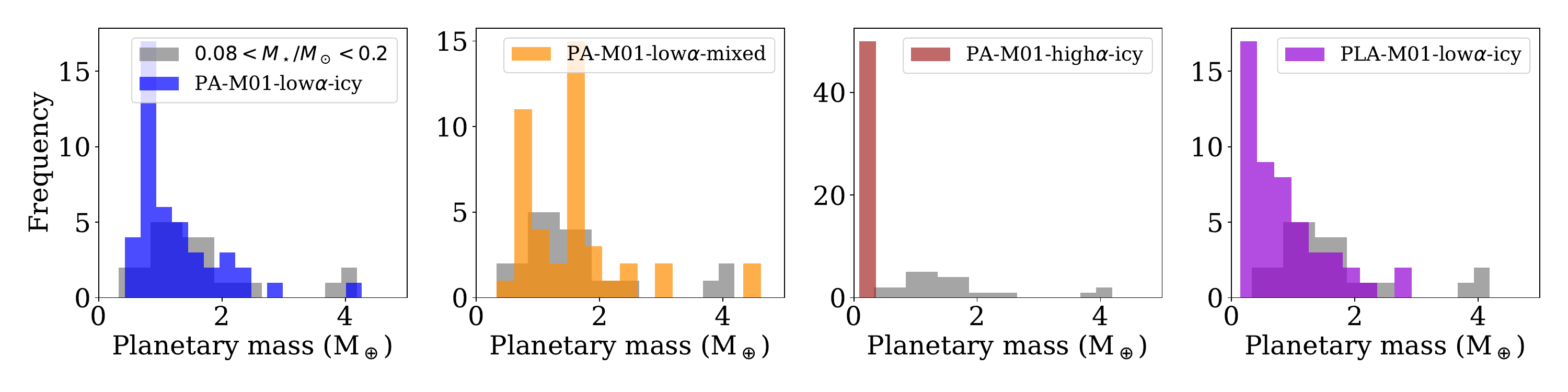}\\   \includegraphics[width=1.\textwidth]{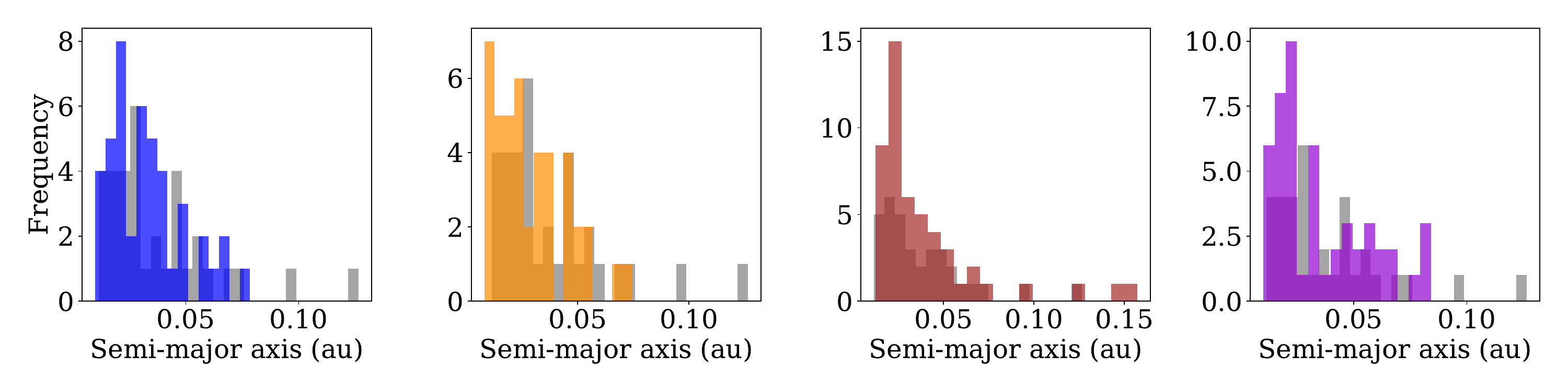}
    \caption{Histograms in planetary mass (top panels) and semi-major axis (bottom panels) of the observed low-mass close-in exoplanets around stars with masses $0.08<M_\star/\rm M_\odot<0.2$ (gray) and the resulting simulated planetary systems in scenario PA-M01-low$\alpha$-icy (blue), PA-M01-low$\alpha$-mixed (orange), PA-M01-high$\alpha$-icy (brown), and PLA-M01-low$\alpha$-icy (violet).}
    \label{fig:histo-semiejesmasas-S1S2S3S4}
\end{figure*}

We focus first on the final simulated planets of the scenarios in which we assumed a star of M$_\star=0.1$~M$_\odot$, with the sample of low-mass close-in exoplanets around stars $0.08<M_\star/\rm M_\odot<0.2$ that have an estimation of their masses. In Figure \ref{fig:exo-sample-S1S2S3S4} we show the planetary masses and semi-major axis of the simulated planets and the confirmed exoplanets. We also show the isolation mass for a star of 0.1 M$_\odot$ at 1 Myr (as the planets migrated inward early in their evolution) and the detection limit in mass for planets around M dwarfs (0.2 M$_\oplus$). Additionally, in Figure \ref{fig:histo-semiejesmasas-S1S2S3S4} we compare the histograms in planetary mass and semi-major axis of the resulting planets of each scenario (around 40 planets each) with the confirmed exoplanet sample (35 planets). In all the scenarios, we generated planets with semi-major axis close-in to the star, which match the observations. On the other hand, as it was explained in Section \ref{sec:accretionhistory}, the scenarios in which we considered a sample of embryos that grow by pebble accretion in a gas disk with low viscosity ($\alpha=10^{-4}$) present the planets with higher masses in compact configurations. Thus, there is an overlap between the simulated and observed planets for masses less than 5 M$_\oplus$ as it can be seen in Figure \ref{fig:histo-semiejesmasas-S1S2S3S4}. Finally, the scenario in which the embryos grow by planetesimal accretion formed a bigger amount of small planets ($M_p<0.5$~M$_\oplus$), while the scenario with a gas disk with a higher viscosity ($\alpha=10^{-3}$) present just small planets ($M_p<0.5$~M$_\oplus$), many of them below the detection limit (see Figure 6). \\
Whether the smallest simulated planets may exist or not still needs to be confirmed in future missions with much higher instrument sensitivity. On the other hand, in order to reproduce the more massive planets, it would be worth it trying to decrease the gas disk viscosity and/or increase the pebble flux.\\

\begin{figure*} 
\sidecaption
\includegraphics[width=12cm]{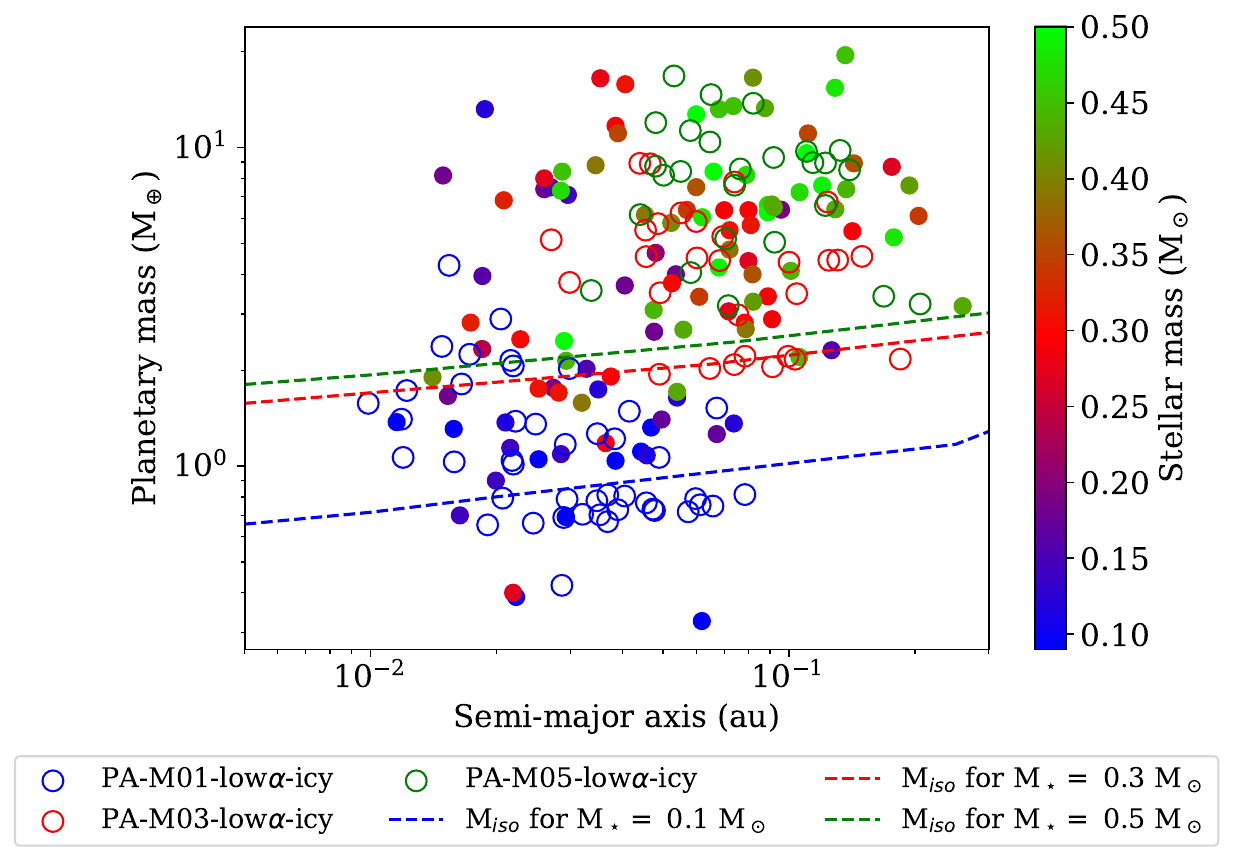}
    \caption{Confirmed low-mass exoplanets around M dwarfs with $a<0.2$~au (solid dots), together with the final simulated planets of the scenarios in which we consider pebble accretion in a disk with low gas disk viscosity for a star of 0.1~M$_\odot$ (PA-M01-low$\alpha$-icy, blue), a star of 0.3 M$_\odot$ (PA-M03-low$\alpha$-icy, red), and a star of 0.5 M$_\odot$ (PA-M05-low$\alpha$-icy, green). The isolation mass for a star of 0.1 M$_\odot$ (blue dashed line), for a star of 0.3 M$_\odot$ (red dashed line), and a star of 0.5 M$_\odot$ (green dashed line) at 1 Myr, are overplotted (see Eq. \ref{eq:miso}).} 
    \label{fig:exo-sample-S1S5S6}
\end{figure*}

 \begin{figure*}    
    \centering
     \includegraphics[width=0.96\textwidth]{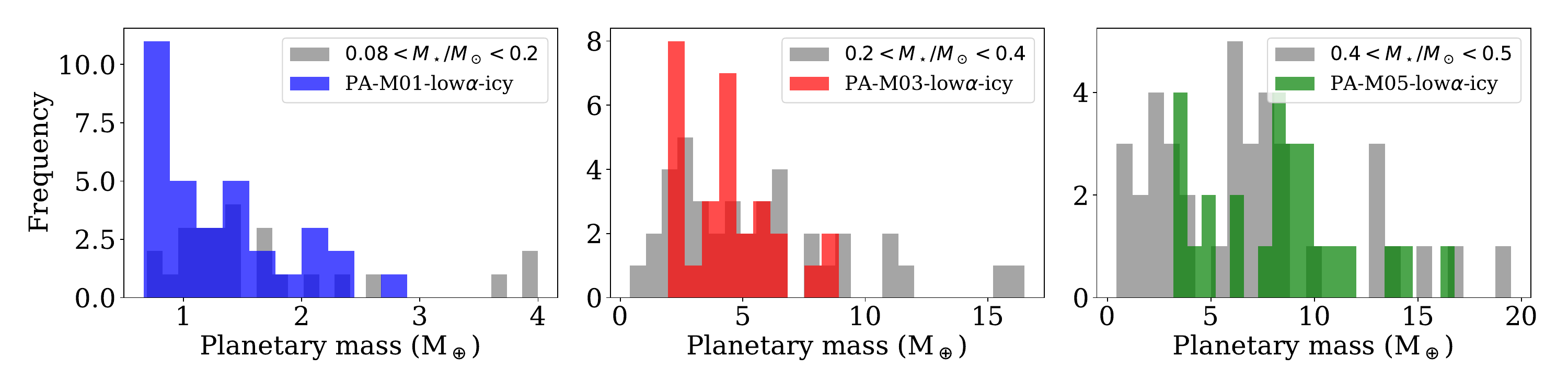}\\
    \includegraphics[width=0.95\textwidth]{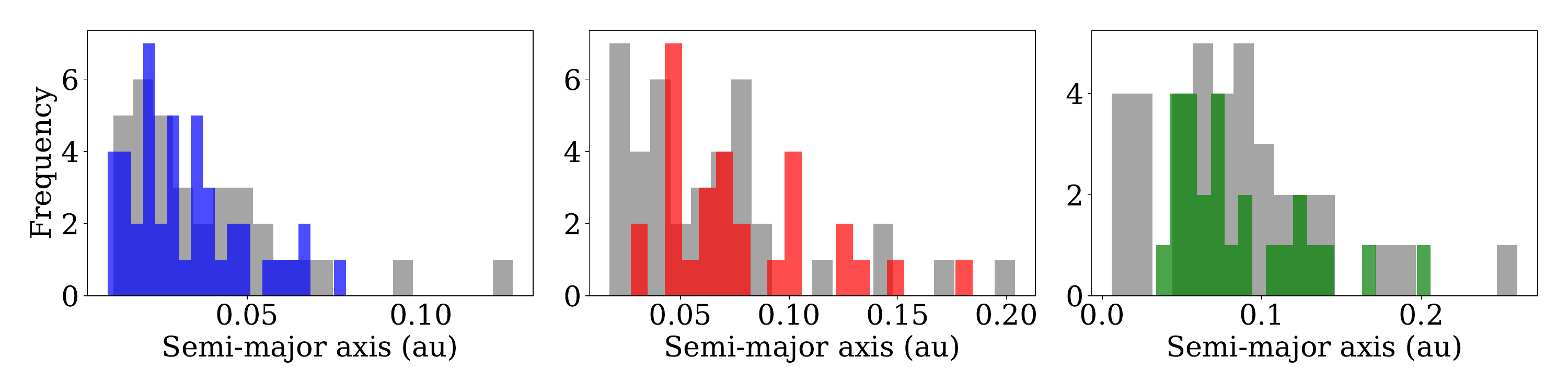}
    \caption{Histograms in planetary mass (top panels) and semi-major axis (bottom panels) of the observed low-mass close-in exoplanets around stars with masses $0.08<M_\star/\rm M_\odot<0.2$ (gray, left panels), with masses $0.2<M_\star/\rm M_\odot<0.4$ (gray, middle panels) and with masses $0.4<M_\star/\rm M_\odot<0.5$ (gray, right panels), and the resulting simulated planetary systems of the scenarios that included pebble accretion and low gas disk viscosity for a star of 0.1~M$_\odot$ (PA-M01-low$\alpha$-icy, blue), of 0.3~M$_\odot$ (PA-M03-low$\alpha$-icy, red), and 0.5~M$_\odot$ (PA-M05-low$\alpha$-icy, green).}
    \label{fig:histo-semiejesmasas-S1S5S6}
\end{figure*}

\begin{figure*} 
\includegraphics[height=8.5cm,width=6.2cm]{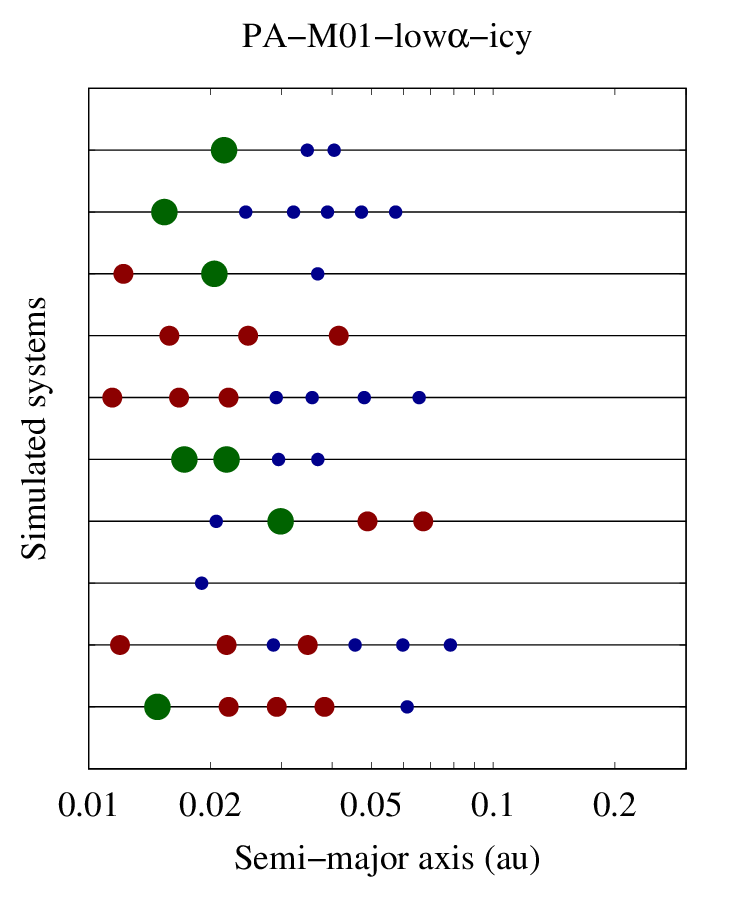}
\includegraphics[height=8.5cm,width=6cm]{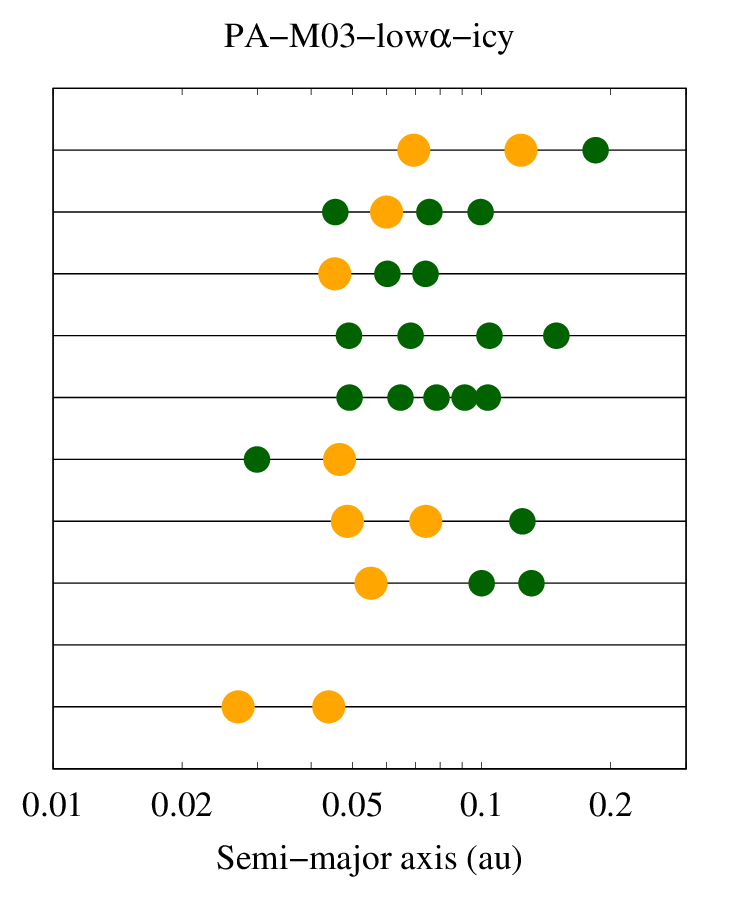}
\includegraphics[height=8.5cm,width=6cm]{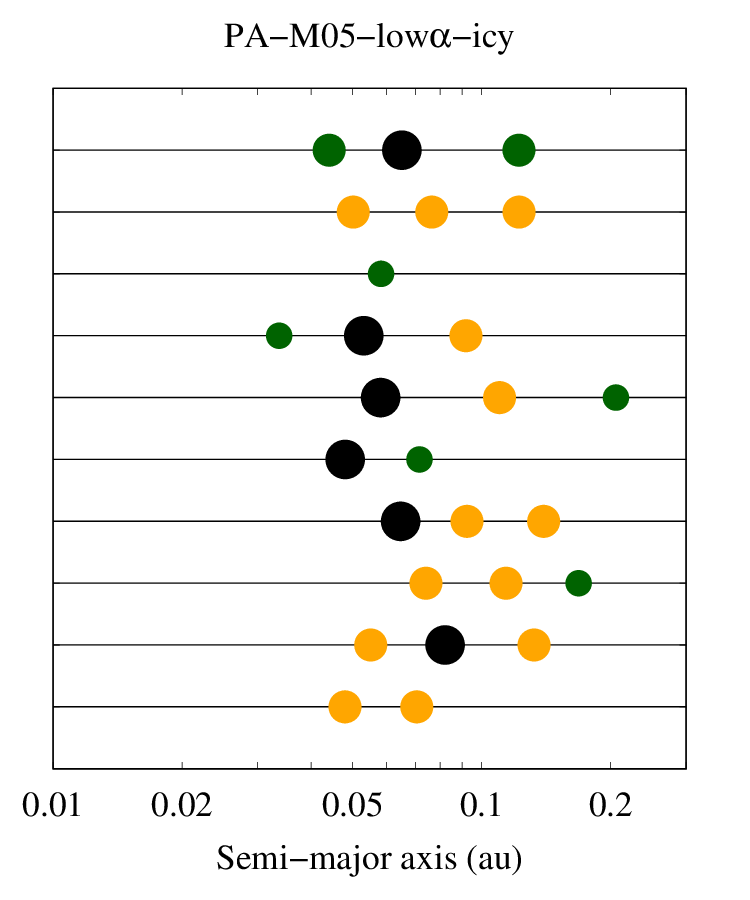}\\
\includegraphics[height=8.5cm,width=6.2cm]{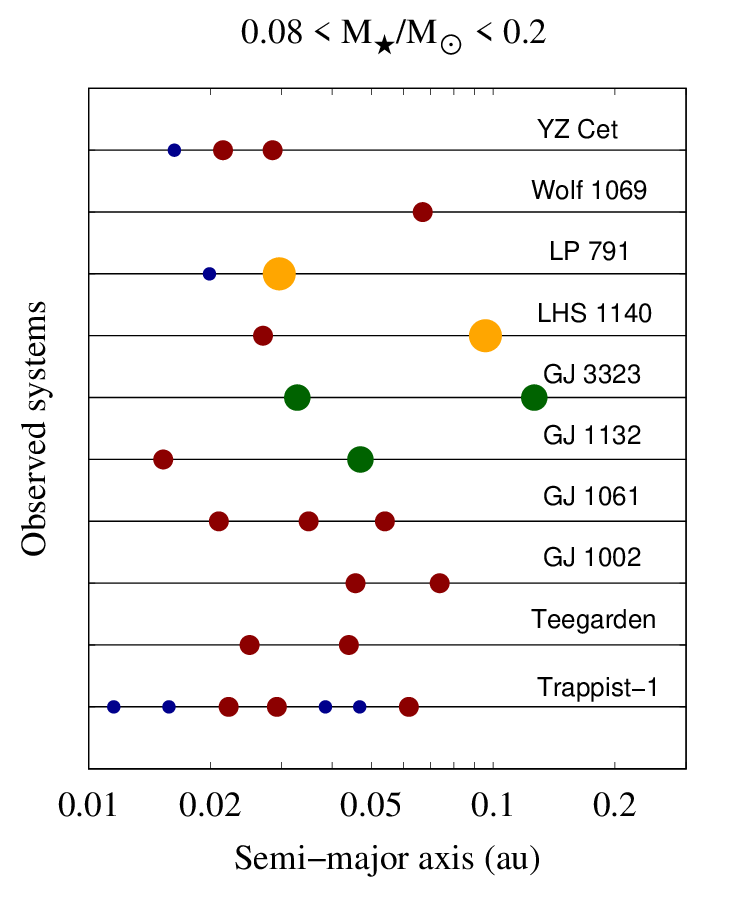}
\includegraphics[height=8.5cm,width=6cm]{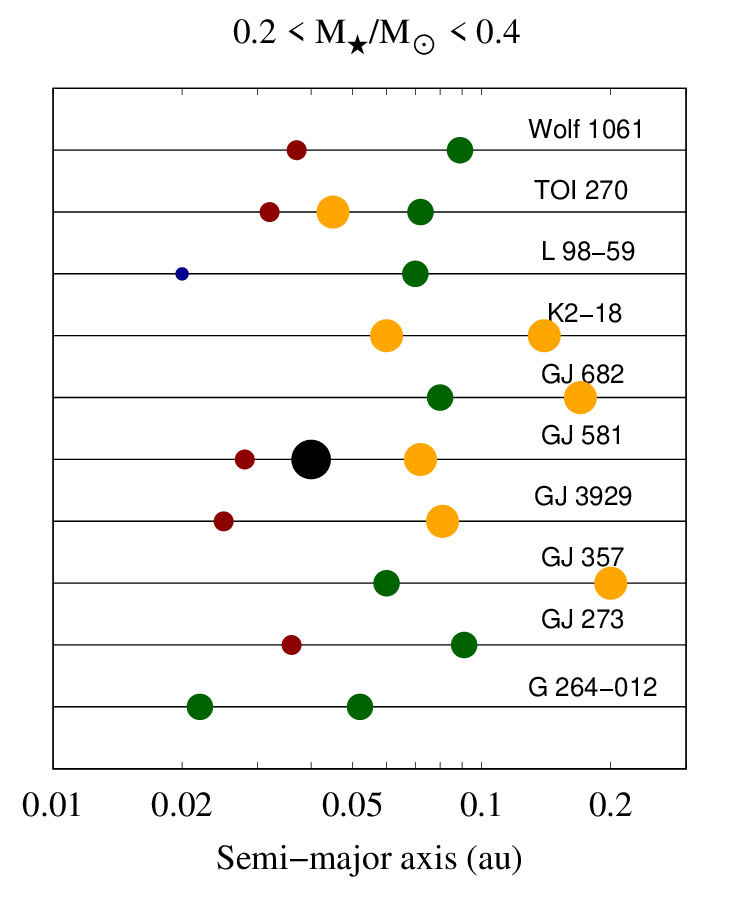}
\includegraphics[height=8.5cm,width=6cm]{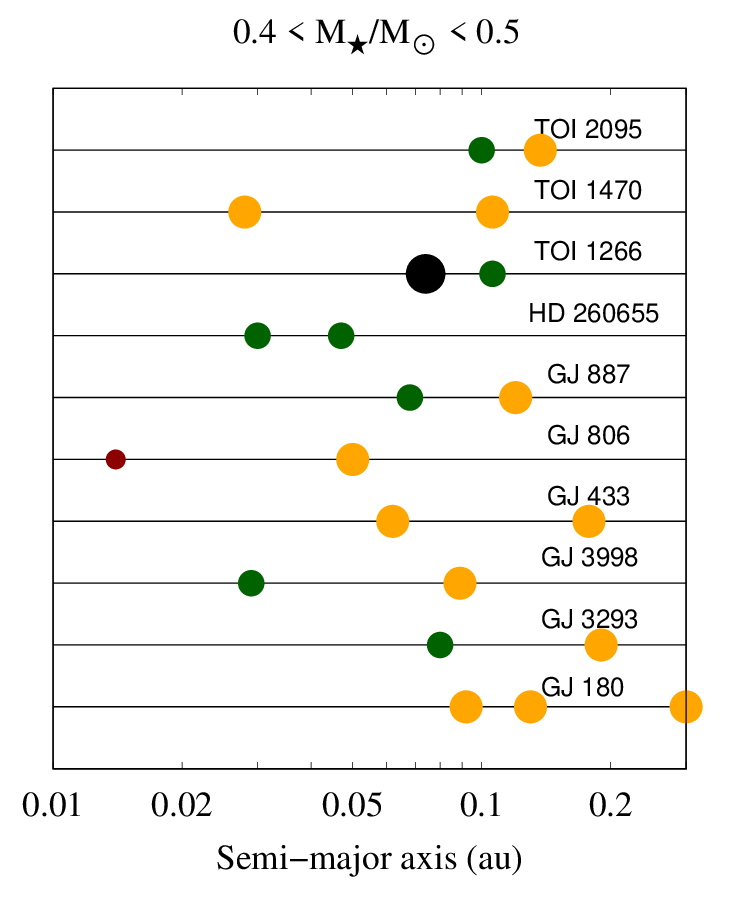}
    \caption{Planetary architectures around M dwarfs. Top panels: Simulated planetary systems of each simulation of the scenarios that include pebble accretion and low gas disk viscosity for a star of 0.1~M$_\odot$ (left), 0.3~M$_\odot$ (middle), and 0.5~M$_\odot$ (right). Bottom panels: Observed planetary systems around different stellar masses: $0.08<M_\star/\rm M_\odot<0.2$ (left), $0.2<M_\star/\rm M_\odot<0.4$ (middle), and $0.4<M_\star/\rm M_\odot<0.5$ (right). Planets are discriminated by their masses (as in Figure \ref{fig:sunmary}): $pM_p<1$~M$_\oplus$ (blue dots), $1<M_p/$M$_\oplus<2$ (red dots), $2<M_p/$M$_\oplus<5$ (green dots), $5<M_p/$M$_\oplus<10$ (orange dots), and $M_p/$M$_\oplus<20$ (black dots). From left to right, panels in the same column present simulated and observed planetary systems around stars with analogous masses for better comparison.}
    \label{fig:arc-S1}
\end{figure*}

\begin{figure}
    \centering    \includegraphics[width=9cm]{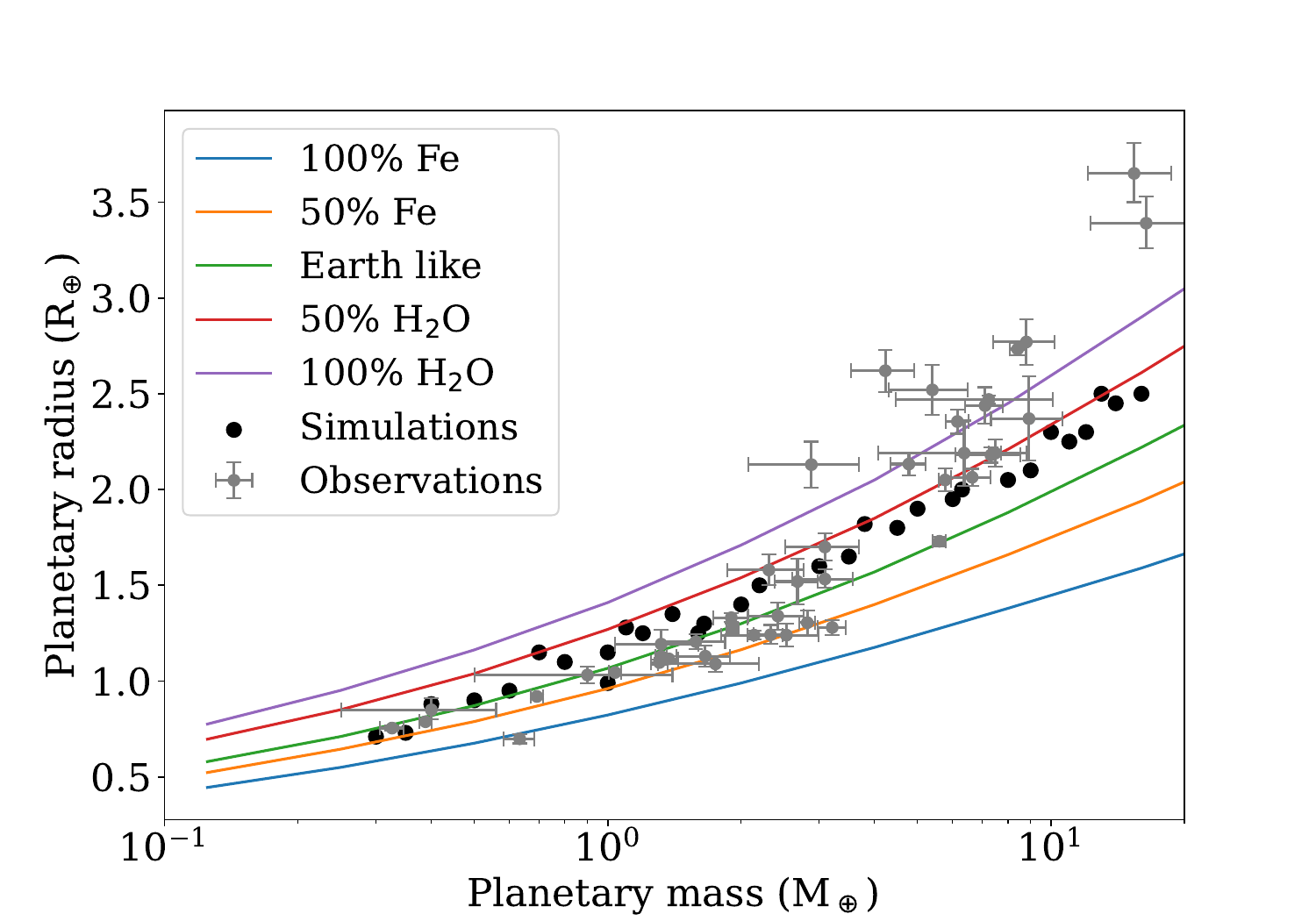}
    \caption{Different planetary bulk compositions (\cite{Zeng2008}): 100$\%$ Fe (blue line), 50$\%$ Fe (orange line), Earth-like composition (green line), 50$\%$ water in mass (red line), and 100 $\%$ water in mass (violet line), in comparison with the sample of low-mass close-in exoplanets around M dwarfs with masses and radius measurements (black dots) together with some simulated planets from different formation scenarios (the radius derived from the amount of water in mass calculated during the accretion process).}
    \label{fig:bulk_composition}
\end{figure}

\begin{figure*} 
\centering
\includegraphics[width=0.95\textwidth]{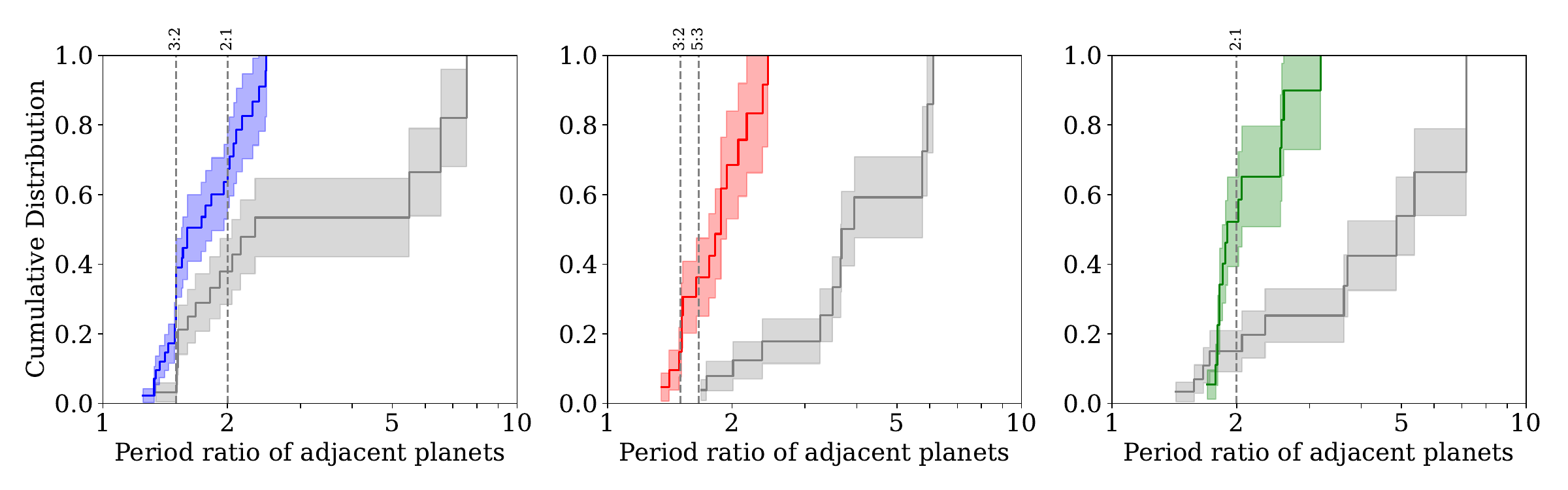}\\
\includegraphics[width=0.95\textwidth]{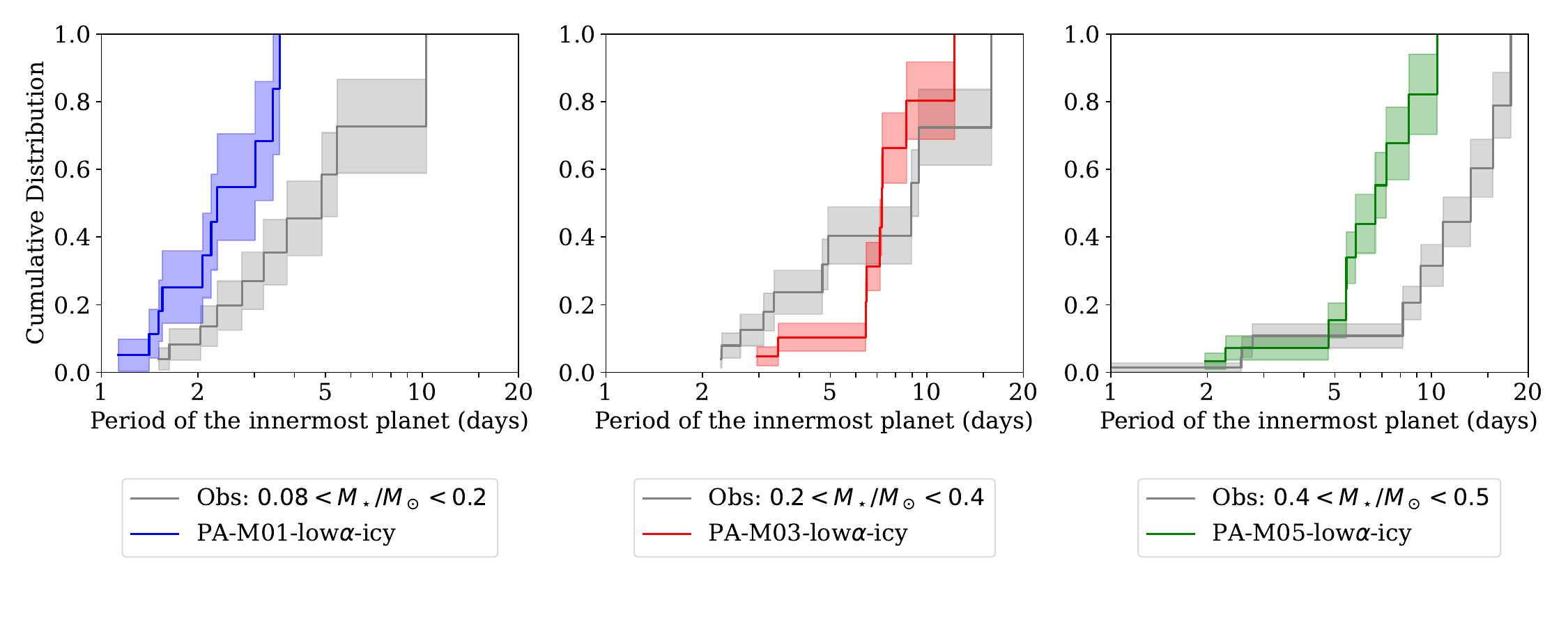}
    \caption{Cumulative distributions of planetary systems around M dwarfs. Left panels: Cumulative distributions of period ratio of adjacent planets (top) and of the period of the innermost planet of each system (bottom) for the simulated planets of the scenario that include pebble accretion and low gas disk viscosity for a star of 0.1 M$_\odot$ (blue line) together with the exoplanet sample around star with $0.08<M_\star/\rm M_\odot<0.2$ (gray line). Middle panels: Analogous distributions, but comparing the simulated planets for a star of 0.3 M$_\odot$ (red line) with the exoplanet sample around stars with $0.2<M_\star/\rm M_\odot<0.4$ (gray line). Right panels: Analogous distributions, but comparing the simulated planets from 0.5 M$_\odot$ (green line) with the exoplanet sample around stars with $0.4<M_\star/\rm M_\odot<0.5$ (gray line). The shadow areas represent the Poissonian errors of each distribution. The representative mean motion resonances are overplotted (top panels).}
   \label{fig:Pratio-Pinner}
\end{figure*}

\subsection{Planetary systems formed by pebble accretion: M$_\star=0.1$~M$_\odot$, M$_\star=0.3$~M$_\odot$, and M$_\star=0.5$~M$_\odot$}

We compared the simulated planets formed in the scenarios that included pebble accretion in a gas disk with low viscosity ($\alpha=10^{-4}$) with the exoplanets around M dwarfs, as such scenarios are the most efficient in forming planets bigger than 1 M$_\oplus$ (see Section \ref{sec:accretionhistory}). In Figure \ref{fig:exo-sample-S1S5S6} we show the planetary mass and semi-major axis of the simulated planets and the confirmed low-mass close-in exoplanets around M dwarfs of different masses. We can see how the simulated planets and the observed exoplanets present, in most of the cases, masses higher than the isolation mass. As we discuss in the previous section, the final masses could be achieved mostly from planet-planet collisions, especially for higher stellar masses. In addition, in Figure \ref{fig:histo-semiejesmasas-S1S5S6} we show the histograms in planetary mass and semi-major axis of the simulated planets in each scenario (around 30 planets each) with three different subsamples of exoplanets, each of them associated with different stellar masses: $0.08<M_\star/\rm M_\odot<0.2$ (35 planets), $0.2<M_\star/\rm M_\odot<0.4$ (39 planets), and $0.4<M_\star/ \rm M_\odot<0.5$ (38 planets). In each scenario, we generate planets close-in to the star that match the observations. In particular, regarding the planetary masses, the simulated planets from the scenario with a star of 0.1 M$_\odot$ overlap the exoplanets with masses between 0.7 and 5 M$_\oplus$, the ones from the scenario with a mass of 0.3 M$_\odot$ overlap the exoplanets with masses between 2 and 10 M$_\oplus$, and the ones from the scenario with 0.5 M$_\odot$ overlap the exoplanets with masses between 3 and 17 M$_\oplus$.\\ 

Secondly, we compared the simulated planetary systems with the observed systems with more than one planet detected. In Figure \ref{fig:arc-S1} we show the planetary architectures (mass and semi-major axis) of the simulated planets of the scenarios in which we include pebble accretion and low gas disk viscosity, together with each observed planetary system around stars of different masses: $0.08<M_\star/\rm M_\odot<0.2$, $0.2<M_\star/\rm M_\odot<0.4$ and $0.4<M_\star/\rm M_\odot<0.5$. We can see how the simulated planetary systems and the observed planetary systems share the same range of semi-major axis, planetary masses and number of planets per mass (see Table 3). We can also see the agreement in the fact that the innermost planets are located further from the star when increasing the stellar mass. In particular, for a star of 0.1~M$_\odot$ we could reproduce planetary systems analogous to Trappist-1 in 20$\%$ of the simulations. For a star of 0.3~M$_\odot$ we can reproduce some of the systems with two or three planets with analogous masses to the observed exoplanets, but in more compact configurations. Lastly, for a star of 0.5~M$_\odot$ we can reproduce a few of the systems in terms of mass and multiplicity, but also in more compact configurations. We highlight that in this case we can see simulated planets with masses higher than 10 M$_\oplus$ that so far are not been detected as part of multi-planetary systems, but as single planets. This could be because of biases in radial velocity and transit techniques in the observed planetary systems. We could be missing smaller planets at further distances from the star (transit technique bias) or small planets between more massive planets (radial velocity technique bias) that cannot be detected. \\
\indent In terms of percentage of water in mass, we can see some differences between the simulated planets and the observed exoplanet population. In Figure \ref{fig:bulk_composition}, we show planetary mass and radius of a sample of simulated planets and the observed exoplanet population with both measurements (mass and radius), together with different bulk compositions from \citet{Zeng2008}. Even though many of the simulated planets with masses $<$5 M$_\oplus$ have higher amount of water in mass than the observed ones, some of the simulated planets with masses between 2 and 3 M$_\oplus$ are matching the observations, as they are also water worlds ($\sim50 \%$ of water in mass). Moreover, some simulated planets of scenario PA-M01-low$\alpha$-mixed present similar compositions as the observed ones, as they are either dry or with a small percentage of water in mass. Thus, in order to better match the bulk composition of rocky exoplanets, in terms of water content, it is crucial to include an initial dry embryo distribution inside the snowline. 
On the other hand, most of the simulated planets with masses $>$5 M$_\oplus$ present less water in mass than the observed planets. In this case, to have a good contrast with observations, it will be crucial to take into account the influence of planetary atmosphere in the water content (see Section \ref{sec:discussion} for further discussion).\\

Additionally, we compared the period ratio of adjacent simulated planets and confirmed exoplanets in multiple systems. In Figure \ref{fig:Pratio-Pinner} we show the cumulative distributions of the period ratio of adjacent planets in each scenario that include pebble accretion and a low gas disk viscosity with the observed exoplanet sample around stars with similar masses. As it can be seen also from Figure \ref{fig:arc-S1}, many of the simulated planetary systems are in more compact configurations than the observed planetary systems, as the period ratio of the simulated planetary systems are on average two times lower than the observed ones. The simulated systems present period ratios of less than 3, while the observed systems present period ratios up to 8. We suggest that this discrepancy may be due to the fact that there are missing planets between the confirmed exoplanets, at least for the case of a star of 0.1~M$_\odot$. For example, currently it is not possible to detect planets with masses less than 3 M$_\oplus$ with orbital periods between 10 and 40 days (or with semi-major axis between 0.04 and 0.1 au) \citep{Carmenes2023}. For the more massive M dwarfs, which host more massive planets, it could be that our model is predicting more compact orbits because we are not considering later collisions among the planets at a Gyr timescale. This is something to be explored in future works.\\ 
From Figure \ref{fig:Pratio-Pinner} we could also infer that in the simulated systems there are some pairs of planets that are in commensurable orbits: $n:m$. Each pair of such adjacent planets satisfied that $nP_i=mP_{i+1}$, with $P_i$ and $P_{i+1}$ the orbital period of each adjacent planet and $n$ and $m$ two integer numbers. In the scenario with a star of 0.1 M$_\odot$, 20$\%$ of the planets are in commensurability 3:2  and 10$\%$ in 2:1. In the scenario with a star of 0.3 M$_\odot$, 15$\%$ of the planets are in 3:2 and almost 10$\%$ in 5:3. Lastly, in the scenario with a star of 0.5 M$_\odot$ just less than 10$\%$ of the planets are in 2:1. In most of the cases, such commensurability was reached after the last collisions among embryos. It happened earlier in the gas phase for a star of 0.1 M$_\odot$ and once the gas dissipated from the disk for the more massive M dwarfs. We can see also how there are more planets in commensurable orbits around the less massive M dwarfs, and the number is decreasing toward the more massive ones. This tendency is also seen in the observed exoplanet samples, even though it seems that our simulations are overestimating the number of planets in commensurable orbits. This trend seems to be in agreement with the lack of resonances found around Sun-like stars regarding the \textit{Kepler} sample, which is also shown from a theoretical perspective when modeling super-Earths around Sun-like stars \citep[e.g.,][]{Izidoro2017,Izidoro2021} .\\

Finally, in Figure \ref{fig:Pratio-Pinner} we also show the cumulative distributions of the period of the innermost planets in each scenario that include pebble accretion and a low gas disk viscosity and the related observed exoplanet sample. For each stellar mass, we obtained simulated innermost planets with periods as low as the confirmed exoplanet sample. For a star of 0.1~M$_\odot$ we could form innermost planets with periods less than 3 days,  however, we are not able to reproduce the ones with periods between 3 and 10 days (or semi-major axis between 0.02 and 0.04 au). For a star of 0.3~M$_\odot$ we could form innermost planets with periods 3 to 12 days, which is in agreement with observations, but we are underestimating the planets with period less than 5 days (or semi-major axis less than 0.04 au). Lastly, for a star of 0.5~M$_\odot$ we could form innermost planets with periods up to 12 days, but we are not able to reproduce the innermost planets with periods from 12 to 20 days (or semi-major axis between 0.07 and 0.11 au). We are also overestimating the planets with periods less than 12 days. Even though our simulated sample does not match perfectly with the observations, the difference between the innermost simulated and observed planets' periods is the only few days or between 0.02 and 0.04 au in semi-major axis. We think that such differences are due to the selection of the initial stellar rotation period that sets the inner edge of the disk. We claim that it could be possible to better reproduce the observations if we assumed different values of the stellar rotation period, given the range in the estimations of the rotational period of the stars, but this is out of the scope of this work.  

\noindent

\section{Discussion}
\label{sec:discussion}

We summarize some points to discuss regarding the simulated planetary systems from the different formation scenarios proposed, the initial parameters that we assumed, and the N-body model used to form them. Additionally, we highlight some facts regarding the comparison between the simulated planetary systems and the low-mass close-in exoplanet population around M dwarfs.

\subsection{Outcomes from the formation scenarios}

The main conclusion from the different formation scenarios that we explored is the fact that close-in super-Earth formation is possible around M dwarfs in compact disks with low gas disk viscosity ($\alpha=10^{-4}$) when the core accretion is driven by an efficient pebble radial drift (pebble efficiency between 30$\%$ and 60$\%$). We found a linear relationship between the super-Earth masses and the stellar mass, in agreement with previous works \citep[e.g.,][]{Liu2019}. However, it is not just the isolation mass that sets the final planetary masses, but also the gravitational interactions among planets that lead to collisions among them, and thus an increment in their masses. Late collisions among planets become more relevant toward the more massive M dwarfs. As we can see in Section \ref{sec:accretionhistory}, only around a star of 0.5 M$_\odot$ some simulated planets have masses higher than 10 M$_\oplus$. Interestingly, such planets had masses below 10 M$_\oplus$ while they were immersed in the gas disk. In every case, they reached higher masses after the gas dissipated from the disk, due to late collisions between planets.\\

    For the scenarios in which we study rocky planet formation around a star of 0.1 solar mass, it seems that the formation of Earth-like and super-Earth-like planets is independent of the initial location of the planetary embryos, in terms of their final masses and semi-major axis. However, the initial location of the embryos is important in terms of water content, as the scenario in which we considered a sample of embryos initially located inside the snowline is the only one that formed dry planets or planets with less than 10$\%$ of water in mass, more consistent with the observations \citep{Luque2022,Rogers2023}. Even though we did not test the impact of different masses for the initial embryo distribution, we expect that it will not change significantly the final planetary mass range that we found: this is shown by \cite{Pan2024} when studying planet formation around a star of 0.1 M$_\odot$ with $\alpha=10^{-4}$, comparing the final planetary masses with both an initial equal mass embryos distribution and an embryo distribution with masses derivate from streaming instability \citep{Liu2020}. However, it would be interesting to see the dependence between the initial mass of the embryos  with the final planetary masses in future works.\\ 
      
    On the other hand, assuming compact protoplanetary dust disks with higher viscosity ($\alpha=10^{-3}$) and lower pebble flux is not leading to an efficient planet formation, as all the simulated planets have masses $< 0.5~M_\oplus$. This is a crucial point as it seems that the migration speed of the embryos immerse in the gas disk as well as the pebble reservoir is directly related to this parameter. We clarify that the pebble flux proposed in this case is two orders of magnitude lower than the one proposed in a disk with $\alpha=10^{-4}$, as the gas density is almost two orders of magnitude lower (in average during the disk lifetime), than the gas density in a disk with lower viscosity. We made this assumption following \cite{LJ2014}, as the pebble flux seems to scale with the gas density, and the gas density increases toward lower values of gas viscosity. However, we are interested in studying the impact of changing the pebble flux, and thus the total pebble mass available for rocky planet assembly in future works to see the impact in the final planetary masses that could be achieved.\\

We found that core accretion by pebbles is efficient in forming planets bigger than Earth, while core accretion by planetesimals in efficient in forming also planets smaller than Earth. For a star of 0.1 M$_\odot$, no planets had masses less than 0.5~M$_\oplus$ when consider pebble accretion, while if we consider planetesimal accretion 55$\%$ of the planets had masses between 0.35 and 0.5 M$_\oplus$. For a star of 0.3~M$_\oplus$ no planets with masses less than 2~M$_\oplus$ could be formed by pebble accretion, but 30$\%$ of the simulated planets by planetesimal accretion had masses between 0.3 and 2 M$_\oplus$. Lastly, for a star of 0.5 M$_\oplus$ no planets with masses less than 3~M$_\oplus$ could be formed by pebble accretion, but 15$\%$ of the simulated planets by planetesimal accretion had masses between 0.5 and 4 M$_\oplus$. We highlight that in the planetesimal formation scenarios all the resulting planets are water-worlds with around 50$\%$ of water in mass, while in the pebble accretion scenarios with an initial embryos distribution outside the snowline there is a spread in water content between 20$\%$ and 50$\%$. \\

We note that in all the scenarios it seems that the final configurations of the planetary systems are related mainly to the planet-disk interactions, and that tidal star-planet interactions are playing a second role in the dynamical evolution of the planets. In order to understand the importance of tidal interactions in the early rocky planet formation stages, we re-run the simulations from the scenarios that included pebble accretion in a gas disk with low viscosity, switching off the tidal effects. We have seen that the resulting planetary systems present the same range in mass and semi-major axis as the ones formed when including tidal effects. However, we found differences in the multiplicity of the systems and the frequency of planetary masses. When we neglected tidal interactions, we always formed planetary systems with higher multiplicity, and thus a greater percentage of smaller planets per simulation. For a star of 0.1~M$_\odot$, 70$\%$ of the simulations present planetary systems with more than 5 planets, while for the star of 0.3~M$_\odot$ the percentage decrease to 60$\%$, and for a star of 0.5~M$_\odot$ to 40$\%$. On the contrary, when including tidal effects for a star of 0.1 M$_\odot$, 40$\%$ of the systems host more than 5 planets, for a star of 0.3 M$_\odot$ just 10$\%$, and for a star of 0.5 M$_\odot$ there is no planetary system with more than 5 planets. The higher multiplicity that we obtained when neglecting tidal effects is reflected in a higher percentage of smaller planets per scenario. We conclude that it is relevant to include tidal effects during rocky planet formation around M dwarfs, as it gives a more accurate estimate of the  number of planets per system and the frequency of planetary masses. Interestingly, it seems that tidal effects enhanced the collisions among embryos, which is highly contributing to the final planetary masses.  \\

\subsection{Simulated planets vs observed exoplanets around M dwarfs}

We highlight that in the formation scenarios in which we considered pebble accretion in a gas disk with low viscosity, the simulated planetary systems match the semi-major axis and masses of the observed low-mass close-in exoplanet sample less than 5~M$_\oplus$ for a star of 0.1~M$_\odot$, between 2 and 10 M$_\oplus$ for a star of 0.3~M$_\odot$ and between 3 and 17 M$_\oplus$ for a star of 0.5~M$_\oplus$. \\
Moreover, we reproduce the inward shift in semi-major axis of close-in exoplanets around M dwarfs in agreement with \cite{Sabotta2021}. As we can see from Figure \ref{fig:arc-S1}, the planetary systems are located closer-in for the star of 0.1~M$_\odot$, and further from the stars of 0.3~M$_\odot$ and 0.5~M$_\odot$. This is most probably related to the location of the inner edge of the disk for each stellar mass that we proposed (see Appendix A), which was based on the estimations of the stellar rotation period \citep{Bouvier2014,Scholz2018}. However, it seems we are overestimating innermost planets with periods less than three days. Thus, we will study the impact of the location of the inner gas disk radius in the location of the innermost planet population in future works.\\
Furthermore, we found that most of the simulated systems are closely spaced. For stars close to 0.1 M$_\odot$ this could be true if we take into account the observational biases toward close-in planets with masses less than 2~M$_\oplus$ \citep[e.g.,][]{Carmenes2023} and radius less than 0.9 R$_\oplus$ with transit technique \citep{Ment2023}. However, it seems that for the more massive M dwarfs we are overestimating compact planetary systems, maybe due to a short integration time for the N-body simulations. Extending the simulations over a few Gyr would let us see if late collisions among planets could take place and if the star-planet tidal interactions have enough impact in the dynamical evolution of the systems to break such compact configurations. Unfortunately, this is out of the scope of this work due to the amount of CPU time required to run the
N-body simulations. We note that other authors also find issues to break compact configurations in planetary systems around Sun-like stars \citep[e.g.,][]{Izidoro2017,Izidoro2021}. We point out that the mechanism that could make systems dynamically unstable is still under debate.\\  

In our simulations, we formed planets with masses between 0.35 and 10 M$_\oplus$ during the gas disk phase. Predominantly, rocky planets are expected to be smaller than $\approx 1.6$ R$_\oplus$ and with masses less than 5 M$_\oplus$ to 10 M$_\oplus$, while bigger and more massive planets are more luckily to have substantial gaseous envelopes \citep[e.g.,][]{Rogers2015}. Moreover, planets with masses between 0.3 and 2 M$_\oplus$ are considered to have a pure rocky composition, while planets between 2 and 10 M$_\oplus$ are, most probably, water worlds \cite{Luque2022}. We note that we could form planets with purely rocky cores just in the simulations that include an initial embryo distribution inside the snowline for a star of 0.1 M$_\odot$. On the other hand, we could form water worlds planets (planets with around 50$\%$ of water in mass) around stars of 0.3 and 0.5 M$_\odot$. In order to better constrain the final core composition of rocky planets it is necessary to include an initial sample of dry embryos located inside the snowline. Moreover, to better constrain the bulk composition of planets with masses higher than 5 M$_\oplus$, it will be important to take into account the fact that water worlds likely have steam atmospheres and this inflates the planetary radius with respect to most bulk models \citep{Venturini2020,Burn2024,Venturini2024}.\\ 

We note that there is a gap in planetary mass that we can see both in the confirmed exoplanet sample around M dwarfs and in the simulated planets from the scenarios that consider pebble accretion in a gas disk with low viscosity (see Figure \ref{fig:exo-sample-S1S5S6}). This gap seems to separate super-Earths and potential mini-Neptunes, which could be associated with the location of the radius valley for M dwarfs \citep{Cloutier2020,Venturini2024}. We note that in some of the simulations, we could form planets with masses bigger than the isolation mass during the gas disk phase, which could have retained a small fraction of gas in the envelope. In order to better constrain the core+envelope compositions of the simulated Super-Earths and potential mini-Neptunes, and confirm the relationship between the mass gap and radius valley, we need to include gas accretion in our model as well. This is something that we will study in future works and explore the probability that photoevaporation may not be necessary to explain the radius valley \citep[e.g.,][]{Berger2023}. In this case, it would also be interesting to study the effect of different disk lifetimes, as it will be related with the amount of gas available in the disk for planet gas accretion. However, this is something we did not test in this work as we did not consider planet gas accretion and most of the low-mass simulated planets formed achieve their final configurations early during the gas-disk phase.\\

Finally, we would like to highlight the likely existence of small, inner dust traps due to the location of the outermost planets in our simulations that halts the inward pebble drift when reaching the isolation mass. Unfortunately, as the location of the outermost planets is around 0.1 au for a star of 0.1 M$_\odot$ and 0.3 au for a star of 0.5 M$_\odot$, such dust traps will remain unresolved, even with the James Webb Space Telescope (JWST). The MINDS JWST team could just detect dust traps that masses bigger than the isolation masses associated with each stellar mass generate indirectly by comparing H$_2$O and CO$_2$ fluxes as the CO$_2$ snowline is further out \citep{Grant2023}. As our outermost planets are located inside both the H$_2$O and CO$_2$ snowline, JWST would not be able to confirm the existence of such small dust traps. However, such small-scale dust traps could exist as well, since they will still halt pebble drift and prevent millimeter-sized dust from disappearing entirely.

\section{Conclusions}
\label{sec:conclusions}

In this work we studied the formation of low-mass close-in planets ($M_p<20$~M$_\oplus$ and $a<0.3$~au) around M dwarfs from compact dust disks ($r_{out}<4$~au) with efficient radial drift to form planetary cores and with dust masses representative for initial disk dust masses from disk population models before radial drift effects ($15<M_{dust}/\rm M_\oplus<35$) \citep{Appelgren2023}.
In order to do that, we developed different sets of N-body simulations that included a sample of embryos growing either by pebble accretion (see Section \ref{sec:pebblemodel}) or planetesimal accretion immersed in a gas disk for 10 Myr and exposed to planet-disk interactions (during the gas disk lifetime) as well as star-planet tidal interactions and general relativistic corrections that include the evolution of the stellar luminosity, radius, and rotational period (see Appendix A and B). We proposed different formation scenarios, and in each of them we ran ten N-body simulations lasting 50 Myr. To start we considered a star of 0.1~M$_\odot$ and assumed different initial distributions of embryos (embryos outside the snowline, and embryos both inside and outside the snowline), two different viscosities of the gas disk ($\alpha=10^{-3}$ and $\alpha=10^{-4}$), and two accretion mechanisms (pebble accretion and planetesimal accretion). We also analyzed rocky planet formation around a star of 0.3~M$_\odot$ and a star of 0.5~M$_\odot$. In these cases we just changed the accretion mechanism as we considered samples of embryos that grow by pebble accretion and samples that grow by planetesimal accretion (see Section \ref{sec:scenarios} for a detailed description of all the formation scenarios). \\

In Figure \ref{fig:sunmary} we showed a summary of the different formation scenarios in terms of the initial conditions for the disk and planetary cores together with the resulting simulated planetary masses in each case. We list below the main conclusions of this work:\\

\begin{itemize}
    \item Close-in super-Earth formation
is possible around M dwarfs from compact dust disks with low pebble scale heights ($\alpha = 10^{-4}$) when the core accretion is driven by
an efficient pebble radial drift.\\
\item The final masses of the simulated close-in rocky planets around M dwarfs depend on the pebble isolation mass and on the collisions among planets.\\
\item The
simulated planetary systems from the formation scenarios in which we considered efficient pebble accretion in a gas disk with low viscosity ($\alpha=10^{-4}$) are the ones that match the semi-major axis and masses of the observed low-mass close-in exoplanet sample around M dwarfs.\\
\item At least for a star of 0.1 M$_\odot$, assuming a gas disk with high viscosity ($\alpha = 10^{-3}$) and low pebble flux does not allow the formation of either Earth-like or super-Earth-like planets.\\
\item Core accretion by planetesimal accretion seems to be efficient in forming planets smaller than the Earth. We note that the existence of such planets could not be confirmed as they cannot be detected with the current technology. \\
\item To reproduce the observed rocky exoplanets with Earth-like composition, it is necessary to start from an initial embryo population located inside the snowline that grows by efficient pebble accretion.\\ 
\end{itemize}

From the good match that we found in our model between our simulated planets and the exoplanet sample of low-mass close-in exoplanets around M dwarfs in terms of masses, semi-major axis and multiplicity, we conclude that rocky planet formation around M dwarfs most probably took place in compact dust disks with efficient radial drift and in a gas disk with low viscosity. This work highlights the disk conditions needed for rocky planet formation to take place around very low-mass stars.\\
    

\begin{acknowledgements}
     This work was performed using the compute resources from the Academic Leiden Interdisciplinary Cluster Environment (ALICE) provided by Leiden University. This work made use of the NASA Exoplanet Archive, which is operated by the California Institute of Technology, under contract with the National Aeronautics and Space Administration under the Exoplanet Exploration Program. G.D.M. acknowledges support from FONDECYT project 11221206, from ANID --- Millennium Science Initiative --- ICN12\_009, and the ANID BASAL project FB210003.
    The results reported herein benefitted from collaborations and/or information exchange within NASA’s Nexus for Exoplanet System Science (NExSS) research coordination network sponsored by NASA’s Science Mission Directorate and project “Alien Earths” funded under Agreement No. 80NSSC21K0593. We also thank the referee for his useful and constructive comments.
\end{acknowledgements}

%
%
\newpage

\bibliographystyle{aa} 
\bibliography{sanchez24} 

\begin{appendix} 
\section{Gas disk model}

\label{sec:diskmodel}

As in \cite{Liu2019} and \cite{Liu2020}, we adopted the disk model from \citet{Ida2016} in which the structure is 
described by the gas surface density profile $\Sigma_{\textrm{g}}$, the disk temperature 
 $T_{\textrm{g}}$ and a gas disk aspect ratio $h_{\textrm{g}}=H_{\textrm{g}}/r$ where $r$ is the radial coordinate in the mid-plane of the disk, and $H_{\textrm{g}}$ is the gas scale height, which depend on the 
 heating process. The model includes two different dominant heating mechanism: 
 the internal viscous dissipation for the inner disk and the irradiation from 
 the central star for the outer disk. For the inner region of the disk 
 $\Sigma_{\textrm{g}}$, $T_\textrm{g}$ and $h_\textrm{g}$ profiles are given by:
 
 \begin{equation}
 \begin{split}
 \Sigma_{\textrm{g},\textrm{vis}}= &2100\left(\frac{\dot{M_{\textrm{g}}}}{10^{-8}~\textrm{M}_\odot ~\textrm{yr}^{-1}}\right)^{3/5}\left(\frac{M_{\star}}{\textrm{M}_{\sun}}\right)^{1/5}\left(\frac{\alpha_{\textrm{g}}}{10^{-3}}\right)^{-4/5}
     \\
     &\left(\frac{r}{\textrm{au}}\right)^{-3/5}~\mathrm{g~cm^{-2}}
\end{split}
 \end{equation}

 \begin{equation}
 \begin{split}
 T_{\textrm{g},\textrm{vis}}= &200\left(\frac{\dot{M_{\textrm{g}}}}{10^{-8}~\textrm{M}_\odot ~\textrm{yr}^{-1}}\right)^{2/5}\left(\frac{M_{\star}}{\textrm{M}_{\sun}}\right)^{3/10}\left(\frac{\alpha_{\textrm{g}}}{10^{-3}}\right)^{-1/5}
     \\
     &\left(\frac{r}{\textrm{au}}\right)^{-9/10}~ \mathrm{K}
\end{split}
 \end{equation}

  \begin{equation}
 \begin{split}
 h_{\textrm{g},\textrm{vis}}= &0.027\left(\frac{\dot{M_{\textrm{g}}}}{10^{-8}~ \textrm{M}_\odot ~\textrm{yr}^{-1}}\right)^{1/5}\left(\frac{M_{\star}}{\textrm{M}_{\sun}}\right)^{-7/20}\left(\frac{\alpha_{\textrm{g}}}{10^{-3}}\right)^{-1/10}
     \\
     &\left(\frac{r}{\textrm{au}}\right)^{1/20}  
\end{split}
 \end{equation}
 where $M_\star$ is the mass of the central object, $\dot{M_\textrm{g}}$ 
 is the gas accretion rate and $\alpha_\textrm{g}$ is the viscous coefficient 
 related with the viscosity $\nu=\alpha_\textrm{g} c_\textrm{s}T_\textrm{g} H_\textrm{g}$, 
 where $c_\textrm{s}$ is the sound speed at the temperature of the disk mid-plane at a 
 given radial distance \citep{Shakura1973}.
 We assumed that the inner disk is optically thick with an average opacity
 $\kappa=1~\mathrm{cm^{2}~g^{-1}}$.

In order to smooth $\Sigma_\textrm{g,vis}$ at the inner edge of the disk we 
multiplied it with the term $\tanh[(r-r_{\rm inner})/(r_{\rm inner}h_{\rm inner})]$, where $r_{\rm inner}$ and $h_{\rm inner}$ are
the radius and aspect radius at the inner edge respectively, as it was suggested 
by \cite{Cossou2014}, \cite{Matsumura2017}, and \cite{Brasser2018}. Thus, the surface density profile in the viscous region can be expressed as follows,
\begin{equation}
 \Sigma_\textrm{g,vis}=\Sigma_\textrm{g,vis} \tanh[(r-r_{\rm inner})/(r_{\rm inner}h_{\rm inner})].\\
\label{eq:density_visc}
\end{equation}

 For the outer region of the disk, the corresponding profiles are given by:
 
 \begin{equation}
 \begin{split}
 \Sigma_{\textrm{g},\textrm{irr}}= &2700\left(\frac{\dot{M_{\textrm{g}}}}{10^{-8} ~\textrm{M}_\odot~ \textrm{yr}^{-1}}\right)\left(\frac{M_{\star}}{\textrm{M}_{\sun}}\right)^{9/14}\left(\frac{\alpha_{\textrm{g}}}{10^{-3}}\right)^{-1}
     \\
     &\left(\frac{L_{\star}}{\textrm{L}_{\sun}}\right)^{-2/7}\left(\frac{r}{\textrm{au}}\right)^{-15/14}~\mathrm{g~cm^{-2}}
\end{split}
 \end{equation}

 \begin{equation}
 T_{\textrm{g},\textrm{irr}}= 150\left(\frac{M_{\star}}{\textrm{M}_{\sun}}\right)^{-1/7}\left(\frac{L_{\star}}{\textrm{L}_{\sun}}\right)^{2/7}\left(\frac{r}{\textrm{au}}\right)^{-3/7}~ \mathrm{K}
 \end{equation}

  \begin{equation}
 h_{\textrm{g},\textrm{irr}}= 0.024\left(\frac{M_{\star}}{\textrm{M}_{\sun}}\right)^{-4/7}\left(\frac{L_{\star}}{\textrm{L}_{\sun}}\right)^{1/7}\left(\frac{r}{\textrm{au}}\right)^{2/7}
 \end{equation}
 where $L_{\star}$ is the luminosity of the central object and for simplicity
 the disk is assumed to be vertical optically thin. 
 
 The boundary $r_\textrm{tran}$ that separates the viscous region from the irradiated region is given by:
 
\begin{equation}
\begin{split}
    r_\textrm{tran}&=1.8\left(\frac{L_\star}{L_\odot}\right)^{-20/33} \left(\frac{M_\star}{M_\odot}\right)^{31/33} \left(\frac{\alpha}{10^{-3}}\right)^{-14/33} \\
    &\left(\frac{\dot{M_\textrm{g}}}{10^{-8}M_\odot yr^{-1}}\right)^{28/33}~\text{au}.
    \end{split}
\end{equation}
\\
In this model, the snowline is set at the radial location in the disk where $T_g\sim170$K and is defined as $r_{snow}\sim max(r_{snow,vis} , r_{snow,irr})$, where:
\begin{eqnarray}
 r_\textrm{snow,vis}&=&1.2\left(\frac{M_{\star}}{\textrm{M}_{\sun}}\right)^{1/3}\left(\frac{\alpha_{\textrm{g}}}{10^{-3}}\right)^{-2/9}\left(\frac{\dot{M_{\textrm{g}}}}{10^{-8}~\textrm{M}_\odot ~\textrm{yr}^{-1}}\right)^{4/9}~au,\\
 r_\textrm{snow,irr}&=&0.75\left(\frac{L_{\star}}{\textrm{L}_{\sun}}\right)^{2/3}\left(\frac{M_{\star}}{\textrm{M}_{\sun}}\right)^{-1/3}~ au.\\
\end{eqnarray}

The evolution of $\dot{M_\textrm{g}}$ in time $t$ is taken from the fit made by \citet{Manara2012} which is based on a large sample of accreting stars in the Orion Nebular Cluster, and it is given by:

\begin{equation}
\begin{split}
    \log\left(\frac{\dot M_{\textrm{g}}}{\textrm{M}_\odot~ \textrm{yr}^{-1}}\right)=&-5.12-0.46\log\left(\frac{t}{\textrm{yr}}\right)-5.75\log\left(\frac{M_{\star}}{\textrm{M}_{\odot}}\right)\\
    &+1.17\log\left(\frac{t}{\textrm{yr}}\right)\log\left(\frac{M_{\star}}{\textrm{M}_{\odot}}\right).
\label{eq:accretion_rate}
\end{split}    
\end{equation}
As it can be seen in Eq. \ref{eq:accretion_rate}, the accretion rate also depends on the stellar mass. For example, at 1 Myr, $\dot M_{\textrm{g}} \propto M_\star^{1.26}$, while at 10 Myr, $\dot M_{\textrm{g}} \propto M_\star^{2.43}$ \citep{Manara2012}, which is in agreement with the disk mass dependence with stellar mass \citep{Appelgren2023}.\\
We set a disk lifetime of $10~\textrm{Myr}$, as very low-mass stars could retain their primordial disks for longer times than more massive stars, up to tens of $\textrm{Myr}$ \citep[e.g.,][]{Damjanov2007,Bayo2012,Downes2015,ManzoM2020}. We assumed that the main responsible for gas dispersion is the accretion onto the star. We ignored the effect of the disk photoevaporation produced by the radiation from the star, as photoevaporation rates of gas disk around VLMS are estimated to be as low as $\sim10^{-11}~M_\odot/\textrm{yr}$ \citep{herczeg2007}, which is lower than the gas accretion rates onto the star during the disk lifetime. \\

\begin{figure}
    \centering \includegraphics[width=8.5cm]{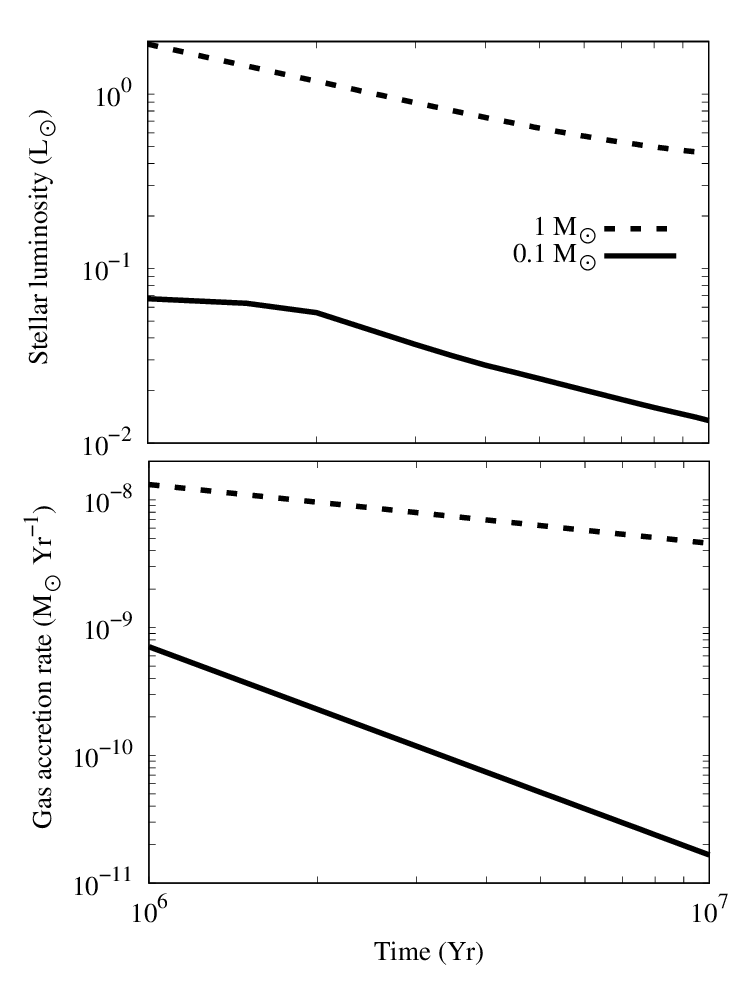}
luminosity    \caption{Evolution of stellar parameters. Top: evolution of the stellar luminosity of a star of 0.1 M$_\odot$ (solid line) and of a star of 1 M$_\odot$ (dash line). Bottom: evolution of the gas accretion rate for the same stellar masses as in the top panel.}
    \label{fig:tasa-Ls}
\end{figure}
\begin{figure} 
\includegraphics[width=9cm]{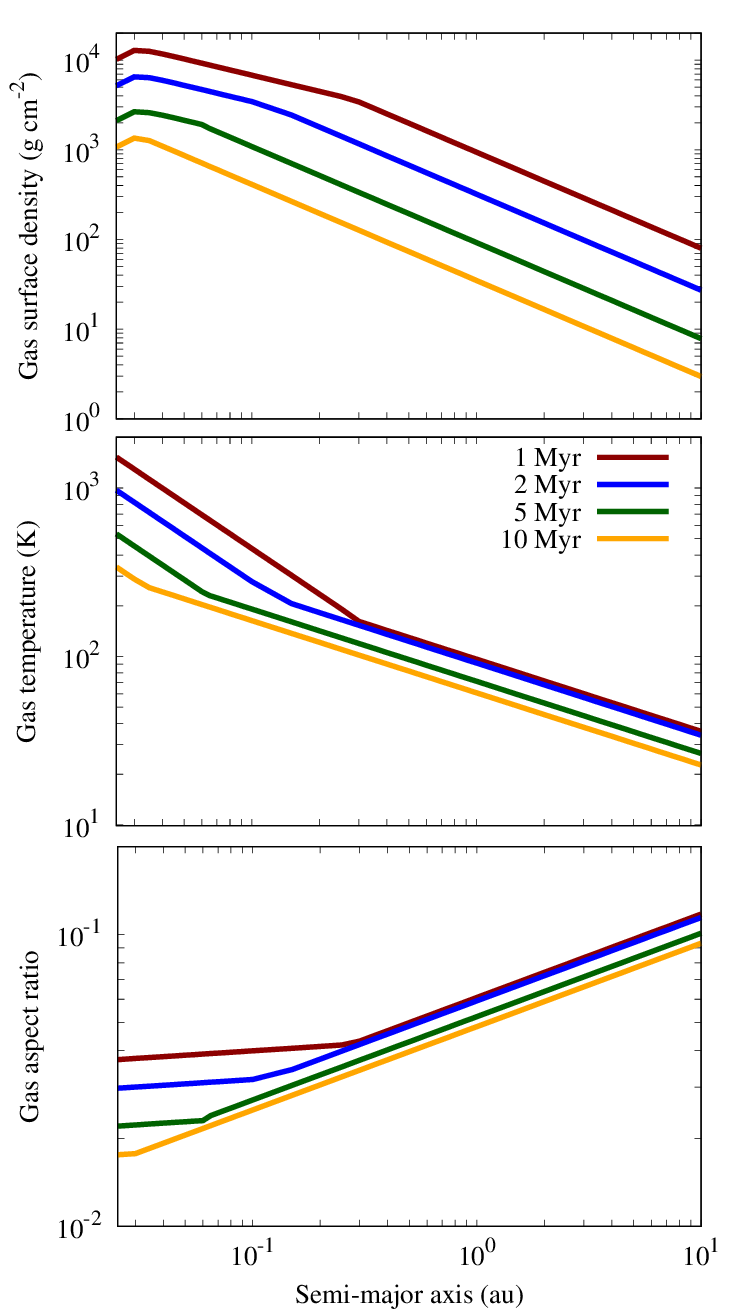}
    \caption{Gas surface density (top panel), gas mid-plane temperature (middle panel), and gas aspect ratio (bottom panel), assuming a host star of 0.1 M$_\odot$ and $\alpha=0.0001$ at different times along the disk lifetime: 1 Myr (red), 2 Myr (blue), 5 Myr (green) and 10 Myr (orange).}
    \label{fig:gas_profiles}
\end{figure}

In this work we consider as standard model the scenario where the central star has a mass $M_\star=0.1$~M$_\odot$. In Figure \ref{fig:tasa-Ls} we show the gas accretion rate and the stellar luminosity of a star of 0.1 M$_\odot$, as the evolution of the gas component of the protoplanetary disk depends on these two parameters. For comparison, we also show the corresponding stellar parameters of a star of 1 M$_\odot$. On the other hand, in Fig. \ref{fig:gas_profiles} we show the evolution of the gas surface density, the gas temperature in the mid-plane and the gas aspect ratio for a star of 0.1 M$_\odot$ along the disk lifetime. \\

 In all the scenarios that we propose, the initial parameters were set at 1 Myr. We fixed the value $r_\textrm{inner}(t)=r_\textrm{inner}(1 Myr)$ thought the disk lifetime. We assumed that the inner edge of the disk is located at the same distance as the co-rotation radius (radius where the stellar rotational period is equal to the orbital rotation period). Thus, for a star of 0.1 M$_\odot$,
considering a initial stellar rotation period $P_\star = 3$~days \citep{Bolmont2011,Scholz2018}, the inner edge of the disk is $r_\textrm{inner}=0.02$~au. In the case of the stars of 0.3 M$_\odot$ and 0.5 M$_\odot$, assuming $P_\star = 10$~days \citep{Bouvier2014}, the inner edge of the disks are $r_\textrm{inner}=0.06$~au and $r_\textrm{inner}=0.07$~au, respectively.

\section{External forces in the N-body code}

 The external forces that we incorporated in \textsc{Mercury} code estimate the acceleration corrections for a sample of embryos due to planet-disk interactions, star-planet tidal interactions and general relativistic corrections. We describe below the different external effects.
 
\subsection{Planet-disk interactions}
\label{sec:planet-disk-interac}

 Planet interactions with the gas disk lead to migration and orbital decay of the sample of embryos, and are crucial in the orbital configurations of planetary systems during the gas phase. In this section, we summarize the considerations assumed for estimating such interactions. We consider the same gas disk model described in Sec. \ref{sec:diskmodel}. The torques that the gas exert on the embryos or planets were computed using the non-isothermal disk model from \citet{Paardekooper2010,Paardekooer2011}, which includes thermal and viscous diffusion. The total torque over each embryo or planet is given by:
\begin{equation}
    \Gamma_\textrm{total} = \Delta_\textrm{L}\Gamma_{\textrm{L}} +  \Delta_\textrm{C}\Gamma_{\textrm{C}},
\label{eq:torquetotal}
\end{equation}
where $\Delta_L$ and $\Delta_C$
are the reduction factors for non-circular nor co-planar planetary orbits. The factors $\Gamma_\textrm{L}$ and $\Gamma_\textrm{C}$ represent, respectively, the Lindblad and corotation torques for a circular and co-planar motion given by:
\begin{equation}
    \Gamma_{\textrm{L}}=(-2.5-1.7\beta+0.1\delta)\frac{\Gamma_{0}}{\gamma_{\textrm{eff}}}
\end{equation}
and
\begin{equation}
\begin{split}
    \Gamma_{\textrm{C}}= & \Gamma_{\textrm{c},\textrm{hs},\textrm{baro}} F(p_{\nu})G(p_{\nu})+(1-K(p_{\nu})) \Gamma_{\textrm{c},\textrm{lin},\textrm{baro}} + \\ & \Gamma_{\textrm{c},\textrm{hs},\textrm{ent}}  F(p_{\nu}) F(p_{\chi})  \sqrt{G(p_{\nu})G(p_{\chi})} + \\ & \sqrt{(1-K(p_{\nu}))(1-K(p_{\chi}))} \Gamma_{\textrm{c},\textrm{lin},\textrm{ent}}.
    \end{split}
    \label{eq:corot_T}
\end{equation}
where $\Gamma_{\textrm{c},\textrm{hs},\textrm{baro}}$ and $\Gamma_{\textrm{c},\textrm{lin},\textrm{baro}}$ are barotropic terms related respectively with the horseshoe drag and the linear corotation torque, and $\Gamma_{\textrm{c},\textrm{hs},\textrm{ent}}$ and $\Gamma_{\textrm{c},\textrm{lin},\textrm{ent}}$ are their corresponding non-barotropic entropy counterparts. These terms are given by:

\begin{equation}
\Gamma_{\textrm{c},\textrm{hs},\textrm{baro}}=1.1\left(1.5-\delta\right)\frac{\Gamma_{0}}{\gamma_{\textrm{eff}}},
\end{equation}

\begin{equation}
\Gamma_{\textrm{c},\textrm{lin},\textrm{baro}}=0.7\left(1.5-\delta\right)\frac{\Gamma_{0}}{\gamma_{\textrm{eff}}},
\end{equation}

\begin{equation}
\Gamma_{\textrm{c},\textrm{hs},\textrm{ent}}=7.9 \epsilon \frac{\Gamma_{0}}{\gamma_{\textrm{eff}}^{2}},
\end{equation}

\begin{equation}
\Gamma_{\textrm{c},\textrm{lin},\textrm{ent}}=\left(2.2-\frac{1.4}{\gamma_{\textrm{eff}}}\right) \epsilon \frac{\Gamma_{0}}{\gamma_{\textrm{eff}}},
\end{equation}
where the scaling torque is $\Gamma_0=\left(M_\textrm{p}/M_\star\right)^2\Sigma_\textrm{g}r^4h_{\textrm{g}}^{-2}\Omega_\textrm{k}^2$, with angular Keplerian velocity $\Omega_\textrm{k}$. The negative of the entropy slope is $\epsilon=\beta-(\gamma-1)\delta$, with $\delta=-d\ln\Sigma_{\textrm{g}}/d\ln r$,  $\beta=-d\ln T_{\textrm{g}}/d\ln r$ and $\gamma=1.4$ the adiabatic index. The effective $\gamma_\textrm{eff}$ is given by:

\begin{equation}
\gamma_{\textrm{eff}}=\frac{2Q \gamma}{\gamma Q+0.5\sqrt{2\sqrt{(\gamma^{2} Q^{2}+1)^{2}-16Q^{2}(\gamma-1)}+2 \gamma^{2} Q^{2} -2}}
\end{equation}

which is related to thermal diffusion by the coefficients $Q=2\chi/3h_\textrm{g}^3r^2\Omega_\textrm{k}$ and $\chi = 16 \gamma (\gamma-1) \sigma T_\textrm{g}^4/[3 \kappa (\rho_\textrm{g}h_\textrm{g} r \Omega_\textrm{k})^2]$, with $\sigma$ the Stefan-Boltzmann constant, $\kappa$ the gas opacity and $\rho_\textrm{g}$ the volumetric gas density $\rho_\textrm{g}=\Sigma_\textrm{g}/
(H_\textrm{g} \sqrt{2\pi})$.

Additionally, the functions $F(p)$, $G(p)$ and $K(p)$ from the Eq. \ref{eq:corot_T} are given by:

\begin{equation}
F(p)=\frac{1}{1+\left(\frac{p}{1.3}\right)^2},
\end{equation}

\begin{equation}
G(p) = \left\lbrace
\begin{array}{lll}
\frac{16}{25}\left(\frac{45\pi}{8}\right)^{3/4} p^{3/2} & \textup{if } p<\sqrt{\frac{8}{45\pi}} \\
 1-\frac{9}{25}\left(\frac{8}{45\pi}\right)^{4/3} p^{-{8/3}} & \textup{if } p\geq\sqrt{\frac{8}{45\pi}} \end{array}
\right.,
\end{equation}

\begin{equation}
K(p) = \left\lbrace
\begin{array}{ll}
\frac{16}{25}\left(\frac{45\pi}{28}\right)^{3/4} p^{3/2} & \textup{if } p<\sqrt{\frac{28}{45\pi}} \\
 1-\frac{9}{25}\left(\frac{28}{45\pi}\right)^{4/3} p^{-{8/3}} & \textup{if } p\geq\sqrt{\frac{28}{45\pi}} \end{array}
\right..
\end{equation}
\\
The functions are evaluated in $p$ which takes the form of $p_\nu$, the saturation parameter associated with viscosity or $p_\chi$ the saturation parameter related with thermal diffusion. Both parameters are given by

\begin{equation}
p_{\nu}=\frac{2}{3}\sqrt{\frac{r^{2}\Omega_{\textrm{k}}x_{\textrm{s}}^{3}}{2\pi\nu}},
\end{equation}
where $x_\textrm{s}=(1.1/\gamma_\textrm{eff}^{0.25})\sqrt{M_\textrm{p}/(M_{\star}h_\textrm{g})}$ is the non-dimensional half-width of the horseshoe region and

\begin{equation}
p_\chi=\sqrt{\frac{r^2\Omega_\textrm{k}x_\textrm{s}^3}{2\pi\chi}}.
\end{equation}
All the previous formulas used to calculate Lindblad and corotation torques are evaluated at the semi-major axis $a$ of the embryo's orbit.\\

 To estimate the reduction factors for non-circular nor co-planar orbits, as well as for the estimation of the acceleration that the embryos or planets suffered due to the gas disk torques, we used the new prescription from \cite{Ida2020}. The authors
 studied the gravitational interactions between the gas and the embryos on the basis of dynamical friction, resulting in reduction factors given by:

 \begin{equation}
    \Delta_\textrm{L}=\left(1+\frac{C_\textrm{P}}{C_\textrm{M}}\sqrt{e_\textrm{rat}^{2}+i_\textrm{rat}^{2}}\right)^{-1},
    \label{eq:deltaLida}
 \end{equation}
 
 \begin{equation}
     \Delta_\textrm{C}=\exp{\left(-\frac{\sqrt{e^{2}+i^{2}}}{e_\textrm{f}}\right)},
\label{eq:deltaCida}    
 \end{equation}
 where $C_\textrm{P}=2.5-0.1\delta+1.7\beta$, $C_\textrm{M}=6(2\delta-\beta+2)$, $e_\textrm{rat}=e/h_\textrm{g}$, $i_\textrm{rat}=i/h_\textrm{g}$ and $e_\textrm{f}=0.5h_\textrm{g}+0.01$.
 
In cylindrical coordinates $(r,\theta,z)$ the acceleration terms for an embryo with a velocity $\vec{v}=(v_\textrm{r},v_\theta,v_\textrm{z})$ are given by:

\begin{equation}
    \textbf{f}_{\textrm{gas}}=-\frac{v_\textrm{r}}{t_\textrm{e}}\hat{e_\textrm{r}}-\frac{(v_\theta-v_\textrm{k})}{t_\textrm{e}}\hat{e_\theta}-\frac{v_\textrm{k}}{2t_\textrm{a}}\hat{e_\theta}-\frac{v_\textrm{z}}{t_\textrm{i}}\hat{e_\textrm{z}},
\end{equation}
where $\hat{e_\textrm{r}}$, $\hat{e_{\theta}}$ and $\hat{e_\textrm{z}}$ are versors in the respective directions. The gas velocity is given by $\vec{v_\textrm{g}}=(0,(1-\eta)v_{\textrm{k}},0)$, with $v_\textrm{k}$ the Keplerian velocity. The $t_\textrm{a}$, $t_\textrm{e}$ and $t_\textrm{i}$ represent, respectively, the damping timescales of the semi-major axis $a$, eccentricity $e$ and inclination $i$ of the embryo's orbit.
Considering that the embryo migration due to its interaction with the gas is non-isothermal and assuming the condition $i<h_\textrm{g}$, the damping timescales can be expressed as follows

\begin{equation}
    t_{\textrm{a}} = -\frac{t_{\textrm{wave}}}{2h_{\textrm{g}}^{2}}\frac{\Gamma_0}{\Gamma_\textrm{total}},
\end{equation}

\begin{equation}
    t_\textrm{e} = \frac{t_\textrm{wave}}{0.78}\left[1+\frac{1}{15}(e_\textrm{rat}^{2}+i_\textrm{rat}^{2})^{3/2}\right],
\label{eq:teIDA}
\end{equation}

\begin{equation}
     t_\textrm{i} = \frac{t_\textrm{wave}}{0.544}\left[1+\frac{1}{21.5}(e_\textrm{rat}^{2}+i_\textrm{rat}^{2})^{3/2}\right],
\label{eq:tiIDA}
\end{equation}
where
\begin{equation}
t_{\textrm{wave}} = \left(\frac{M_{\star}}{M_{\textrm{p}}}\right)\left(\frac{M_{\star}}{\Sigma_{\textrm{g}}r^{2}}\right)h_{\textrm{g}}^{4}\Omega_{\textrm{k}}^{-1},
\label{eq:twave}
\end{equation}
is the timescale from \citet{Papaloizou2000} and \citet{Tanaka2004} in which all physical parameters are evaluated at the semi-major axis of the orbit.\\

\begin{figure}
    \centering\includegraphics[width=9cm]{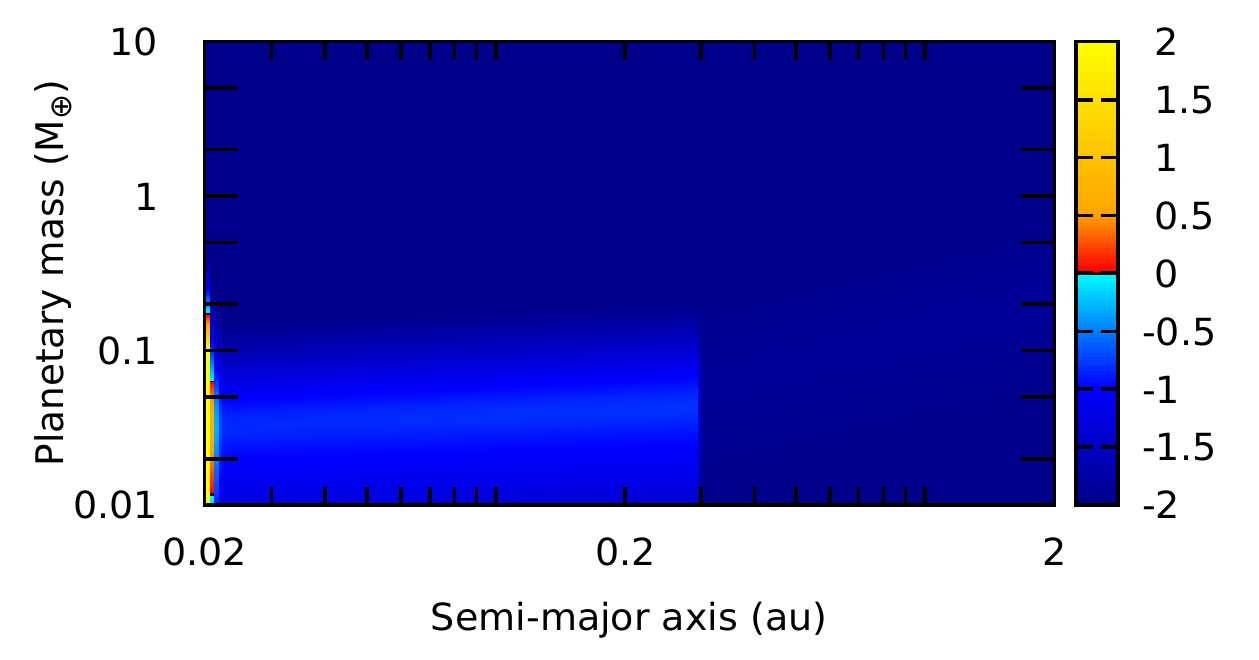}
    \caption{Normalized gas disk total torque (see Eq. B.1) map at 1 Myr for a host star of 0.1 M$_{\odot}$, $\alpha=0.0001$, for a sample of protoplanetary embryos masses and semi-major axis, with $e=0.01$. }
    \label{fig:example-torque}
\end{figure}

For our standard model, we show in Figure \ref{fig:example-torque} the normalized total toques $\Gamma_{\textit{total}}/\Gamma_0$ associated with a host star of 0.1 M$_{\odot}$ with $\alpha=10^{-4}$, for a range of planetary embryos masses and semi-major axis, assuming  $e=0.01$ and considering the gas disk model at 1 Myr. It can be seen that except for a narrow area close to the inner edge of the disk and masses lower than 0.3 M$_{\oplus}$ where the total torques are positive, for the rest of the combinations of mass and semi-major axis, the total torque reminds negative.  For a planet in a quasi-circular orbit, this means that a negative torque indicates an inward migration of the planet, while a positive torque indicates an outward migration. This pattern in the sign of the total torque remind the same at different times along the disk lifetime. The difference is that the area where the torques are positive is even narrower and closer to the inner edge of the disk for even lower values of planetary masses, until it almost disappears at 10 Myr.

\subsection{Star-planet tidal interactions and general relativistic correction}

\begin{figure*}
    \centering
    \includegraphics[width=9.15cm]{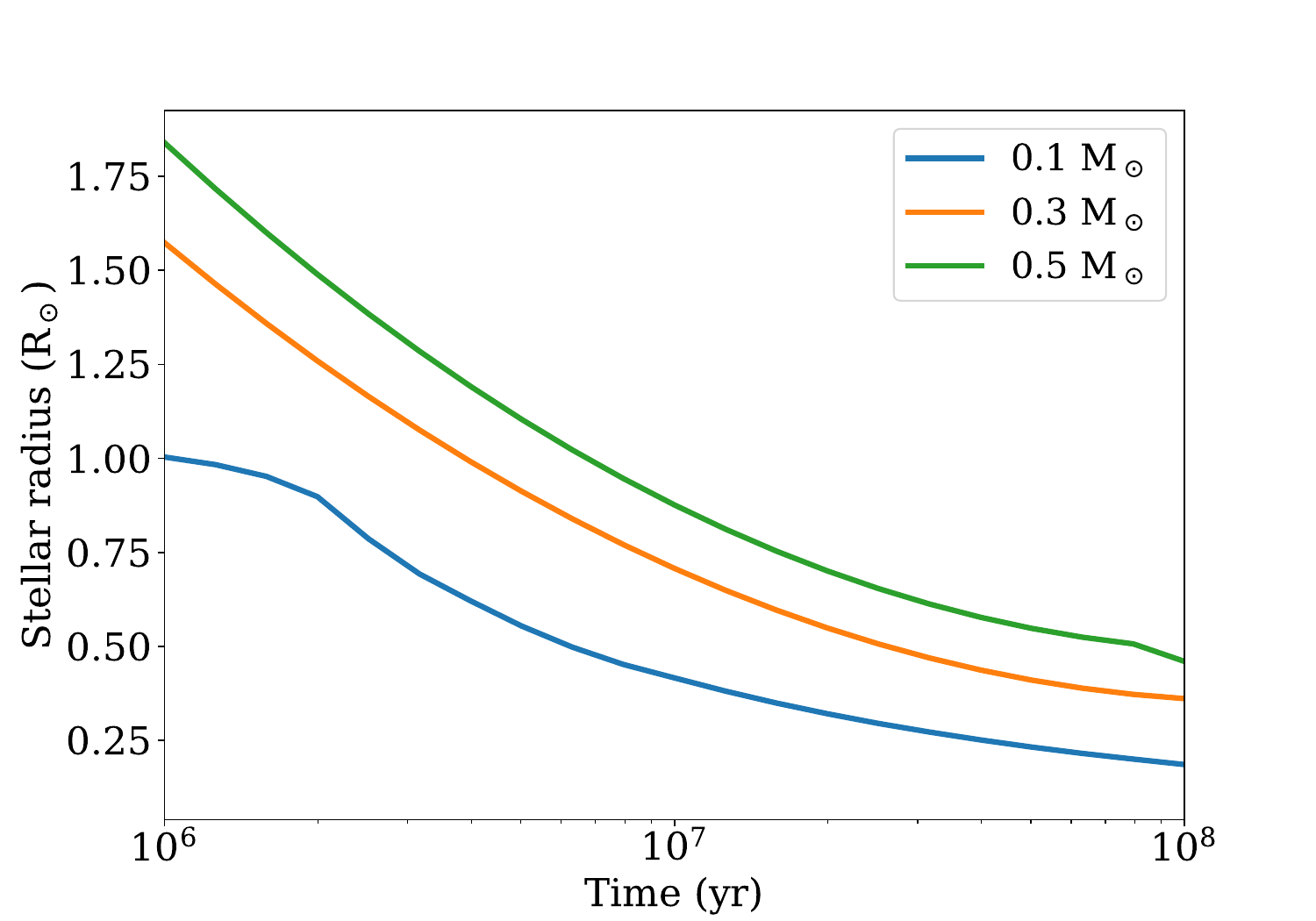}
    \includegraphics[width=9.15cm]{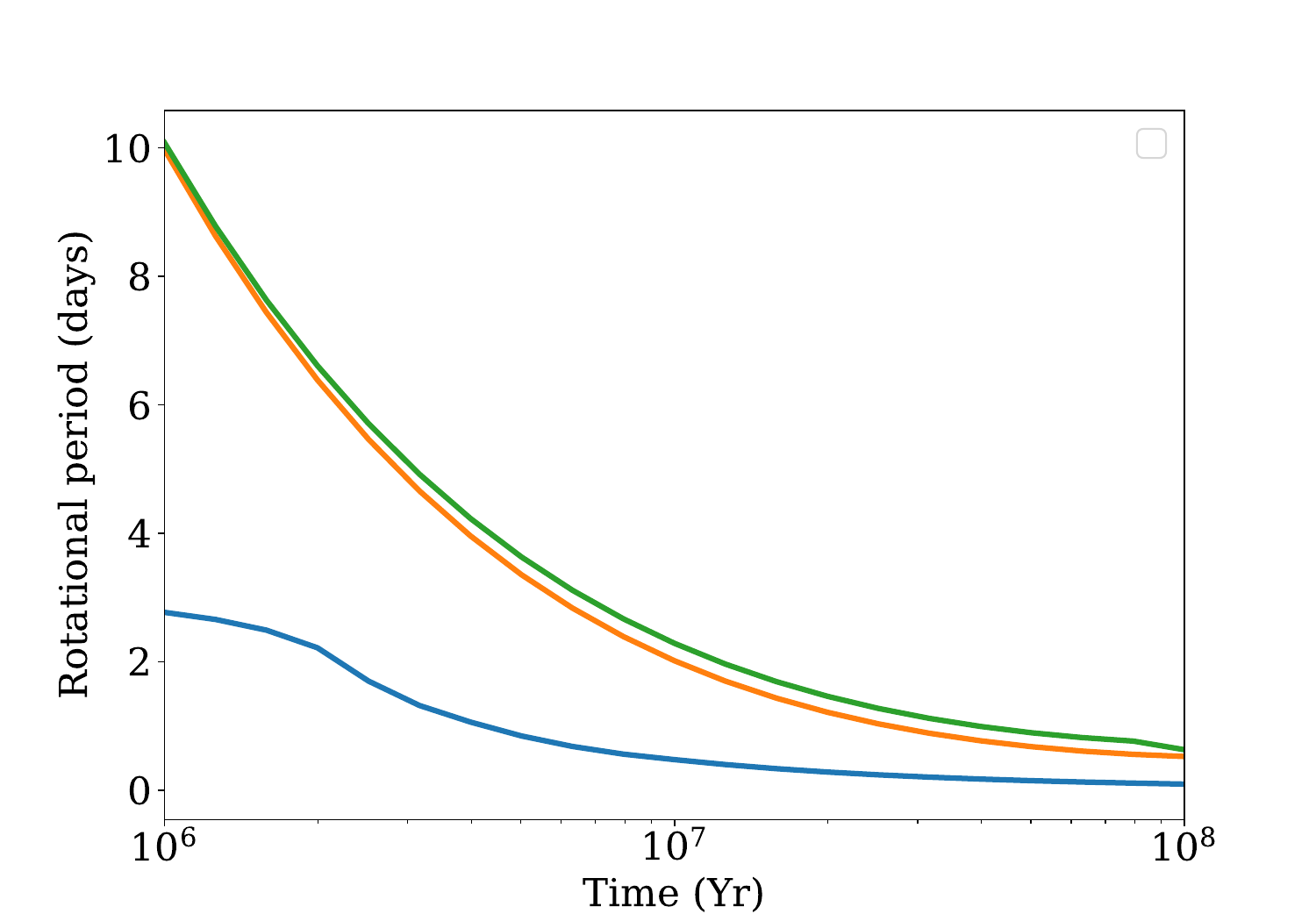}
    \caption{Evolution of the stellar radius (left) and the stellar rotational period (right) of a star of 0.1 M$_\odot$ (blue lines), a star of 0.3 M$_\odot$ (orange lines) and 0.5 M$_\odot$ (green lines).}
    \label{fig:evRyP}
\end{figure*}

Tidal interactions between the planets (or embryos) and the star are necessary to be taken into account when studying planet formation around very low-mass stars as they have an impact in the orbital evolution of close-in planets leading to the precession of their argument of periastron $\omega$ as well as their semi-major axis $a$ and eccentricity $e$ decays \citep[e.g.,][]{Bolmont2011,Sanchez2020}. It is also relevant to include the evolution of some physical parameters of the star, such as stellar radius and rotational period of the star as they change significantly during the first tens of million years \citep{Baraffe2015}.
The star-planet interactions considered in our simulations follow the equilibrium tide model from \citet{Hut1981}
and \citet{Eggleton1998}. We included tidal distortions and dissipation terms, considering
the tide raised by the host star on each embryo or planet and by each embryo or planet on the star and neglected the tide between embryos or planets as follows,

\begin{equation}
  \textbf{f}_{\textrm{tide}-\omega} = -3\frac{\mu}{r^8}\left[k_{2,\star}\left(\frac{M_\mathrm{p}}{M_\star}\right)R_\star^5 + k_{2,\mathrm{p}}\left(\frac{M_\star}{M_\mathrm{p}}\right)R_\mathrm{p}^5\right] \textbf{r},
\end{equation}

\begin{align*}
\textbf{f}_{\textrm{tide-ae}} = & -3\frac{\mu}{r^{10}} \left[
  \frac{M_{\textrm{p}}}{M_\star} k_{2,\star} \Delta \mathrm{t}_\star R_\star^{5}\left(2\textbf{r}(\textbf{r} \cdot \textbf{v}) + r^{2}(\textbf{r} \times \Omega_\star + \textbf{v})\right)\right]
\end{align*}
\begin{equation}
-3\frac{\mu}{r^{10}} \left[\frac{M_\star}{M_{\textrm{p}}}k_{2,\textrm{p}} \Delta \mathrm{t}_{\textrm{p}} R_\textrm{p}^{5}
       \left(2\textbf{r}(\textbf{r}\cdot\textbf{v}) + r^{2}(\textbf{r} \times \Omega_\textrm{p} + \textbf{v})\right)\right],
\end{equation}
where $k_{2,\star}=0.307$ and $k_{2,\textrm{p}}=0.305$ are the potential Love numbers of
degree 2 of the star and the embryos, respectively. For the star we assumed the Love number of a M dwarf and for the embryos, the Love number estimated for the Earth \citep{Bolmont2015}. The variable \textbf{r} is the position vector of the embryo with respect to the host star,
$\mu=G(M_\star + M_\textrm{p})$, $G$ is the gravitational constant and $M_\star$, $R_\star$,
$M_\textrm{p}$ and $R_\textrm{p}$ are, respectively, the masses and radius of the star and each embryo under the approximation that these objects can instantaneously
adjust their equilibrium shapes to the tidal force and considering only up to the
second-order harmonic distortions \citep{Darwin1908}. The velocity vector of the embryo $\textbf{v}$, $\Delta \mathrm{t}_\star$
and $\Delta \mathrm{t}_\mathrm{p}$ are the time-lag model constants
for the star and each embryo, respectively. The factors
$k_{2,\star}\Delta \mathrm{t}_\star$ and $k_{2,\mathrm{p}}\Delta \mathrm{t}_\mathrm{p}$
are related with the dissipation factors by
\begin{equation}
  k_{2,\mathrm{p}} \Delta \mathrm{t} _\mathrm{p} =  \frac{3R_\mathrm{p}^5\sigma_\mathrm{p}}{2G} \\
   k_{2,\star} \Delta \mathrm{t}_{\star} = \frac{3R_{\star}^5\sigma_{\star}}{2G}
\end{equation}  
with the dissipation factor for each embryo
$\sigma_\mathrm{p}=8.577\times10^{-43}~\mathrm{k^{-1} m^{-2} s^{-1}}$,
the same dissipation factor estimated for the Earth \citep{Neron1997},
and the dissipation factor of the host star
$\sigma_\star=2.006\times10^{-53}~\mathrm{k^{-1} m^{-2} s^{-1}}$, the same factor for a M dwarf
\citep{Hansen2010}.\\
We include the evolution of the stellar rotational velocity ($\Omega_\star$) following the prescription proposed by \citet{Bolmont2011} with the initial values taken from the observations within their work and \cite{Scholz2018} (see Appendix A),
and the evolution of the stellar radius fitting the models from \citet{Baraffe2015} for each stellar mass. In Figure \ref{fig:evRyP} we show the evolution of the stellar radius and rotational period of the stars ($P_\star=1/\Omega_\star$) of 0.1, 0.3 and 0.5 M$_\odot$. We also fixed
each embryo at pseudo-synchronization ($\Omega_\textrm{p}$) following \cite{Hut1981}. \\

We also incorporated in the simulations the acceleration corrections derived from the General
Relativity Theory  \citep{Einstein1916} proposed by \citet{Anderson1975} as follows,
\begin{equation}
  \textbf{f}_{\mathrm{GR}} = \frac{GM_\star}{r^3c^2}\left[\left(\frac{4GM_\star}{r} - \textbf{v}^2\right)\textbf{r}+4(\textbf{v}.\textbf{r})\textbf{v}\right],
  \label{eq:grav}
\end{equation}

with $c$ the speed of light. The corrections in the accelerations of the planets are mainly dependent on the stellar mass of the host star and lead to the precession of their periastron.\\

We refer to
\citet{Sanchez2020} and \citet{Sanchez2022} for a detailed description of planet-disk interactions as well as tidal and relativistic corrections, and their associated orbital decay timescales.

\section{Test simulations}
\label{sec:testsim}
\begin{figure}
    \centering    \includegraphics[width=8.5cm]{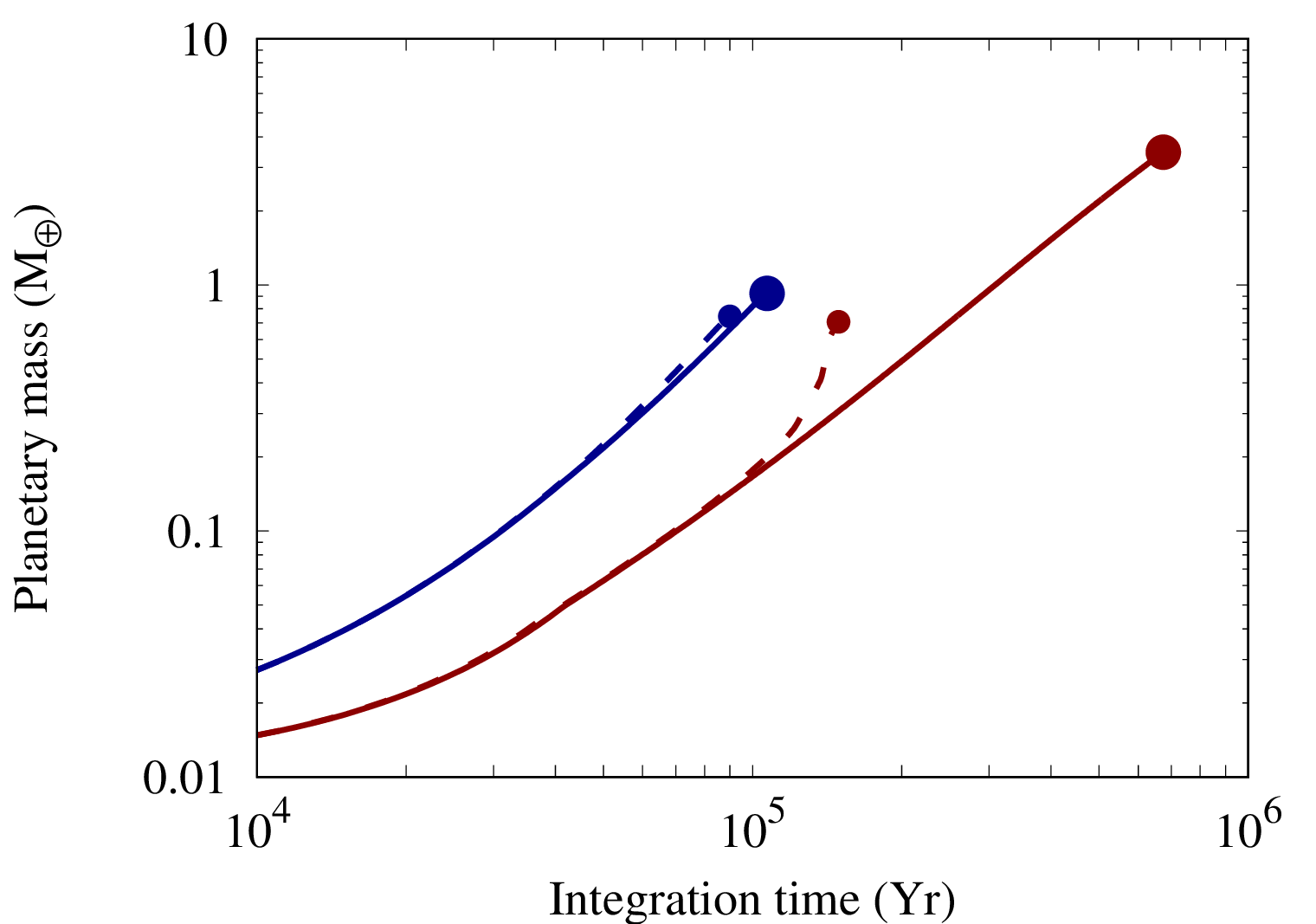}
    \caption{Evolution of planetary mass by pebble accretion of four planetary seeds. The initial mass of the planets is 0.01 M$_{\oplus}$ and all of them are in circular orbits. Two of them are initially located at $a=0.1$~au (blue lines) and the other two at $a=1$~au (red lines). Two of them are affected by gas disk interactions (dotted lines) and two of them are not affected by such interactions (solid lines). The red and blue dots represent the isolation mass reached by each planet.}
    \label{fig:pebbleaccretion-eg}
\end{figure}
\begin{figure}
    \centering    \includegraphics[width=8.5cm]{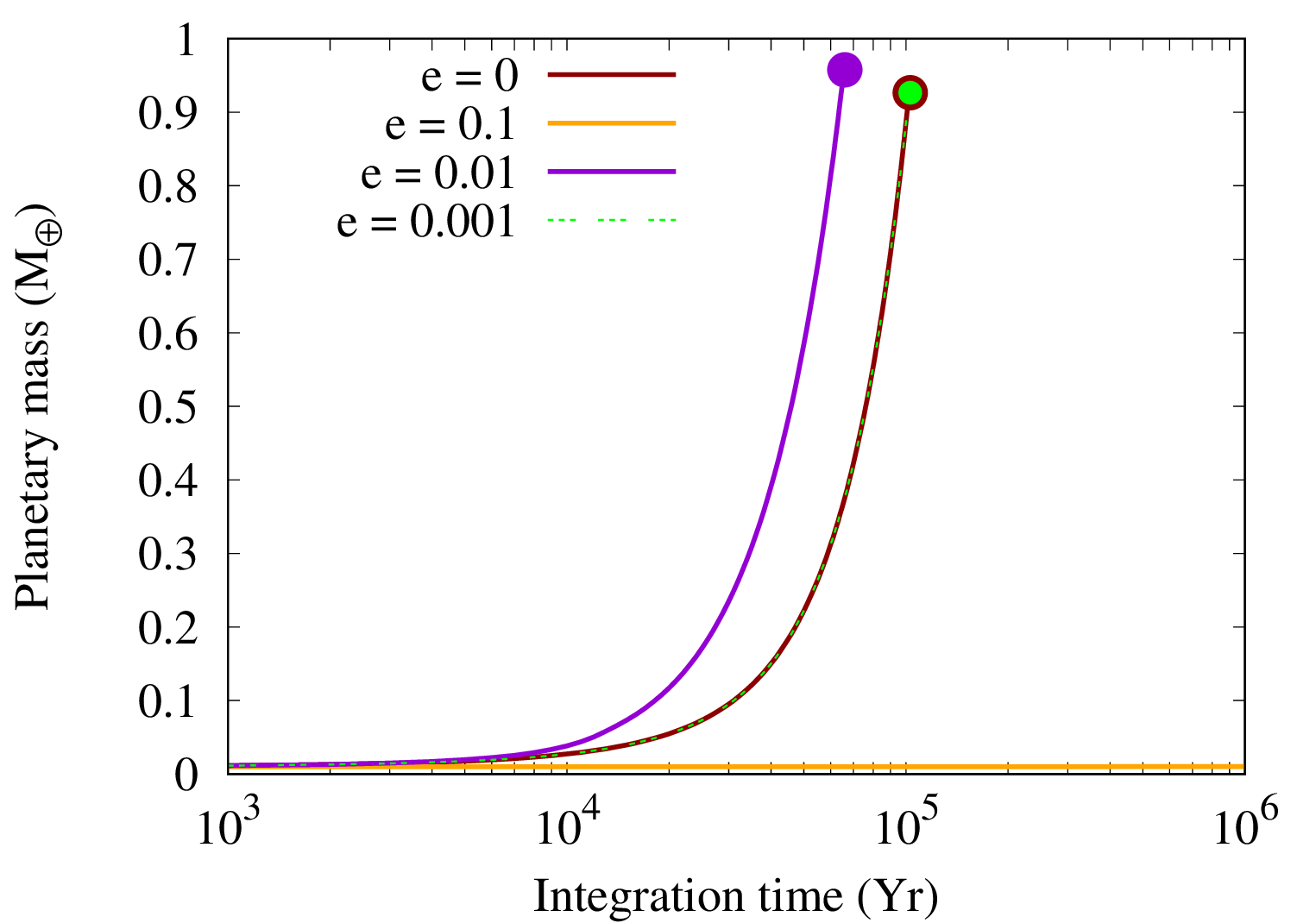}
    \caption{Evolution of planetary mass by pebble accretion for four planets of 0.01 M$_{\oplus}$ at $a=0.1$ for different eccentricities: $e=0$ (red line), $e=0.1$ (orange line), $e=0.01$ (violet line), and  $e=0.001$ (green dotted line).}
    \label{fig:pebbleaccretion-eg2}
\end{figure}
As a first test, we need to validate the pebble model (see Section \ref{sec:pebblemodel}) that was incorporated in the main routine of the N-body code \textsc{Mercury} \cite{Chambers1999} (see Appendix A for the gas disk model and Appendix B for the prescriptions used to treat planet-disk interactions). To do that, we tested if the isolation mass of a planet is reached in the expected time for different semi-major axis and eccentricities. 

First, we tested the isolation mass that could be reached regarding different semi-major axis. Thus, in Figure \ref{fig:pebbleaccretion-eg} we show the planetary mass evolution of four initial seeds of 0.01 M$_{\oplus}$, two of them initially located at $a=0.1$ au and the other two at $a=1$ au. All of them are in circular orbits and evolve by accreting pebbles with different efficiencies (following either Eq. (17) or (18)) associated with each of their locations in a disk of 1 Myr around a star of 0.1 M$_\odot$. Two of the planetary seeds are also affected by planet-disk interactions (see Section \ref{sec:planet-disk-interac}), which allow them to migrate inward at early stages. This makes them reach the isolation masses (see Eq. (24)) faster but with lower values than the ones the other two seeds that were not affected by the interactions with the gas disk could reach. The seed initially located at $a=0.1$~au not affected by planet-disk interactions reached an isolation mass of 1 M$_\oplus$ at 100,000 yr, while the one that was affected by such interactions reached a slightly smaller mass at 90,000 yr. On the other hand, the seed initially located at $a=1$~au not affected by the gas disk interactions reached an isolation mass of 3.5 M$_\oplus$ at $\sim$700.000 yr, while the one that was affected by the interactions reached a much lower value, $\sim1$ M$_\oplus$, at 150,000 yr. It can be seen that depending on the initial location of the planets in the disk, the isolation mass that can be reached is different, leading to more massive planets at further distances from the star. Moreover, if planet-disk interactions are taken into account, it will affect the isolation mass and the period of time when it could be reached.\\
Afterward, we analyzed the impact of the eccentricity in the efficiency of pebble accretion. Thus, in Figure \ref{fig:pebbleaccretion-eg2} the planetary mass evolution by pebble accretion is shown for a planet with initial 0.01 M$_{\oplus}$ located at $a=0.1$~au for different eccentricities: 0, 0.001, 0.01 and 0.1. It can be seen that the evolution for $e=0.001$ is equivalent to the one of a planet in a circular orbit, but the evolution is enhanced almost by a factor 2 if the planet has an $e=0.01$. On the contrary, if the eccentricity of the planet is $e=0.1$ the pebble accretion efficiency decay notably, and the planet takes several Myr to reach the isolation mass. This if because in the case of the highest eccentricity proposed in this example, the planet is accreting always in the 3D regime ($\epsilon_{3D} \sim 10^{-7}$, see Eq. (20)) while for the lower values of eccentricity, even though at the beginning the planets are accreting in the 3D regime, they quickly switch to the 2D regime ($\epsilon_{2D} \sim 10^{-2}-10^{-1}$, see Eq. (19)). This shows the importance of the value of the eccentricity in the efficiency of pebble accretion.



\end{appendix}

\end{document}